\documentclass{aa}

\usepackage[varg]{txfonts}

\usepackage{graphicx}
\usepackage{commath}
\usepackage{mathabx}
\usepackage{rotating}
\usepackage{multirow}
\usepackage{lscape}
\usepackage{booktabs}
\usepackage{natbib}
\bibpunct{(}{)}{;}{a}{}{,}

\begin{document}

\title{Demographics of disks around young very low-mass stars and brown dwarfs in Lupus}

\author{E. Sanchis\inst{1,2}
  \and L. Testi\inst{1, 3, 4}
  \and A. Natta\inst{4, 5}
  \and C. F. Manara\inst{1} 
  \and B. Ercolano\inst{2, 3}
  \and T. Preibisch\inst{2}
  \and T. Henning\inst{6}
  \and S. Facchini\inst{1} 
  \and A. Miotello\inst{1} 
  \and I. de Gregorio-Monsalvo\inst{7, 8} 
  \and C. Lopez\inst{8}
  \and K. Mu\v{z}i\'c\inst{9}
  \and I. Pascucci\inst{10, 11}
  \and A. Santamar\'ia-Miranda\inst{7, 12}
  \and A. Scholz\inst{13}
  \and M. Tazzari\inst{14}
  \and S. van Terwisga\inst{15}
  \and J. P. Williams\inst{16}
  }

\offprints{E. Sanchis, \email{esanchis@eso.org}}

\institute{European Southern Observatory, Karl-Schwarzschild-Strasse 2, D-85748 Garching bei M{\"u}nchen, Germany
  \and Universit{\"a}ts-Sternwarte, Ludwig-Maximilians-Universit{\"a}t M{\"u}nchen, Scheinerstrasse 1, D-81679 M{\"u}nchen, Germany
  \and Excellence Cluster Origins, Boltzmannstrasse 2, D-85748 Garching bei M{\"u}nchen, Germany
  \and INAF/Osservatorio Astrofisico di Arcetri, Largo E. Fermi 5, I-50125 Firenze, Italy
  \and School of Cosmic Physics, Dublin Institute for Advanced Studies, 31 Fitzwilliams Place, Dublin 2, Ireland
  \and Max Planck Institute for Astronomy, K{\"o}nigstuhl 17, D-69117, Heidelberg, Germany
  \and N\'ucleo Milenio Formaci\'on Planetaria - NPF, Universidad de Valpara\'iso, Av. Gran Breta\~na 1111,  Valpara\'iso, Chile
  \and Atacama Large Millimeter/Submillimeter Array, Joint ALMA Observatory, Alonso de C\'ordova 3107, Vitacura 763-0355, Santiago, Chile
  \and CENTRA, Faculdade de Ci\^{e}ncias, Universidade de Lisboa, Ed. C8, Campo Grande, 1749-016 Lisboa, Portugal
  \and Lunar and Planetary Laboratory, The University of Arizona, Tucson, AZ $85721$, USA
  \and Earths in Other Solar Systems Team, NASA Nexus for Exoplanet System Science, USA
  \and European Southern Observatory, 3107, Alonso de C\'ordova, Santiago de Chile
  \and SUPA, School of Physics \& Astronomy, University of St Andrews, North Haugh, St Andrews, KY$16$ $9$SS, UK
  \and Institute of Astronomy, University of Cambridge, Madingley Road, CB$3$ $0$HA Cambridge, UK
  \and Leiden Observatory, Leiden University, PO Box $9513$, $2300$ RA Leiden, The Netherlands
  \and Institute for Astronomy, University of Hawaii, Honolulu, HI $96822$, USA
  }

\date{Received ... / Accepted ...}

\abstract{We present new $890$ $\mu m$ continuum ALMA observations of 5 brown dwarfs (BDs) with infrared excess in Lupus I and III -- which, in combination with 4 BDs previously observed, allowed us to study the mm properties of the full known BD disk population of one star-forming region. 
Emission is detected in 5 out of the 9 BD disks. 
Dust disk mass, brightness profiles and characteristic sizes of the BD population are inferred from continuum flux and modeling of the observations. 
Only one source is marginally resolved, allowing for the determination of its disk characteristic size. 
We conduct a demographic comparison between the properties of disks around BDs and stars in Lupus. 
Due to the small sample size, we cannot confirm or disprove if the disk mass over stellar mass ratio drops for BDs, as suggested for Ophiuchus. 
Nevertheless, we find that all detected BD disks have an estimated dust mass between $0.2$ and $3.2$ $M_{\Earth}$; 
these results suggest that the measured solid masses in BD disks can not explain the observed exoplanet population, analogous to earlier findings on disks around more massive stars. 
Combined with the low estimated accretion rates, and assuming that the mm-continuum emission is a reliable proxy for the total disk mass, we derive ratios of $\dot{M}_{\mathrm{acc}} / M_{\mathrm{disk}}$ significantly lower than in disks around more massive stars. If confirmed with more accurate measurements of disk gas masses, this result could imply a qualitatively different relationship between disk masses and inward gas transport in BD disks.
}

\maketitle

%%%%%%%%%%%%%%%%%%%%%%%%%%%%%%%%%%%%%%%%%%%%%%%%%%
%%%%%%%%%%%%%%%%% BODY OF PAPER %%%%%%%%%%%%%%%%%%

\section{Introduction}\label{sec:intro}
Sub/mm wavelengths observations are particularly useful to study dust properties in protoplanetary disks, since the dust thermal emission of the outer disk, where the bulk of the dust mass resides, can be probed at these wavelengths \citep{testi+2014,andrews2015}. Demographic studies based on (sub)-mm-wavelengths surveys of the Class II population from nearby star-forming regions  \citep{ansdell+2016,barenfeld+2016,pascucci+2016,cox+2017,cieza+2018,cazzoletti+2019}, have found positive correlations between various disk properties: disk mass with stellar mass \citep[$M_{\mathrm{disk}}$-$M_{\star}$, ][]{andrews+2013,ansdell+2016,pascucci+2016}, disk size with luminosity \citep{andrews+2010,tazzari+2017A,tripathi+2017,andrews2018A}, and mass accretion rate onto the central star with the disk mass \citep[$\dot{M}_{\mathrm{acc}}$-$M_{\mathrm{disk}}$, ][]{manara+2016,mulders+2017,rosotti+2017}.

These relations are poorly constrained in the brown dwarf (BD) and very low-mass (VLM) stars regime, since these surveys focused primarily on disks around more massive stars. Therefore, observations at (sub-)mm wavelengths targeting BDs and VLM stars are necessary in order to extend these demographic studies, and to investigate their formation mechanisms and ongoing physical processes in their disks.

General interest on BDs and VLM stars has increased substantially thanks to the recent exoplanet discoveries around VLM objects. The most thrilling cases are Trappist-1 \citep{gillon+2017}, a $\sim 0.085$ $M_{\odot}$ VLM star that hosts 7 rocky planets in a packed orbital configuration, and Proxima B \citep{angladaescude+2016}, an Earth-like planet found in our closest neighbor ($M_{\star} = 0.12$ $M_{\odot}$), at a distance of only $\sim 1.3$ pc from us. These and other discoveries \citep[e.g., $2$M$1207$b and $2$M J$044144$b,][]{chauvin+2004,todorov+2010} suggest that planets orbiting BDs and VLMs may be a common outcome of their formation.

The study of the early stages of BDs and VLM stars is crucial to understand the viability of planet formation around these objects and to determine the properties of the potential planetary systems that may form. In \cite{klein+2003}, millimeter emission of dust from disks around BDs was detected for the first time. Like stars, BDs are often found surrounded by a protoplanetary disk in their early stages \citep{comeron+1998,NattaTesti2001,scholz2008}, where planet formation is expected to take place. 
The disk fraction for stellar and BD populations are found to be similar \citep{luhman2012}. Disk accretion \citep{jayawardhana+2003,ScholzEisloeffel2004,muzerolle+2003,muzerolle+2005} and outflows \citep{natta+2004,whelan+2005} also occur in the early stages of BDs, analogous to those around more massive stars.

In this work we conducted a systematic survey of BD disks in the Lupus star-forming region, observing the full known sample of BD disks from a single region with the Atacama Large Millimeter/submillimeter Array (ALMA) in the same band for the first time. Previous ALMA observations of BD disks studied incomplete samples of the known BD population of other regions \citep{testi+2016,vanderplas+2016,wardduong+2018}.

Dust disk masses, dust emission distribution profiles and dust disk characteristic sizes are determined from these observations. The last two properties are inferred from interferometric modeling of the dust disk emission. 
The characteristic size of the dusty disks is crucial to constrain the ongoing disk evolution processes (e.g., radial drift, grain growth). However, its determination is not straightforward. Firstly, the disk emission needs to be sufficiently resolved. For disks around BDs and VLM stars, this is only possible using state-of-art facilities, like ALMA, that allow for the high resolution and sensitivity required at these wavelengths. Besides, a general size definition is needed for a reliable comparison between observations and theoretical models. In this work we use the radius enclosing $68 \%$ of the object's emission distribution; this definition is representative of the physical size of the object \citep{tripathi+2017}, and independent of the model used to fit the observations. Another important disk property that can be derived from (sub-)mm observations is the disk mass. For the formation of rocky planets, the dust mass in disks should be larger than the mass of the resulting planets. However, comparing the mass derived from disk emission with the results from exoplanetary surveys, there is an apparent lack of material to produce the known planetary systems \citep{GreavesRice2010,williams2012,NajitaKenyon2014,mulders+2015,pascucci+2016,testi+2016,manara+2018}.

The inferred disk properties of the young BD population are compared to the properties of disks around stars in the same region, with the aim of testing whether the known relations for stars hold for disks around BDs. 
In addition to the BD observations, other seven T Tauri star (TTS) disks are characterized and modeled in this work for the first time.

The work is organized as follows: the target selection is described in Section~\ref{sec:sampleselection}. A summary of the observations and the data processing can be found in Section~\ref{sec:obsresults}. Section~\ref{sec:modeling} provides a description of the modeling employed for the disk properties determination, together with the modeling results. The demographic comparison of the inferred properties between BD and stellar disks, and the planet formation implications from the measured dust masses of the BD disks are discussed in Section~\ref{sec:discussion}, and the main conclusions of this work are included in Section~\ref{sec:conclusions}.

\section{Sample selection}\label{sec:sampleselection}
The list of selected targets of the Lupus BD disks survey (Cycle $5$; PI: L. Testi, Project ID: 2017.1.01243.S) encompasses all the known BDs in the Lupus region I-IV that were not observed previously with ALMA Band 7. Our population of BD disks in the Lupus star-forming region consists of all the known objects from the region census \citep{merin+2008,muzic+2014, muzic2015} that show excess emission in at least two mid-infrared bands (Spitzer IRAC/MIPS), that have been spectroscopically classified as spectral type (SpT) M6 or later, and with estimated mass $\leq0.09$ $M_{\odot}$ (down to $\sim 0.02$ $M_{\odot}$). 
Eleven sources in Lupus satisfied these selection criteria, seven were the targets for the new observations; the remaining four had been already observed in the Lupus disks survey (Cycle $2$; PI: J. Williams, Project ID: 2013.1.00220.S). Based on radial velocity analysis and X-shooter spectra, two sources in the sample, IRAS $15567$-$4141$ and SSTc$2$d J$160034.4$-$422540$, have been recently excluded from being Lupus members, and are likely background giants \citep{frasca+2017,alcala2017}; in agreement with the poorly constrained parallaxes from Gaia DR$2$ \citep{gaiacollaboration2018}. Therefore, these will not be discussed further in this paper; only five new targets are discussed. All the studied BDs are isolated systems, except SONYC-Lup3-7, which might form a very wide ($\sim 7^{\prime\prime}$) binary system with SONYC-Lup3-6, but this last object has no confirmed membership to the region \citep{muzic+2014}.

In Table~\ref{tab:targets}, we list all the sources analyzed in this work: the known BD population (5 objects from the new observations and 4 from previous observations), together with the $7$ disks around stars observed in the Lupus completion survey (Cycle $5$; PI: S.E. van Terwisga, Project ID: 2016.1.01239.S). The last two objects in the table are those that were observed but later excluded from the Lupus census \citep{frasca+2017,alcala2017}. 
The names, sky position and main stellar properties of the central stars are included in the Table~\ref{tab:targets}. 
The stellar properties shown in the table (SpT, effective temperature $T_{\mathrm{eff}}$, extinction in V-band $A_{V}$, stellar luminosity $L_{\star}$ and $M_{\star}$) were reported by \cite{alcala2014,alcala2017} and \cite{muzic+2014}. The methodology for the stellar luminosity derivation between these studies differs, nevertheless the agreement between the two methods is very good, as shown in \cite{manara+2016a}. $L_{\star}$ have been adjusted accounting for updated distances by Gaia DR$2$ \citep[distance estimated as the inverse of the parallax,][]{gaiacollaboration2018}. The stellar mass is derived from the pre-main sequence (MS) evolutionary models of \cite{baraffe+2015}, estimated from the position in the Hertzsprung–Russell (HR) diagram. For objects with estimated mass >$1.4$ $M_{\odot}$ and objects laying above the $1$ $\mathrm{Myr}$ isochrone, the tracks from \cite{siess+2000} are used instead. Stellar mass uncertainties are computed with a Monte Carlo approach \citep[as described in][]{alcala2017}, which takes into account the associated uncertainties of the stellar properties $L_{\star}$ and $T_{\mathrm{eff}}$ used to infer the mass.

\begin{table*}
\caption{Protoplanetary disks from Lupus modeled in this study.}             
\label{tab:targets}      
\centering                          
\small

\begin{tabular}{lcccccccccc}        

\hline\hline                 
Object & $\alpha$ & $\delta$  & Lupus & Distance & SpT & $T_{\mathrm{eff}}$ & $A_V$ & $L_{\star}$ & $M_{\star}$ & Notes \\ 
  & (J$2000$) & (J$2000$) & cloud & [pc] &  & [K] & [mag] & [$L_{\odot}$] & [$M_{\odot}$] &  \\ 
\hline  
\multicolumn{11}{c}{\it BDs from this survey:} \\ 
\hline

 J$154518.5$-$342125$ & $15$:$45$:$18.53$ & -$34$:$21$:$24.8$ & I & $152\pm4$ & M$6.5$ & $2935$ & $0.0$ & $0.04$ & $0.09\pm0.02$ & 1 \\ 

 SONYC-Lup3-7 & $16$:$08$:$59.53$ & -$38$:$56$:$27.6$ & III & $150 \pm 6$ & M$8.5$ & $2600$ & $0.0$ & $0.01$ & $0.02\pm0.01$ & 1, 4 \\ 

 Lup706 & $16$:$08$:$37.30$ & -$39$:$23$:$10.8$ & III & $158.5$ & M$7.5$ & $2795$ & $0.0$ & $0.002$ & $0.05\pm0.01$ & 1, 3, 5 \\ 

 AKC$2006$-$18$ & $15$:$41$:$40.82$ & -$33$:$45$:$19.0$ & I & $149 \pm 8$ & M$6.5$ & $2935$ & $0.0$ & $0.01$ & $0.07\pm0.02$ & 1 \\ 

 SONYC-Lup3-10 & $16$:$09$:$13.43$ & -$38$:$58$:$04.9$ & III & $158.5$ & M$8.8$ & $2650$ & $1.5$ & $0.003$ & $0.03\pm0.01$  & 2, 3 \\

\hline                                  
\multicolumn{11}{c}{\it BDs from Lupus disks survey \citep{ansdell+2016}:} \\ 
\hline

 Lup818s & $16$:$09$:$56.29$ & -$38$:$59$:$51.7$ & III & $157 \pm 3$ & M$6$ & $2990$ & $0.0$ & $0.02$ & $0.09\pm0.02$ & 1 \\ 

 J$161019.8$-$383607$ & $16$:$10$:$19.84$ & -$38$:$36$:$06.8$ & III & $159 \pm 3$ & M$6.5$ & $2935$ & $0.0$ & $0.04$ & $0.09\pm0.02$ & 1 \\ 

 J$160855.3$-$384848$ & $16$:$08$:$55.29$ & -$38$:$48$:$48.1$ & III & $158 \pm 3$ & M$6.5$ & $2935$ & $0.0$ & $0.05$ & $0.09\pm0.02$ & 1 \\ 

 Lup607 & $16$:$08$:$28.10$ & -$39$:$13$:$10.0$ & III & $175 \pm 6$ & M$6.5$ & $2935$ & $0.0$ & $0.05$ & $0.10\pm0.02$ & 1 \\

\hline
\multicolumn{11}{c}{\it Disks from Lupus completion survey:} \\ 
\hline
Sz$102$ & $16$:$08$:$29.71$ & -$39$:$03$:$11.0$ & III & $158.5$ & K$2$ & $4900$ & $0.7$ & $0.01$ & - & 1, 3, 5, 6 \\ 

 V$1094$ Sco & $16$:$08$:$36.18$ & -$39$:$23$:$02.5$ & III & $154 \pm 1$ & K$6$ & $4205$ & $1.7$ & $1.15$ & $0.86 \pm 0.18$ & 1 \\ 

 GQ~Lup & $15$:$49$:$12.10$ & -$35$:$39$:$05.1$ & I & $152 \pm 1$ & K$6$ & $4205$ & $0.7$ & $1.48$ & $0.85 \pm 0.17$ & 1 \\ 
 
 Sz$76$ & $15$:$49$:$30.74$ & -$35$:$49$:$51.4$ & I & $160 \pm 1$ & M$4$ & $3270$ & $0.2$ & $0.18$ & $0.23 \pm 0.04$ & 1 \\ 
 
 Sz$77$ & $15$:$51$:$46.96$ & -$35$:$56$:$44.1$ & I & $155 \pm 1$ & K$7$ & $4060$ & $0.0$ & $0.59$ & $0.75 \pm 0.15$ & 1 \\ 
 
 RXJ$1556.1$-$3655$ & $15$:$56$:$02.10$ & -$36$:$55$:$28.3$ & II & $158 \pm 1$ & M$1$ & $3705$ & $1.0$ & $0.26$ & $0.5 \pm 0.14$ & 1 \\ 
 
 EX~Lup & $16$:$03$:$05.49$ & -$40$:$18$:$25.4$ &  III & $158 \pm 1$ & M$0$ & $3850$ & $1.1$ & $0.76$ & $0.56 \pm 0.13$ & 1 \\

\hline                                   
\multicolumn{11}{c}{\it Observed objects rejected from being members of Lupus:} \\ 
\hline

 IRAS $15567$-$4141$ & $16$:$00$:$07.42$ & -$41$:$49$:$48.4$ & - & - & - & - & - & - & - & 7 \\ 

 J$160034.4$-$422540$ & $16$:$00$:$34.40$ & -$42$:$25$:$38.6$ & - & - & - & - & - & - & - & 7 \\

\hline
\end{tabular}

\tablefoot{\\
\tablefoottext{1}{Stellar properties from \cite{alcala2014,alcala2017} and adjusted to the new Gaia DR$2$ parallaxes.}\\
\tablefoottext{2}{Stellar properties from \cite{muzic+2014}, and adjusted to the Gaia DR$2$ parallaxes.}\\
\tablefoottext{3}{Gaia parallax unknown, mean distance of Lupus region considered.}\\
\tablefoottext{4}{Two sets of stellar properties \citep{alcala2017,muzic+2014}.}\\
\tablefoottext{5}{Sub-luminous object \citep[see][]{alcala2014,alcala2017}.}\\
\tablefoottext{6}{No estimation of the stellar mass since it falls below the zero-age main sequence in the HR diagram \citep[see][]{alcala2017}.}\\
\tablefoottext{7}{Categorized as background sources \citep{frasca+2017,alcala2017}.}
}
\end{table*}

The BD disk population is compared to the sample of young stellar objects (YSOs) in Lupus that have a protoplanetary disk and estimated stellar mass >$0.09$ $M_{\odot}$. Thanks to the inclusion of the 7 stellar disks from the Lupus completion survey, we have the largest sample of stellar disks in the Lupus clouds (I-IV) observed with ALMA in Band 7. The stellar disk population is assembled from different census of the region from 2MASS, Spitzer, and Herschel surveys \citep{hughes+1994,comeron+2008,merin+2008,mortier+2011,dunham+2015,bustamante+2015}. $82$ of these disks were observed with ALMA in Band 7 in the original Lupus disks survey (ALMA Cycle $2$; PI: J. Williams), IM~Lup and Sz~91 were observed separately (Cycle $2$; PI: I. Cleeves, Project ID: 2013.1.00226.S and Cycle $2$; PI: H. Canovas, Project ID: 2013.1.00663.S), and the seven remaining belong to the Lupus completion survey (Cycle $5$; PI: S.E. van Terwisga).

\begin{figure}
  \resizebox{\hsize}{!}{\includegraphics{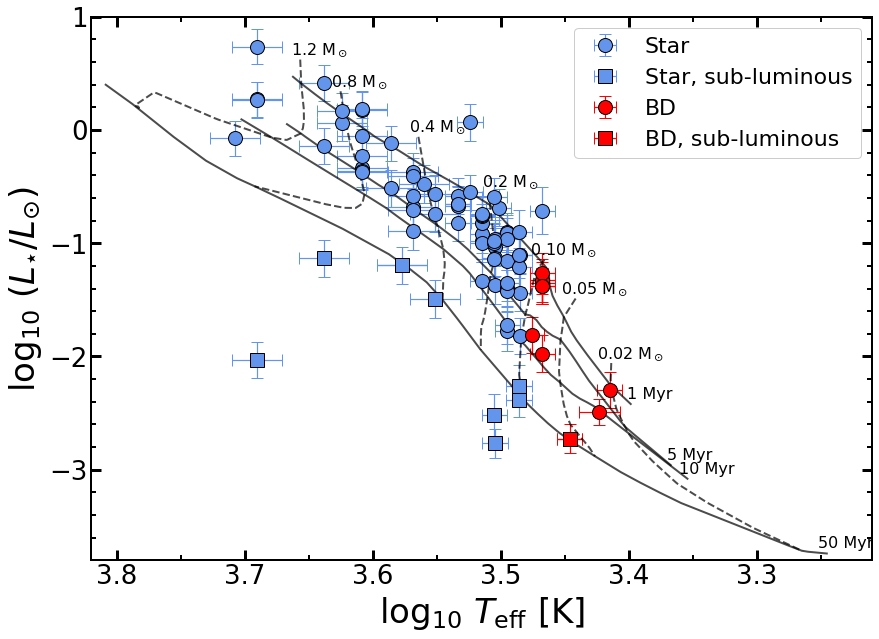}}
  \caption{Hertzsprung–Russell diagram for the studied BD and stellar populations from Lupus. The stellar luminosity is taken from the literature and re-adjusted to the new distance associated to the parallaxes from Gaia DR$2$ \citep{gaiacollaboration2018}. The BD population is shown in red, and the stellar population is indicated in blue. The pre-MS tracks of \cite{baraffe+2015} are overlaid in the figure. Objects with luminosities that would correspond to older ages than expected are considered sub-luminous and marked as squares. A number of points lay on top of each other (e.g., 4 BDs near the 1 Myr and the $0.1 M_{\odot}$ lines.)}
  \label{fig:hrd}
\end{figure}

Therefore, the BD and stellar disks samples of our demographic study consist of $9$ and $91$ sources respectively. The HR-diagram for the studied disk population is shown in Figure~\ref{fig:hrd}, using the stellar properties from previous studies as described before. 
The BD disks are marked in red, and the stellar disk population in blue. Sub-luminous sources, those with luminosities lower than those expected for YSOs of age $\sim 3$ Myr \citep[likely due to gray obscuration][]{alcala2014,alcala2017}, are illustrated with square symbols. From the re-adjusted $L_{\star}$ using the more accurate Gaia DR$2$ parallaxes, $2$ objects (J$16081497$-$3857145$ and J$16085373$-$3914367$) are now added to the list of sub-luminous objects of the region.

\section{Observations}\label{sec:obsresults}
ALMA observed our targets on 2018 April 1 and April 2 with $44$, and $42$ $12$ m antennas respectively. The baselines ranged between $15.1$ and $704.1$ m for the array configuration of the first day, and between $15.1$ and $629.2$ m for the configuration of the second day. Four spectral windows were set for the continuum observation of the targets, centered at $334.432$, $336.321$, $345.889$ and $347.821$ GHz and bandwidths of $2$, $1.875$, $1.875$ and $0.938$ GHz respectively (total receivers bandwidth of $\sim6.688$ GHz). 
The calibrators for the observations were J$1517$-$2422$ for flux and passband, and J$1610$-$3958$ for the complex gain calibration, the same in both executions. The flux density scale accuracy is expected to be of 10$\%$ for observations of the Lupus BD disks survey. 12 scans of $60$-$62$ seconds duration each were performed for every target, for a total integration time of more than $12$ minutes per source.

\begin{figure*}
  \resizebox{\hsize}{!}{\includegraphics{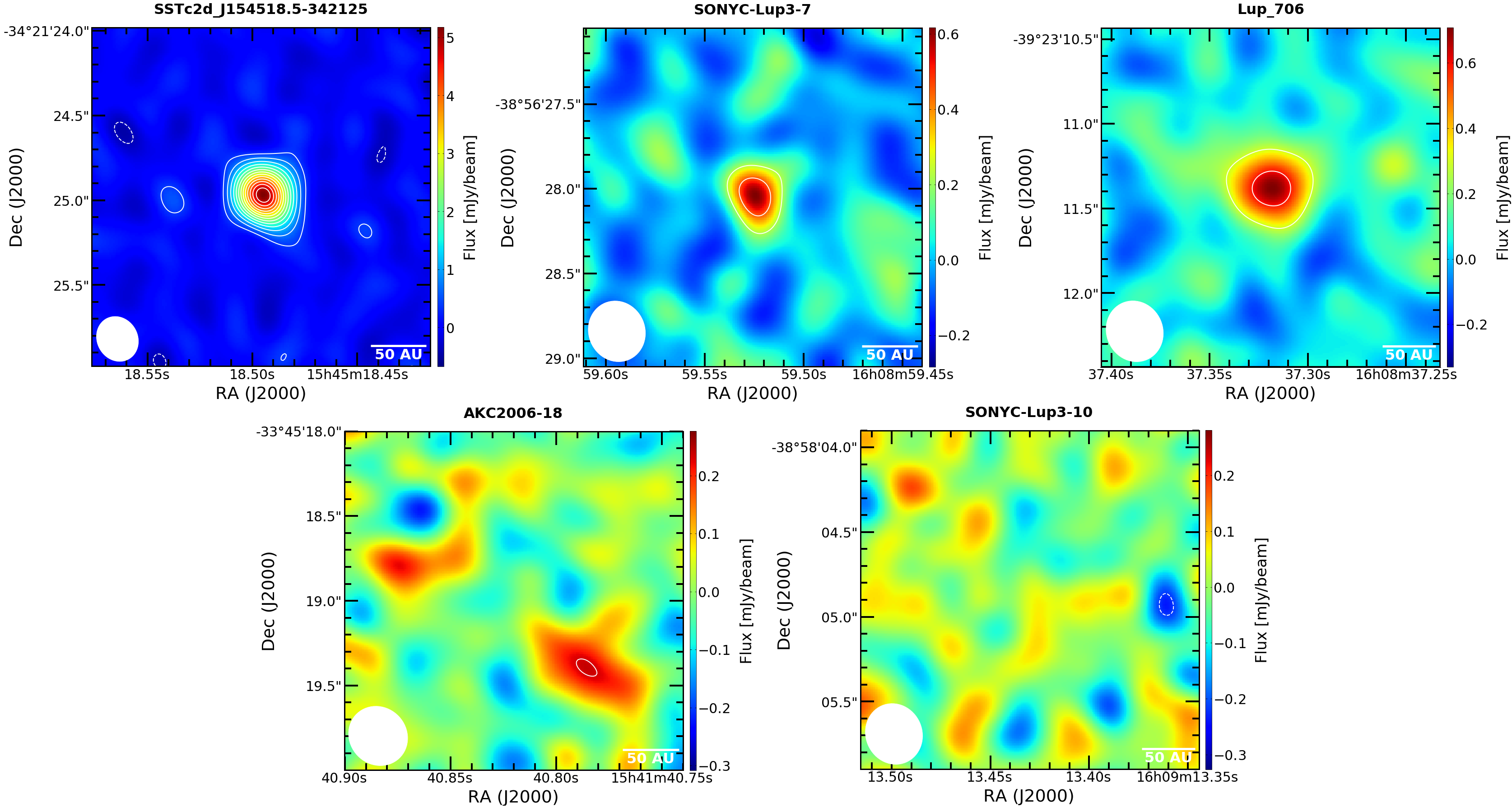}}
  \caption{Dust continuum images at $890$ $\mu m$ of the Lupus BDs disks survey from ALMA Band $7$ observations. The beam size FWHM is $0.27^{\prime\prime} \times 0.24^{\prime\prime}$ for the J$154518.5$-$342125$ map (robust parameter of $-1$), and $0.36^{\prime\prime} \times 0.33^{\prime\prime}$ for the rest of the maps (robustness $=+0.5$). The average beam position angle is $PA = 28^{\circ}$. The contours are drawn at increasing (or decreasing) 3$\sigma$ intervals as solid (dashed) lines.}
  \label{fig:obscontinuum}
\end{figure*}

The CASA 5.3.0 software has been used for the interferometric visibilities calibration and imaging. The continuum maps are produced using the channels free from spectral line emission, with Briggs weighting of the visibilities ($-1.0$ and $+0.5$ robustness for resolved and unresolved objects respectively). 
None of the sources are bright enough to perform self-calibration. The full width at half maximum (FWHM) of the synthesized beam is 
$0.27^{\prime\prime} \times 0.24^{\prime\prime}$ for robust parameter $=-1$, and $0.36^{\prime\prime} \times 0.33^{\prime\prime}$ for robustness $=+0.5$, 
with average position angle ($PA$) of $28^{\circ}$. The continuum maps of the 5 BD disks observed are shown in Figure~\ref{fig:obscontinuum}. The sensitivity for the BD disks survey is improved by a factor of $\sim 3$ with respect to the previous Lupus disks surveys, this allowed us to detect fainter emission. Three BD disks are detected, J$154518.5$-$342125$, SONYC-Lup3-7 and Lup706, with respective signal-to-noise (S/N) of $42$, $8$ and $7$. Emission is not detected from the two other disks (AKC$2006$-$18$ and SONYC-Lup3-10) nor from the two background objects (IRAS $15567$-$4141$ and J$160034.4$-$422540$).

The main results of the observations are reported in Table~\ref{tab:fluxes}. This table includes the total disk flux, peak intensity and the rms of the image. These values from the observations are obtained using identical methodology to the results presented in \cite{ansdell+2016} for the original Lupus disks survey. 
The continuum flux is inferred from the \textit{uvmodelfit} task in CASA: emission is fitted with an elliptical gaussian in case that resulting FWHM along the major axis from the fit is at least five times its uncertainty, otherwise the emission is fitted as a point source. For sources with resolved structure, the flux is obtained from a curve of growth method with increasing circular apertures, centered at the peak emission of the object. The rms is computed from a 4"-9" radius annulus centered on the detected emission, or on the expected source position if no disk emission is detected. The flux upper-limits of the non-detected BD disks are displayed in the $F_{\mathrm{cont}}$ column of the table. In this work, upper-limits of non-detected BD and stellar disks are computed as three times the rms level above the measured flux within the beam size are (centered at the expected source position), this corresponds to a $99.87\%$ confidence level. This differs slightly from the upper-limits in \cite{ansdell+2016}, considered to be $3\sigma$.

\begin{table}
\caption{Inferred fluxes for all the studied disks.}             
\label{tab:fluxes}      
\centering              
\begin{tabular}{lccr}   
\hline\hline            
Object & $F_{\mathrm{cont}}$ & $I_{\mathrm{peak}}$ & rms  \\
 & [$\mathrm{mJy}$] & [$\mathrm{mJy} / \mathrm{beam}$] & [$\mathrm{mJy} / \mathrm{beam}$] \\    
\hline 
\multicolumn{4}{c}{\it BDs from this survey:} \\ 
\hline
J$154518.5$-$342125$ & $5.65$ & $5.18$ & $0.12$  \\
SONYC-Lup$3$-$7$ & $0.52$ & $0.62$ & $0.07$  \\
Lup$706$ & $0.79$ & $0.71$ & $0.10$  \\
AKC$2006$-$18$ & $< 0.23$ & - & $0.08$  \\
SONYC-Lup$3$-$10$ & $< 0.24$ & - & $0.07$  \\

\hline 
\multicolumn{4}{c}{\it BDs from Lupus disks survey:} \\ 
\hline
Lup$818$s & $7.44$ & $7.44$ & $0.24$  \\
J$161019.8$-$383607$ & $< 1.24$ & - & $0.23$  \\
J$160855.3$-$384848$ & $1.81$ & $1.81$ & $0.26$  \\
Lup$607$ & $< 0.95$ & - & $0.24$  \\

\hline
\multicolumn{4}{c}{\it Disks from Lupus disks completion survey:} \\ 
\hline
Sz$102$ & $14.69$ & $7.81$ & $0.34$ \\
V$1094$ Sco & $553.17$ & $23.43$ & $0.37$ \\
GQ~Lup & $78.00$ & $42.36$ & $0.34$ \\
Sz$76$ & $12.00$ & $4.88$ & $0.35$ \\
Sz$77$ & $4.88$ & $3.88$ & $0.20$ \\
RXJ$1556.1$-$3655$ & $57.52$ & $16.70$ & $0.35$ \\
EX~Lup & $41.78$ & $9.80$ & $0.35$ \\

\hline
\end{tabular}

\tablefoot{
From top to bottom: the $9$ protoplanetary disks around BDs ($5$ from the new Lupus BDs disks survey) and $7$ additional disks from the Lupus disks completion survey. The columns are  the total disk flux ($F_{\mathrm{cont}}$), peak intensity ($I_{\mathrm{peak}}$) of each object and the rms of their continuum maps.
}

\end{table}

We have also derived the flux densities and image rms for the sources observed with ALMA in Band $7$ in the Lupus completion survey \citep[see also][]{vanterwisga2018A,vanterwisga+2019}. For these sources we have followed the same procedure as for the BDs disks. As reported by \cite{vanterwisga2018A}, the flux calibrator used for these disks in Band $7$ was highly variable, thus the absolute flux densities are very uncertain. To alleviate this problem, \cite{vanterwisga2018A} compared the flux of one of these sources (GQ~Lup) to a previous observation with high S/N and reliable flux calibration \citep{macgregor+2017}, obtaining a fluxes ratio of $1.3\pm0.009$ between both observations. Since GQ~Lup and the rest of the disks of the Lupus completion survey were observed at the same day, we applied that factor to the measured fluxes of all these sources. The results are included in Table~\ref{tab:fluxes}. For the observations results from the Lupus disks survey (ALMA Project ID: 2013.1.00220.S), we refer to \cite{ansdell+2016}.

\section{Modeling}\label{sec:modeling}
Previous work characterizing interferometric observations of protoplanetary disks modeled the continuum emission with either physical or empirical models. A physical model commonly used is the two-layer approximation model \citep{ricci+2014,testi+2016,tazzari+2017A}. 
Although this model can successfully describe the spectral enegrgy distribution (SED) of TTS disks \citep{ChiangGoldreich1997,dullemond+2001}, the model assumes a simplified physical structure of the disk. In order to provide an observational characterization of the emission, we prefer to fit empirical analytical functions to the emission profile, to allow a more straightforward comparison of the disk properties.

We model the extended disks around stars with the Nuker profile, used in \cite{andrews2018A} to characterize the Lupus disks observed with ALMA in Band $7$ from \cite{ansdell+2016}. 
By using the Nuker profile we ensure the homogeneity on the characterization of the Lupus disks sample, which is a key aspect of the demographic study discussed in the next section. 
We follow the modeling described in \cite{tripathi+2017}, using the \cite{lauer+1995} formulation:
\begin{equation}\label{eq:nuker}
I_{\nu} (\rho) = I_{0} \cdot {\bigg( \frac{\rho}{\rho_{t}} \bigg)}^{-\gamma} \cdot { \bigg[ 1 + { \bigg( \frac{\rho}{\rho_{t}} \bigg) }^{\alpha} \bigg] }^{(\gamma - \beta) / \alpha}  \mathrm{,}
\end{equation}
where ${\rho}_{t}$ is the transition radius that sets the boundary between the inner and outer regimes of the radial profile, $\gamma$ and $\beta$ are the inner and outer disk slopes and $\alpha$ is a factor that determines the smoothness of the transition between both regimes. The disk is assumed to be azimuthally symmetric. The total number of parameters used to model the extended disks with the Nuker profile are $9$: ${\rho}_{t}$, $\gamma$, $\beta$, $\alpha$, the total disk flux density $F_{\mathrm{tot}}$, and four additional geometrical parameters connected to the observation: inclination of the disk in the sky towards the observer ($i$, $0^\circ$ face-on disk, $90^\circ$ edge-on), the position angle in the sky plane ($PA$, defined east of north), and right ascension and declination off-sets to the phase center of the observations ($\Delta \mathrm{RA}$ and $\Delta \mathrm{Dec}$).

A simple parametrized gaussian function has been used to fit the BD disks, to reduce the number of free parameters, since the emission maps of these objects are extremely compact; only one BD disk is marginally resolved. This function can be used to model moderate-resolution observations to characterize the disk brightness profile and its size, 
as shown in Appendix~\ref{sec:appendix_radiustest} (see Figure~\ref{fig:severalfits}) where the Gaussian profile together with other models were used to fit the disk emission around RXJ$1556.1$-$3655$: the resulting profiles are alike and the values of the chosen size definition are indistinguishable. 
The expression of the gaussian function used to model the BDs surface brightness profile is:
\begin{equation}\label{eq:gaussian}
I_{\nu} (\rho) = I_{0} \cdot \exp{ \bigg[-0.5 \cdot { \bigg( \frac{\rho}{\sigma} \bigg) }^{2} \bigg] }  \mathrm{,}
\end{equation}
where $\rho$ is the projected radius in the sky in arcsec, $I_{0}$ is a normalization factor, and $\sigma$ is the standard deviation of the gaussian profile. The disk is assumed to be azimuthally symmetric. The free parameters to model the BD disks are $6$ in total, two from the gaussian model ($I_{0}$ and $\sigma$), together with the observational parameters analogous to the Nuker model ($i$, $PA$, $\Delta \mathrm{RA}$ and $\Delta \mathrm{Dec}$).

To perform the fits, the \texttt{Galario} \citep{tazzari+2018} and \texttt{emcee} \citep{foremanmackey+2013} packages have been used. For a detailed description of the methodology we refer to \cite{tazzari+2016,tazzari+2017A}. To run the affine invariant Markov chain Monte Carlo (MCMC) \cite{GoodmanWeare2010}, $200$ walkers have been used, in order to investigate the parameter space for each disk ($\gtrsim 20$-$30$ walkers for each parameter); the computation ran for $20000$ steps per walker, this guaranteed convergence in all fitted disks.

\subsection{Size definition}\label{sec:sizedef}
An appropriate definition of the disk size is necessary for a proper characterization of disks from observations, and for its comparison to theoretical models. An approach commonly used is to extract the radius from the models used to fit the emission. The transition radius $\rho_{t}$ from the Nuker profile definition (Equation~\ref{eq:nuker}) provides misleading information of the disk size, since systems with very different architectures and extents might have similar values of $\rho_{t}$. When fitting to a power-law with an exponential cut-off, an analogous problem arises if the cut-off radius $\rho_{c}$ from \cite{guilloteau+2011} parametrization is used. 
The outer radius $R_{\mathrm{out}}$ was used in \cite{ricci+2014} and in \cite{testi+2016} to fit BDs disks observations, defined as the outermost radius of their modeled surface density.

All these definitions may be useful in specific studies, but are not suitable for a general characterization of disk sizes from their emission. 
A more convenient size definition is the radius enclosing a certain fraction of the total disk emission. 
This definition with a fraction of $68 \%$ of the total disk emission has been used in many recent works \citep[e.g., ][]{tripathi+2017,andrews2018A,facchini+2019,long+2019,manara+2019}, in \cite{tazzari+2017A} the $95 \%$ of the total disk emission was used. To avoid confusion with the different terminology used in the literature, we simply refer to them as $68 \%$ ($R_{68\%}$) and $95 \%$ ($R_{95\%}$) flux radii. In the nomenclature along this work, $R$ refers to the radius in the system reference frame (typically in $\mathrm{au}$), while $\rho$ stands for the projected radius in the sky plane in arcsec.

We tested both $R_{68\%}$ and $R_{95\%}$ to determine the quality of each radius as the characteristic size for the disk emission (details in Appendix~\ref{sec:appendix_radiustest}). For this test we fitted the same disk to various models and inferred $R_{68\%}$ and $R_{95\%}$ for each model. This test shows that the dispersion on $R_{68\%}$ is much smaller than for $R_{95\%}$. Thus, we consider $R_{68\%}$ as the most reliable size definition for our sample. Nevertheless, in the modeling results of the disks (Table~\ref{tab:fitresults}), we include both $R_{68\%}$ and $R_{95\%}$ for completeness.

Additionally, we fitted several disks that were previously modeled in \cite{andrews2018A}, in order to test the proper functioning of our modeling tool. This additional test is also included in Appendix~\ref{sec:appendix_radiustest}; the resulting $\rho_{68\%}$ from this work and from \cite{andrews2018A} are in very good agreement.

\subsection{Dust disk masses}\label{sec:diskmasses}
To ease the comparison with the existing surveys of disks around stars in the Lupus clouds \citep{ansdell+2016}, we provide an estimate of the disk dust masses using the simplifying assumption of optically thin emission and using an average temperature of $20$ K. We note that these assumptions may lead to underestimating the disk mass in case the emission is optically thick or the average temperature is lower the assumed value.

For the dust mass determination, assumptions on the dust temperature and opacity are needed. 
A dependence of the dust temperature with the stellar luminosity was first proposed from theoretical grounds \citep[e.g.,][]{yorke+1993,sonnhalter+1995}. More recently, from radiative transfer modeling of mm observations, \cite{andrews+2013} proposed a single-valued mean temperature for each disk that could be used to estimate the disk mass for objects with $L_{\star} \in [0.1, 100]$ $L_{\odot}$. Soon after, \cite{vanderplas+2016} suggested a more flattened relation for VLM objects. 
In \cite{ballering+2019}, the correlation of disk temperature with the stellar luminosity was derived using simplistic radiative transfer models from SED fitting of Taurus disks. \cite{daemgen+2016} and \cite{tazzari+2017A} showed that, depending on assumptions on disk size and vertical structure, similar $T_{\mathrm{dust}}$ even with very different luminosities are compatible with the data. 
From the different studies, it is unclear whether there is a simple and general relation between dust temperature and stellar properties. Using additional relations of the temperature with other stellar properties as the proposed by \cite{andrews+2013,vanderplas+2016} might introduce spurious results in our analysis, or even erase possible relations between different disk properties.

We therefore compute the dust mass of each BD and stellar disk assuming a constant dust opacity of $\kappa_{890\mu m} = 2$ $\mathrm{cm}^2$ $\mathrm{g}^{-1}$, following previous ALMA Band $7$ observations for VLMs and BDs \citep{ricci+2014,testi+2016}, and an averaged dust temperature of $T_{\mathrm{dust}} = 20$ K, as in \cite{pascucci+2016,ansdell+2016,ansdell+2018}. This value is the median temperature for protoplanetary disks in the Taurus region \citep{andrewswilliams2005}. Nevertheless, in Appendix~\ref{sec:appendix_otherrelations} we show the main demographic results of this work using the dust temperature and stellar luminosity relations from \cite{andrews+2013,vanderplas+2016}.

\subsection{Modeling results}\label{sec:modelresults}
\begin{figure}
  \resizebox{\hsize}{!}{\includegraphics{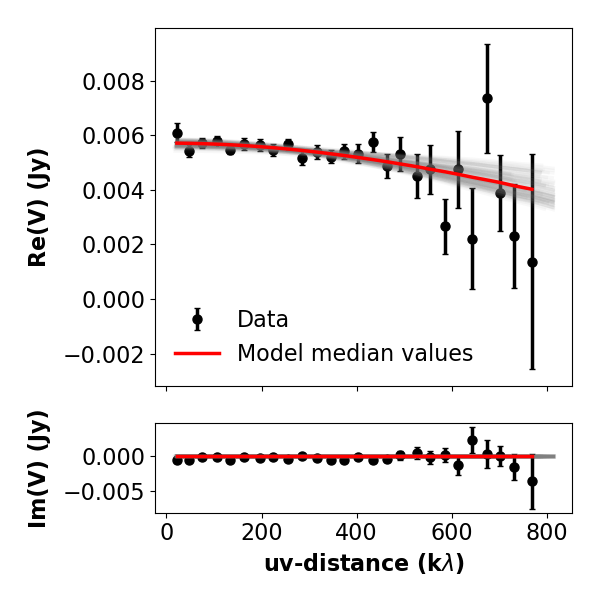}}
  \caption{Observed and model visibilities of J$154518.5$-$342125$, plotted as real and imaginary part as a function of the baseline (in k$\lambda$). The data from the observations are plotted as black data points with error bars, the model with the lowest ${\chi}^{2}$ is shown as solid red curve, and a random set of models from the parameter space investigation are drawn as gray curves. Figure made with the \texttt{uvplot} Python package \cite{uvplot_mtazzari}.}
  \label{fig:j1545uvfit}
\end{figure}

\begin{figure}
  \resizebox{\hsize}{!}{\includegraphics{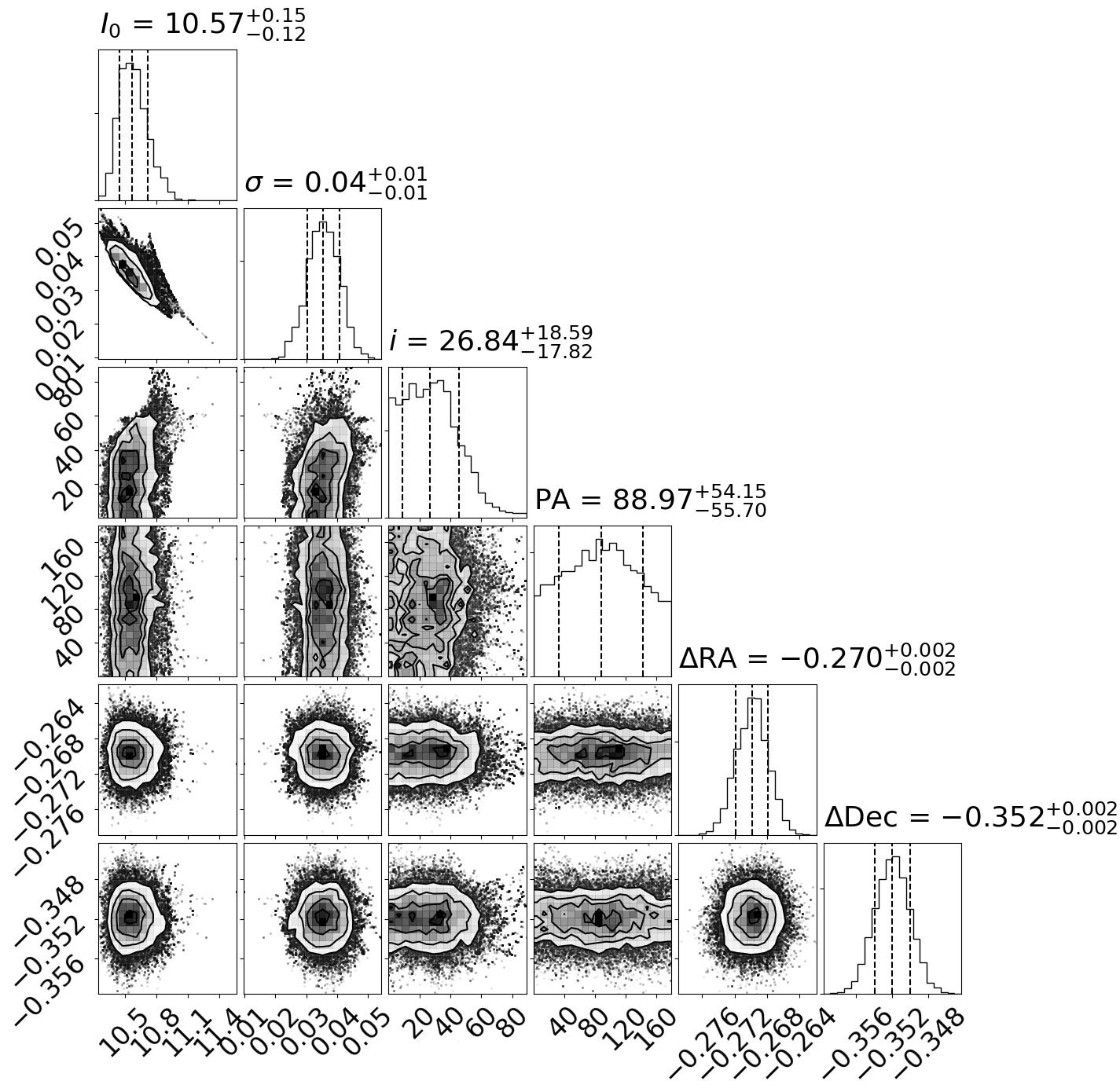}}
  \caption{1D and 2D histograms of the free parameters used to model J$154518.5$-$342125$ ALMA observations, resulting from the MCMC analysis. The marginalized PDFs of the parameters are plotted in the top panels, including the 16th, 50th and 84th percentiles as vertical dashed lines. The $I_{0}$ and $\sigma$ parameters are connected to the gaussian model used, as defined in Equation~\ref{eq:gaussian}; $i$, $PA$, $\Delta \mathrm{RA}$ and $\Delta \mathrm{Dec}$ are geometrical parameters linked to the observations.}
  \label{fig:j1545corner}
\end{figure}

\begin{figure}
  \resizebox{\hsize}{!}{\includegraphics{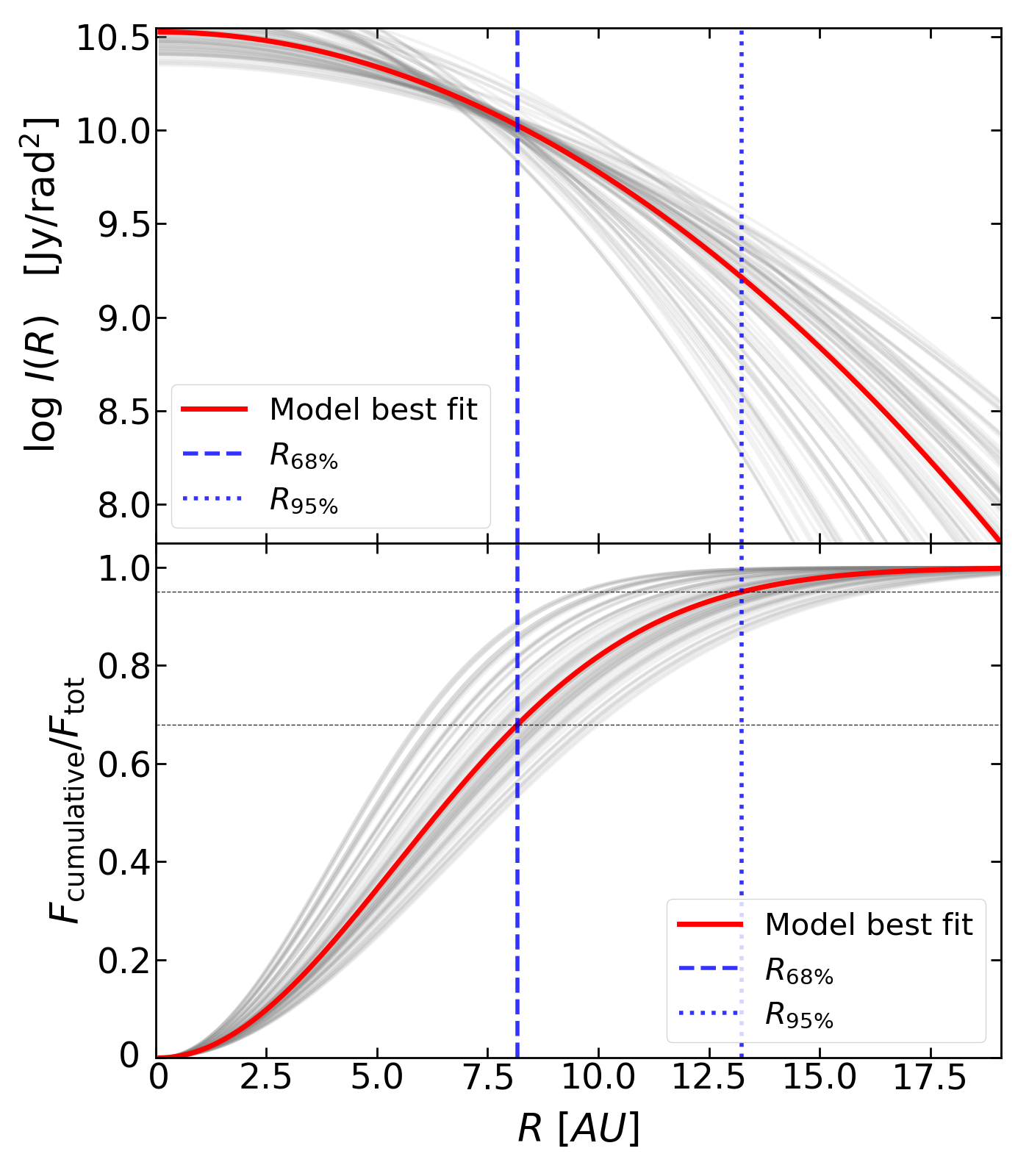}}
  \caption{Radial brightness profile (top panel) and the associated cumulative flux (bottom) for J$154518.5$-$342125$ disk, resulting from the gaussian model used to fit the observed visibilities. The lowest ${\chi}^{2}$ model and a random subset of models are drawn in both panels as a red and thin gray curves respectively.}
  \label{fig:j1545fluxes}
\end{figure}

The results of the modeling and the derived properties of dust mass and disk size are presented in this section. The results for J$154518.5$-$342125$ are shown in Figures~\ref{fig:j1545uvfit}, \ref{fig:j1545corner}, \ref{fig:j1545fluxes}, \ref{fig:j1545obsmodres}, while the results for the remaining disks which fits converged are in Appendix~\ref{sec:appendix_fits}.

In Figure~\ref{fig:j1545uvfit} we show the real and imaginary part of the observed and modeled visibilities as a function of baseline. The visibilities were first centered using $\Delta \mathrm{RA}$ and $\Delta \mathrm{Dec}$ from the model with lowest ${\chi}^{2}$, then de-projected taking $i$ and $PA$ \citep[for a detailed description, see][]{tazzari+2017A,tazzari+2018}.

The posterior probability distribution functions (PDFs) of the free parameters (of the gaussian model, or from the Nuker profile), are shown in the top sub-panels of Figure~\ref{fig:j1545corner}. Vertical dashed lines represent the median values of each histogram, used as the best fitted parameter value, and the 16th and 84th percentiles, used to infer the lower and upper values of the uncertainties. From the figure, $I_0$, $\sigma$ and the sky plane off-sets are well determined, 
while the inclination and position angle are both degenerated, and their values are only poorly constrained. The rest of the sub-panels in the figure show the 2D histograms for the pairs of parameters, which indicate possible correlations between the different parameters. 
In a few disks of the Lupus completion survey, the smoothness parameter ($\alpha$), and/or the inner and outer slopes ($\beta$, $\gamma$) of the Nuker profile can not be appropriately constrained from moderate resolution observations (see corner plots of the fits in Appendix~\ref{sec:appendix_fits}). 
This limitation does not affect the characterization of their disk sizes.

\begin{figure}
  \resizebox{\hsize}{!}{\includegraphics{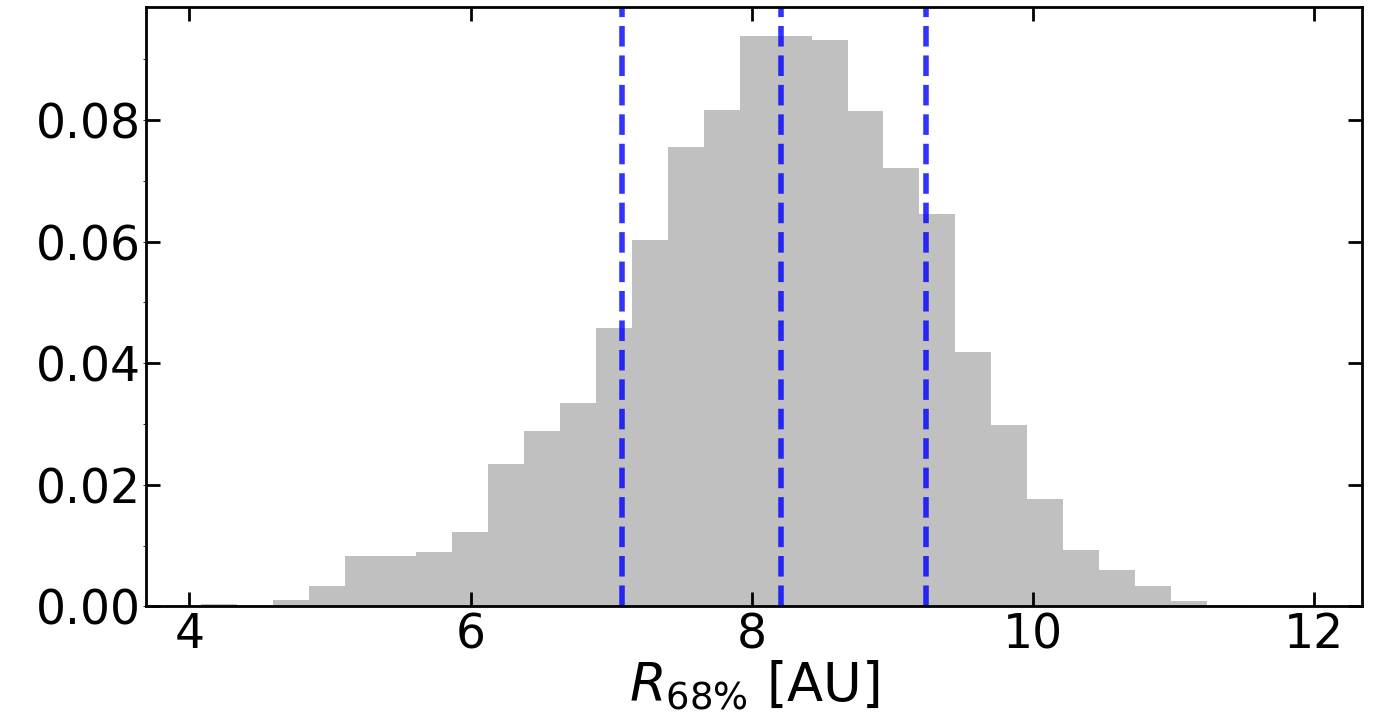}}
  \caption{
  PDF of the $R_{68\%}$ radius for the disk around J$154518.5$-$342125$. $R_{68\%}$ is computed for each model of the parameter space investigation, the value of the radius is taken as the median of the PDF, while upper/lower error are the median $\pm 1\sigma$ of the distribution. The three values are represented as vertical dashed lines.
  }
  \label{fig:j1545r68}
\end{figure}

The modeled emission distribution of the disk is shown in the top panel of Figure~\ref{fig:j1545fluxes}. On the bottom panel, we show the normalized cumulative flux $f_{\mathrm{cumul}}$, derived from Equation~\ref{eq:fcumul} for the respective models of the top panel. In both plots, the inferred values of $R_{68\%}$ and $R_{95\%}$ radii are included as vertical dashed and dotted lines. These radii are computed for each model of the MCMC; the final values of $R_{68\%}$ and $R_{95\%}$ are the median of their respective PDFs, with upper and lower errors as the median $\pm 1\sigma$ (example of a ${R}_{68 \%}$ PDF in Figure~\ref{fig:j1545r68}).

\begin{figure*}
  \resizebox{\hsize}{!}{\includegraphics{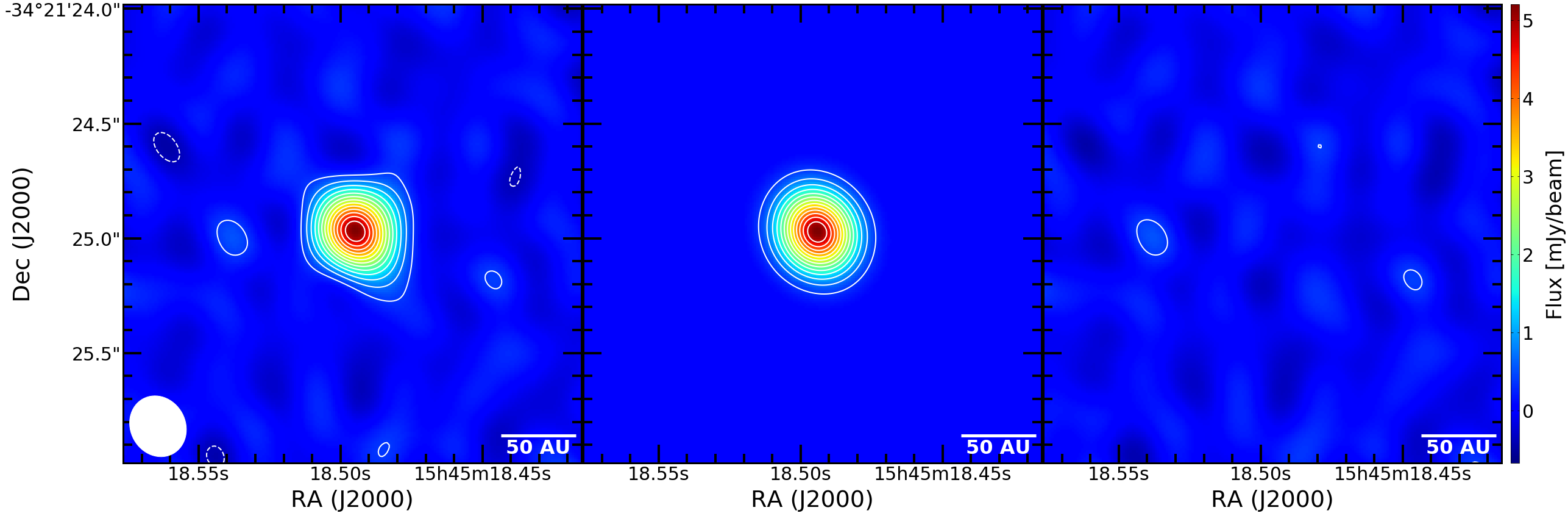}}
  \caption{Observed (left), model (center) and respective residuals (right panel) for the continuum emission of J$154518.5$-$342125$ observed with ALMA. The modeled emission map is reconstructed from the synthetic visibilities with the lowest ${\chi}^{2}$ from the interferometric modeling. The contours are drawn at increasing (or decreasing) 3$\sigma$ intervals as solid (dashed) lines.}
  \label{fig:j1545obsmodres}
\end{figure*}
In Figure~\ref{fig:j1545obsmodres}, we show the reconstructed continuum emission of the source in the sky plane: in the left panel the observed emission, in the central panel the reconstructed emission corresponding to the model with lowest ${\chi}^{2}$, and in the right panel the residuals. Residuals are at noise level on all fitted disks, indicating that the model represents the observation faithfully.

The results of the free parameters from the fits and the derived disk properties can be found in Table~\ref{tab:fitresults}. The values shown are the median of their respective PDF, lower (upper) uncertainties are obtained from the range between median and $16$th ($84$th) percentiles of their posterior distribution. The missing values are for the cases in which the fit did not find a clear convergence. A note on the source Sz~$102$, from our fit results, gray obscuration due to its inclination ($\sim60^{\circ}$) would not explain its sub-luminous nature. Strong episodic accretion as suggested by \cite{BaraffeChabrier2010} could explain its luminosity: this effect reduces the radius of the star, increasing its temperature and resulting in different pre-MS path and a relatively lower luminosity. Another viable explanation would be a misalignment of the inner and the outer disk.
\begin{sidewaystable*}
\caption{Results from the modeling of the studied disks, together with inferred sizes ($R_{68\%}$ and $R_{95\%}$, in $\mathrm{au}$), and total disk dust mass ($M_{\mathrm{dust}}$).}\label{tab:fitresults}
\centering
\small
\begin{tabular}{lcccccccccccr} 
\hline\hline             
Object & ${\log}_{10} (I_{0})$ & $\sigma$ &  &  &  & $i$ & $PA$ & $\Delta \mathrm{RA}$ & $\Delta \mathrm{Dec}$ & ${R}_{68\%}$ & ${R}_{95\%}$ & $M_{\mathrm{dust}}$  \\
 & [$\mathrm{Jy} / \mathrm{sr}$] & [arcsec] &  &  &  & [deg] & [deg] & [arcsec] & [arcsec] & [$\mathrm{au}$] & [$\mathrm{au}$] & [$M_{\Earth}$] \\
\hline
  \multicolumn{13}{c}{\it BDs from this survey:}\\ 
\hline
J$154518.5$-$342125$ & $10.56_{-0.11}^{+0.16}$ & $0.035_{-0.005}^{+0.005}$ &  &  &  & $25.8_{-17.0}^{+18.7}$ & $90.2_{-59.8}^{+54.7}$ & $-0.270_{-0.002}^{+0.002}$ & $-0.352_{-0.002}^{+0.002}$ & $8.2_{-1.2}^{+1.0}$ & $13.2_{-2.0}^{+1.7}$ & $2.3 \pm 0.2$ \\
SONYC-Lup$3$-$7$ & $9.84_{-0.49}^{+0.55}$ & $0.03_{-0.01}^{+0.02}$ &  &  &  & - & - & $-0.07_{-0.02}^{+0.02}$ & $-0.24_{-0.02}^{+0.02}$ & <$17.5$ & <$28.3$ & $0.21 \pm 0.02$ \\ 
Lup$706$ & $8.70_{-0.15}^{+0.17}$ & $0.14_{-0.03}^{+0.04}$ &  &  &  & - & $88.3_{-38.5}^{+33.9}$ & $-0.12_{-0.02}^{+0.02}$ & $-0.46_{-0.02}^{+0.02}$ & <$56.4$ & <$91.4$ & $0.35 \pm 0.04$ \\ 
AKC$2006$-$18$ & - & - &  &  &  & - & - & - & - & - & - & <$0.06$ \\
SONYC-Lup$3$-$10$ & - & - &  &  &  & - & - & -  & - & -  & - & <$0.08$ \\

\hline
  \multicolumn{13}{c}{\it BDs from Lupus disks survey:} \\
\hline
Lup$818$s & $10.99_{-0.36}^{+0.41}$ & $0.03_{-0.01}^{+0.01}$ &  &  &  & - & - & $-0.06_{-0.01}^{+0.01}$ & $-0.207_{-0.005}^{+0.005}$ & <$12.1$ & <$19.7$ & $3.2 \pm 0.3$ \\
J$161019.8$-$383607$ & - & - &  &  &  & - & - & - & - & - & - & <$0.45$ \\
J$160855.3$-$384848$ & $10.15_{-0.47}^{+0.59}$ & $0.04_{-0.02}^{+0.03}$ &  &  &  & - & - & $-0.16_{-0.02}^{+0.02}$ & $-0.43_{-0.02}^{+0.02}$ & <$21.7$ & <$35.2$ & $0.79 \pm 0.08$ \\
Lup$607$ & - & - &  &  &  & - & - & - & - & - & - & <$0.38$ \\

\\ \\

\hline\hline             
Object & ${\rho}_{t}$ & $\gamma$ & $\beta$ & ${\log}_{10}\alpha$ & $F_{\mathrm{tot}}$ & $i$ & $PA$ & $\Delta \mathrm{RA}$ & $\Delta \mathrm{Dec}$ & ${R}_{68\%}$ & ${R}_{95\%}$ & $M_{\mathrm{dust}}$   \\
 & [arcsec] & [-] & [-] & [-] & [mJy] & [deg] & [deg] & [arcsec] & [arcsec] & [$\mathrm{au}$] & [$\mathrm{au}$] & [$M_{\Earth}$]  \\
\hline
  \multicolumn{13}{c}{\it Disks from Lupus disks completion survey:}  \\
\hline
Sz$102$ & $0.55_{-0.26}^{+0.45}$ & $-1.2_{-1.2}^{+1.2}$ & $15.2_{-4.0}^{+3.3}$ & $-0.1_{-0.2}^{+0.2}$ & $27.6_{-2.0}^{+2.4}$ & $57.6_{-2.9}^{+2.9}$ & $14.0_{-2.9}^{+3.0}$ & $-0.134_{-0.002}^{+0.001}$ & $0.130_{-0.002}^{+0.002}$ & $20.1_{-1.0}^{+1.1}$ & $43.6_{-5.4}^{+4.9}$ & $6.4 \pm 0.6$ \\

V$1094$ Sco & $0.18_{-0.01}^{+0.01}$ & $0.38_{-0.03}^{+0.03}$ & $1.25_{-0.01}^{+0.01}$ & $1.7_{-0.3}^{+0.2}$ & $1038_{-4}^{+5}$ & $56.2_{-0.1}^{+0.3}$ & $110.9_{-0.2}^{+0.2}$ & $-0.172_{-0.001}^{+0.001}$ & $0.098_{-0.001}^{+0.001}$ & $201.5_{-0.4}^{+0.4}$ & $301_{-1}^{+1}$ & $260 \pm 26$ \\

GQ~Lup & $0.21_{-0.04}^{+0.04}$ & $0.2_{-0.4}^{+0.3}$ & $14.8_{-4.5}^{+3.3}$ & $0.5_{-0.1}^{+0.2}$ & $158.7_{-2.5}^{+2.5}$ & $60.6_{-0.4}^{+0.5}$ & $-11.6_{-0.5}^{+0.5}$ & $0.0301_{-0.0002}^{+0.0003}$ & $0.1284_{-0.0003}^{+0.0004}$ & $18.4_{-0.2}^{+0.3}$ & $29.2_{-1.0}^{+0.8}$ & $31.6 \pm 3.2$ \\

Sz$76$ & $0.45_{-0.05}^{+0.12}$ & $1.38_{-0.05}^{+0.04}$ & $12.6_{-5.5}^{+5.1}$ & $1.2_{-0.5}^{+0.5}$ & $17.2_{-1.9}^{+2.4}$ & $38.9_{-10.3}^{+7.6}$ & $113_{-12}^{+11}$ & $0.001_{-0.003}^{+0.003}$ & $0.003_{-0.003}^{+0.003}$ & $40.9_{-3.5}^{+4.4}$ & $75_{-8}^{+17}$ & $6.1 \pm 0.6$ \\

Sz$77$ & - & - & - & - & $11.1_{-3.5}^{+4.0}$ & $61.7_{-21.9}^{+10.3}$ & $148_{-17}^{+14}$ & $0.008_{-0.003}^{+0.004}$ & $-0.007_{-0.004}^{+0.004}$ & <$55.5$ & <$136$ & $2.1 \pm 0.2$ \\

RXJ$1556.1$-$3655$ & $0.30_{-0.04}^{+0.03}$ & $0.4_{-0.1}^{+0.1}$ & $15.9_{-3.8}^{+2.8}$ & $0.6_{-0.1}^{+0.1}$ & $85.5_{-1.3}^{+1.3}$ & $49.4_{-0.7}^{+0.7}$ & $58.1_{-0.8}^{+0.8}$ & $-0.061_{-0.001}^{+0.001}$ & $0.081_{-0.001}^{+0.001}$ & $29.4_{-0.3}^{+0.3}$ & $44.9_{-1.1}^{+1.1}$ & $24.3 \pm 2.4$ \\

EX~Lup & $0.23_{-0.01}^{+0.01}$ & $0.38_{-0.06}^{+0.05}$ & $14.4_{-3.5}^{+3.9}$ & $1.4_{-0.3}^{+0.4}$ & $50.0_{-0.9}^{+0.9}$ & $30.8_{-1.6}^{+1.5}$ & $69.2_{-2.9}^{+2.9}$ & $-0.032_{-0.001}^{+0.001}$ & $-0.007_{-0.001}^{+0.001}$ & $29.8_{-0.4}^{+0.4}$ & $39.0_{-1.0}^{+1.4}$ & $18.5 \pm 1.9$ \\

\hline
\end{tabular}
\tablefoot{\\
The first $9$ objects are the full list of known BDs in Lupus, fitted to a gaussian model. The six free parameters are: normalization factor of the emission profile $\log I_{0}$, standard deviation of the gaussian profile $\sigma$, inclination $i$, position angle $PA$, and right ascension and declination off-sets to the phase center of the observations $\Delta \mathrm{RA}$ and $\Delta \mathrm{Dec}$.\\
The last $7$ disks were fitted to a Nuker profile. The nine free parameters of these fits are: transition radius ${\rho}_{t}$, inner and outer slopes $\gamma$ and $\beta$, smoothness parameter $\alpha$, total disk flux density $F_{\mathrm{tot}}$, and the geometrical parameters of the observation $i$, $PA$, $\Delta \mathrm{RA}$ and $\Delta \mathrm{Dec}$.
}

\end{sidewaystable*}

\subsubsection{Disk size results}\label{sec:sizeresults}
The radii enclosing $68 \%$ and $95 \%$ of the total flux (${R}_{68 \%}$ and ${R}_{95 \%}$) are specified in Table~\ref{tab:fitresults}. The values are inferred from their respective PDF (example in Figure~\ref{fig:j1545r68}), as derived from the model parameters results. The radii of the detected BDs disks are unfortunately poorly determined, due to the compactness of the sources, combined with a low S/N of their continuum emission at this waveband. 
Only for J$154518.5$-$342125$ we can properly quantify its size, since its continuum emission is detected with enough S/N and is marginally resolved. 
We denote as marginally resolved disk the case where the emission is of similar spatial scale as the beam size in the image plane, but where the observed visibilities can be fitted by a gaussian function with well constrained $\sigma$. 
For all the other BD sources we provide upper limits of their sizes as $95\%$ confidence level, inferred from the PDF of $R_{68\%}$ and $R_{95\%}$. On the other hand, we have determined the emission distribution size for six out of seven disks of the Lupus completion survey; for the remaining one (Sz$77$) we provide an upper limit.

Previous ALMA observations of BD disks in other regions showed that most of the objects were too compact to be resolved \citep{vanderplas+2016,testi+2016}. 
Using different methodology to define and derive disk radii, \cite{ricci+2014} and \cite{testi+2016} showed that some BD disk radii (R) in Taurus may extend beyond $R \gtrsim 80$ $\mathrm{au}$, while in $\rho$-Oph BD disks seem to all have $R \lesssim 25$ $\mathrm{au}$.

Nevertheless, we should bear in mind that the few BD disks with well determined sizes are among the brightest and most massive of the BD population of their respective regions (Lupus, Taurus and Ophiuchus); likely they are not representative of the BD population.

\subsection{Total dust mass results}\label{sec:dustmassresults}
The total disk dust mass (last column in Table~\ref{tab:fitresults}) is computed from the assumptions detailed in Section~\ref{sec:diskmasses}. As all detections have good S/N (Table~\ref{tab:fluxes}), the main uncertainty when comparing samples observed at different times is the flux calibrator uncertainty ($\sim 10 \%$, see Section~\ref{sec:obsresults}). Dust mass upper limits for non-detected disks around BDs and stars are obtained from the respective continuum flux upper limits as described in Section~\ref{sec:obsresults}.

The total dust mass for the detected BD disks range between $0.2$ and $3.2$ $M_{\Earth}$. 
This means that our sources are within the lightest protoplanetary disks known up to date. In particular, SONYC-Lup$3$-$7$ is the BD disk with the lowest dust mass estimate, independent of the prescription used for the dust disk mass determination. 
Comparing our dust mass results of BD disks in Lupus to the results of BD disks in other regions, our results are found to be similar. In \cite{testi+2016}, a sample of $17$ BD disks in the $\rho$ Ophiuchus region were observed and their dust mass estimates are within $0.5$ and $6.3$ $M_{\Earth}$, with the same assumptions of dust temperature and opacity as for our results. The dust masses of Taurus disks around BD and VLMs \citep{wardduong+2018} range between $\sim 0.25$ and $\sim 16.7$ $M_{\Earth}$, using the same temperature and opacity values as in this work.

\section{Discussion}\label{sec:discussion}

\subsection{Comparison to disks around T Tauri stars}\label{sec:comparison}
We have performed a demographic comparison between the BD and stellar disk populations of Lupus. For this analysis we use the derived disk properties to test whether the known relations for disks around stars stand for BD disks. 
The observational dataset of both populations has been obtained from the same facility (ALMA, Band 7), and the derivation of the properties has been conducted with homogeneous methodology for the entire disk population. In this manner we eradicate systematic errors due to the mixing of diverse datasets handled with different methods.

In addition, we have updated the relations between disk properties in Lupus with the largest sample of disks in the region observed with ALMA in Band 7, thanks to the incorporation of the seven stellar disks from the Lupus completion survey to the stellar population. In Table~\ref{tab:valueslinearregression} we summarize all the correlations discussed along this section.

\subsubsection{Correlation between $M_{\star}$ and $M_{\mathrm{dust}}$}\label{sec:mstarmdust}
\begin{figure}
  \resizebox{\hsize}{!}{\includegraphics{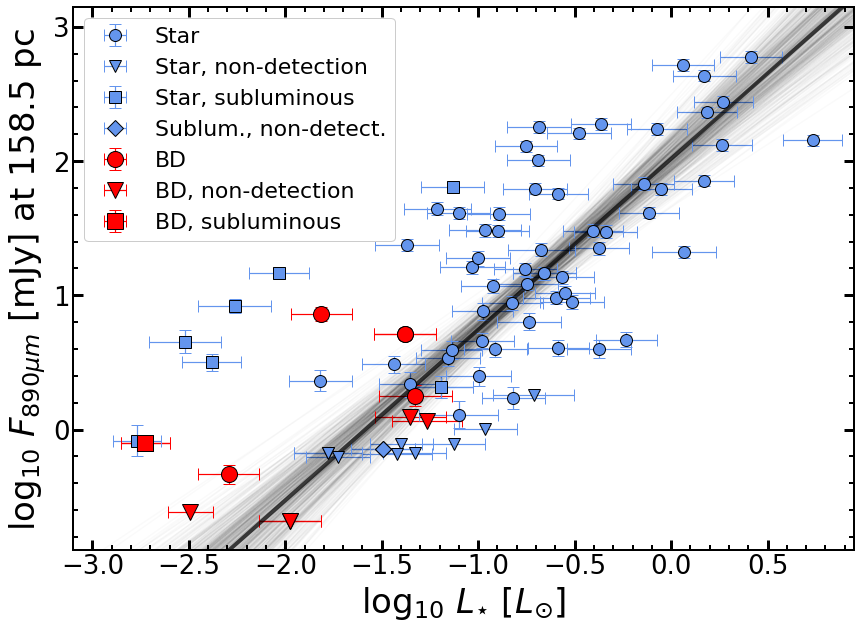}}
  \caption[Relation between the stellar luminosity and the measured disk flux at $890$ $\mu m$ of the Lupus population.]{$890$ $\mu m$ fluxes versus stellar luminosity for the Lupus population. Stellar population is shown in blue; the BD disk population is plotted in red. Detected sources from ALMA observations are represented as circles; upper limits of non-detections are shown as triangles; sub-luminous objects as inferred from X-Shooter spectra are plotted as squares. The linear regression shown has been obtained from the entire population, excluding sub-luminous sources.}
  \label{fig:lstarfcont}
\end{figure}

As a previous step, we show in Figure~\ref{fig:lstarfcont} the relation of the respective observables of $M_{\star}$ and $M_{\mathrm{dust}}$, that is, the stellar luminosities $L_{\star}$ and the fluxes at $890$ $\mu m$ wavelength (scaled to a distance of $158.5$ pc). 
From the figure, there is a continuity of the correlation for any range of $L_{\star}$, and it holds for the BD population. 
The linear regression shown in the figure is for the entire population (stars and BDs), obtained following the Bayesian method described in \cite{kelly2007} \footnote{implemented with the \texttt{linmix} Python package, \texttt{https://linmix.readthedocs.io/en/latest/index.html}}. Uncertainties and upper limits are taken into account, while sub-luminous objects are excluded for the fit.

\cite{testi+2016} found potential evidence of BD disks being less massive than stellar disks, based on the analysis of an incomplete sample of BD disks in Ophiuchus. Our Lupus sample allow us to check whether similar results hold in this star forming region. In Figure~\ref{fig:lstarfcont}, there is no obvious trend for BDs to have very significantly smaller $890$ $\mu m$ fluxes than stars with similar luminosities. To make a quantitative statement, we follow a similar procedure as in \cite{testi+2016} based on a statistical comparison of the two populations, and analyze whether the distribution of the $M_{\mathrm{dust}} / M_{\star}$ ratios in the sample of BD disks is consistent with being drawn from the same distribution as for the stars.
\begin{figure}
  \resizebox{\hsize}{!}{\includegraphics{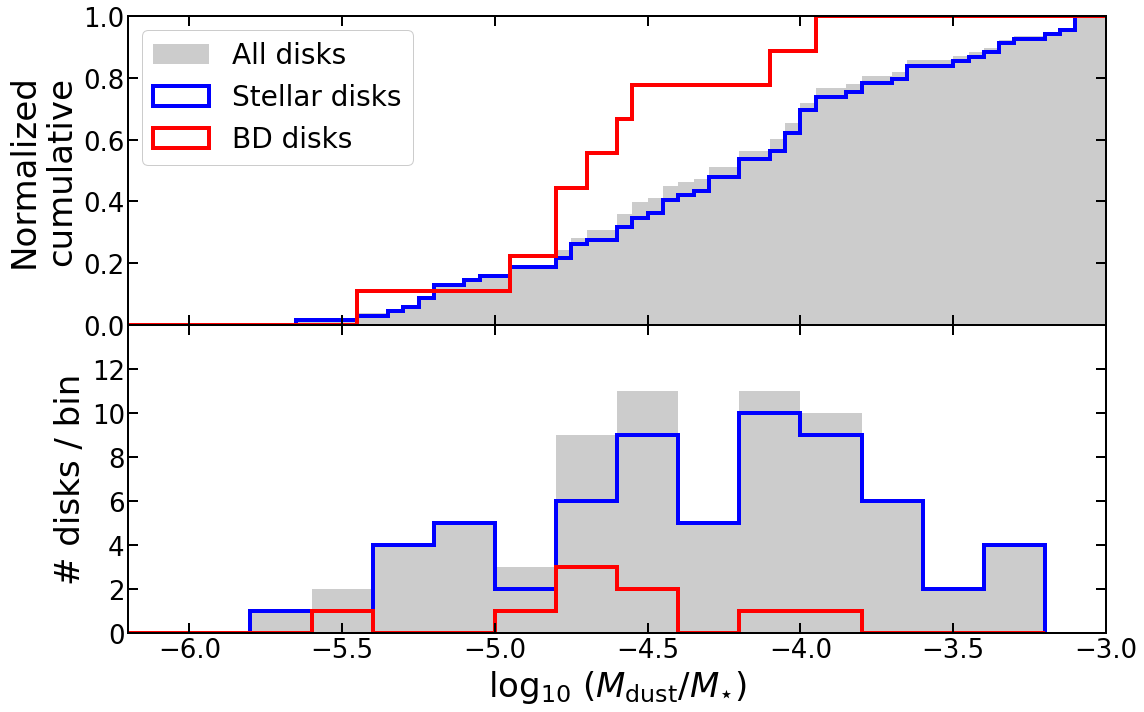}}
  \caption[]{Statistical comparison between BD and TTS disks populations in the Lupus star forming region. The histograms and cumulative distributions of the dust mass-stellar mass ratio is shown for both populations. The results summing up both populations are also included (shown as gray). A bin-size of $0.2$ has been used for the histogram of the populations.}
  \label{fig:populations}
\end{figure}

Figure~\ref{fig:populations} shows the cumulative distributions and the histograms of the values of the $M_{\mathrm{dust}} / M_{\star}$ ratios for the Lupus samples. Dust mass of each object is derived following Section~\ref{sec:diskmasses}, $M_{\star}$ as described in Section~\ref{sec:sampleselection}. 
The histogram shows that the values of the BD ratios are similar to the stellar population ratios, unlike the Ophiuchus sample in \cite{testi+2016}. We have performed the Anderson-Darling test\footnote{using \texttt{scipy.stats} Python module, \texttt{https://docs.scipy.org/doc/scipy/reference/stats.html}} to study the null hypothesis that the two samples are drawn from the same underlying population, 
obtaining a probability of $6 \%$ that the BD and stellar disk populations are drawn from the same distribution. Although it is a low percentage, it is below 2$\sigma$ significance. Moreover, the likelihood increases to $\sim 80$-$90 \%$ if we use the dependence of $T_{\mathrm{dust}}$ with $L_{\star}$ from \cite{andrews+2013,vanderplas+2016}. Thus, the data are consistent with the null hypothesis to be correct. 
Our analysis of the Lupus sample do not show a statistically different fraction of dust mass around BDs as compared to stars. We caution that our sample of VLM stars and BDs in Lupus is very limited and that the results of \cite{testi+2016} was based on highly incomplete and inhomogeneous samples. Further studies with larger and/or unbiased samples are needed to conclusively assess this point.

\begin{figure}
  \resizebox{\hsize}{!}{\includegraphics{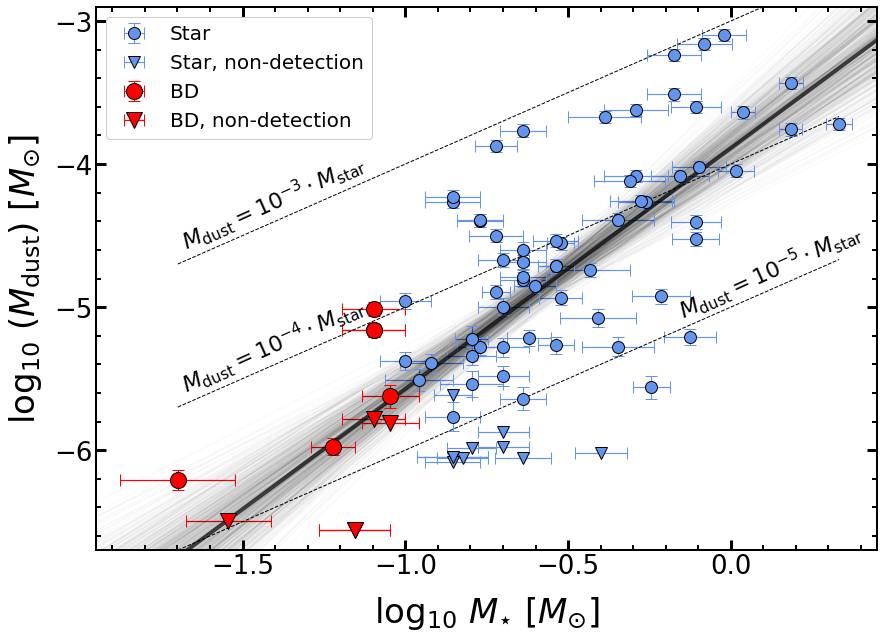}}
  \caption[Lupus disk demographics: stellar mass vs. total dust mass.]{
  Disk dust masses versus central object mass for the BD (red) and stellar (blue) populations in Lupus. Detected sources from ALMA observations are represented as circles; upper limits of non-detections are shown as triangles. Dust mass uncertainties and upper limits as described in Section~\ref{sec:dustmassresults}. Uncertainties of $M_{\star}$ are $1\sigma$. Linear regression shown for the entire disk population (stars and BDs).}
  \label{fig:mstarmdust}
\end{figure}
The result of the previous analysis is also confirmed by inspecting the dependence of $M_{\mathrm{dust}}$ on $M_{\star}$ (Figure~\ref{fig:mstarmdust}). 
In Appendix~\ref{sec:appendix_otherrelations}, this dependence is shown for the other $T_{\mathrm{dust}}$ prescriptions. 
The linear regression result is consistent to \cite{ansdell+2016,pascucci+2016} when using the same assumptions of dust opacity and temperature. 
The slope ($\alpha$) and intercept ($\beta$) for the stellar population are $\alpha = 1.73 \pm 0.25$ and $\beta = -3.88 \pm 0.14$ respectively (inferred using \texttt{linmix} package, and including upper limits of ALMA non-detections). 
As consequence of incorporating the BD population into the fit, there is a substantial reduction of the uncertainty of $\alpha$ and $\beta$, thanks to the extension of the mass range over one order of magnitude: the slope and intercept become $1.69 \pm 0.19$ and $-3.89 \pm 0.13$. If we compute a linear regression only taking into account the BD sample, we obtain a slope and intercept that is in agreement with the stellar fit, although the uncertainties in this case are large due to the short range in both axes of the BD population.

\subsubsection{Disk size-luminosity relation}\label{sec:mdustr68}
\begin{figure}
  \resizebox{\hsize}{!}{\includegraphics{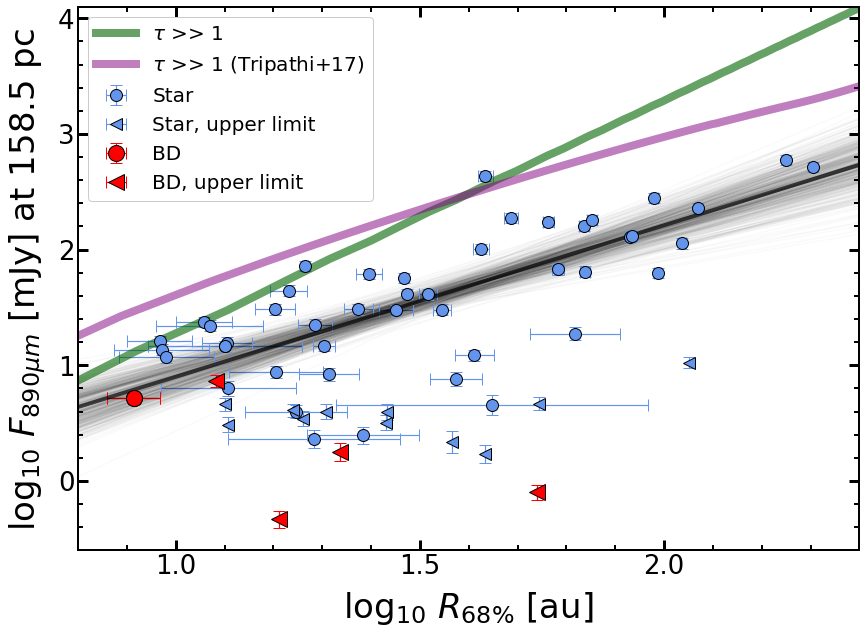}}
  \caption[Lupus disk demographics: effective radii vs. continuum flux normalized at the average distance of Lupus.]{
  ALMA $890$ $\mu m$ fluxes, scaled to a common distance, versus observed size ($R_{68\%}$, see text), for stars (blue) and BDs (red) in the Lupus star forming region. Upper limits of disk sizes that could not be determined accurately from our disk modeling methodology are shown as triangles, they represent the 95$\%$ confidence level of the disk size. $R_{68\%}$ error bars account for $1\sigma$ from the mean value, flux error bars are associated to the $10\%$ flux calibration uncertainty.}
  \label{fig:r68mdust}
\end{figure}

The existence of a correlation between the disk luminosity and its size was first shown from pre-ALMA observations \citep{andrews+2010,pietu+2014}. For the Lupus disk population, this dependence was confirmed in \cite{tazzari+2017A} and \cite{andrews2018A}. 
In Figure~\ref{fig:r68mdust}, we show the updated relation for the Lupus disk population, including the seven new measurements from this paper (one BD disk and six disks around stars). 
The linear regression shown in the figure is obtained for the stellar disk population, excluding the upper-limits of the disks with poorly constrained sizes.

The BD disk with well constrained $R_{68\%}$ (J$154518.5$-$342125$) is in very good agreement with the relation for stars. The result suggests that this BD disk is as a scaled down equivalent of the very extended disks around stars that show sub-structure. Nevertheless, since its central object mass is near the BD/VLM boundary, this result might not be representative of the full BD population. Higher angular resolution observations of the BD population are needed in order obtain reliable estimates of their sizes. 
The estimated size upper limits of the other BD disks provide little constrains on the relation. The compactness of the BD disks might be indicative that BD disks follow the size-luminosity relation of stars, as suggested by \cite{hendler+2017} from SED fitting of disk observations, and also from the results in $\rho$-Ophiuchus \cite{testi+2016}. 
If these objects follow the same relation as stars, their $R_{68\%}$ would range between $1$ and $10$ $\mathrm{au}$.

There is now evidence of optically thick emission in the inner ($\lesssim 50$ $\mathrm{au}$) regions of disks around stars \citep{huang+2018,zhu+2019}. Likewise, BD disks in Lupus might be optically thick, as suggested by their compact continuum emission. 
In Figure~\ref{fig:r68mdust}, we show two optically thick (optical depth $\tau>>1$) fiducial models, the first one (green) assuming a constant $T_\mathrm{dust}$ of $20$ K, and a second model (purple) with radial dependence of $T_\mathrm{dust} [\mathrm{K}] \approx 30 \cdot {(\frac{L_{\star}}{L_{\odot}})}^{0.25} \cdot {(\frac{R}{10})}^{-0.5}$ \citep[][]{andrews+2013}. The emission of these models are described by $I_{\nu}(R) = \mathcal F B_{\nu}(T_{\mathrm{dust}})$, where $\mathcal F$ is a filling factor that describes the fraction of the disk emission distribution that is optically thin: $\mathcal F = 1$ if the disk emission is optically thick, $0<\mathcal F<1$ for a partially optically thick disk \citep[analogous to][]{tripathi+2017,andrews2018A}. The optically thick curves in the figure are built considering a set of disks with increasing outer size. 
Objects laying on the line are compatible with being fully optically thick. Additionally, in optically thick disks, the inferred $R_{68\%}$ trace the location of large grains rather than the physical outer radius of the disk \citep{rosotti+2019}.

The only BD with determined dust disk size (J$154518.5$-$342125$) lays below both fiducial models. Its dust emission can be understood as optically thin with a fraction of the disk emission distribution being optically thick. If its disk emission is partially optically thin, a portion of dust is not observed, thus the inferred dust mass is underestimated. 
The upper-limits of the remaining BD disks are far below the optically thick models, although their exact positions in the $R_{68\%}$-$F_{890\mu m}$ are unknown.

\subsubsection{Correlation between $\dot{M}_{\mathrm{acc}}$ and $M_{\mathrm{dust}}$}\label{sec:mdustmacc}
A linear correlation between mass accretion rate onto the central object ($\dot{M}_{\mathrm{acc}}$) and the disk mass is expected if disks evolve viscously \citep[e.g.,][]{dullemond+2006,NattaTesti2007,lodato+2017}. Observational evidence for this correlation was first reported by \cite{manara+2016} for the Lupus disks, and \cite{mulders+2017} in the Chamaeleon I region.

\begin{figure}
  \resizebox{\hsize}{!}{\includegraphics{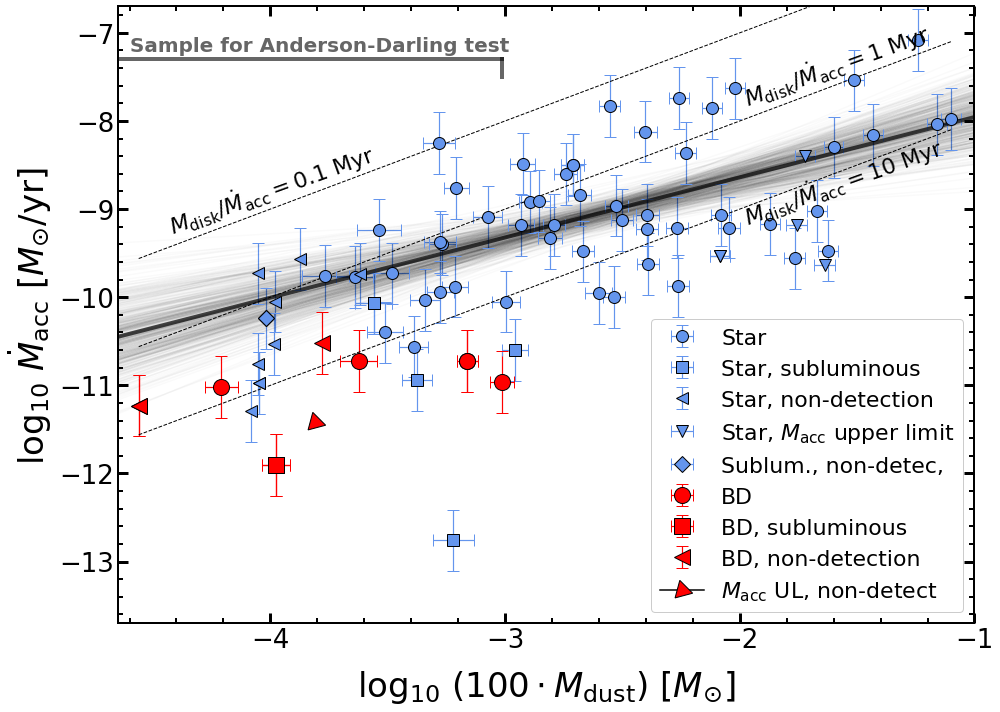}}
  \caption[Lupus disk demographics: mass accretion rate vs. disk mass.]{Relation between the total disk mass (obtained assuming $\kappa_{890\mu m} = 2$ $\mathrm{cm}^2$ $\mathrm{g}^{-1}$, and $T_{\mathrm{dust}} = 20$ K) and the mass accretion rate onto the central object.}
  \label{fig:mdustmacc}
\end{figure}

The $\dot{M}_{\mathrm{acc}}$-$M_{\mathrm{dust}}$ relation for Lupus disks is shown in Figure~\ref{fig:mdustmacc}. 
The x-axis of the figure is an estimate of the total disk mass, based on our derivation of the disk dust mass and assuming a gas-to-dust ratio of $100$. 
The $\dot{M}_{\mathrm{acc}}$ and its uncertainty is taken from the X-Shooter observations presented by \cite{alcala2014,alcala2017}. The accretion rate values have been recomputed with the new accretion luminosities that correspond to the parallaxes from Gaia DR$2$. 
One BD disk (SONYC-Lup3-10) was not characterized from X-Shooter observations. Although the $H_{\alpha}$ emission line is known \citep{muzic+2014}, we have excluded this BD from the analysis in order to have a fully homogeneous sample for our statistical study.

The linear regression for the stellar population in Figure~\ref{fig:mdustmacc} has been obtained excluding BDs (due to the different BD disks behavior compared to stellar disks, demonstration below), non-detections from ALMA, upper limits of $\dot{M}_{\mathrm{acc}}$ and sub-luminous sources. The resulting slope is $\alpha = 0.69 \pm 0.14$, while the intercept is $\beta = -7.26 \pm 0.36$. When using the same assumptions of dust temperatures and opacities, the linear regression is consistent with the results presented in \cite{manara+2016}. 
BDs have systematically lower accretion rates than stars for the same disk mass. This is also seen in the relation between the more direct observed properties (in Appendix~\ref{sec:appendix_laccfcont}), and is independent to the considered prescription of the dust temperature (results using other prescriptions in Appendix~\ref{sec:appendix_otherrelations}).

\begin{table*}
\caption{Linear regression results of all the investigated disk properties correlations.}
\label{tab:valueslinearregression}      
\centering                          
\begin{tabular}{lccccr}      
\hline\hline                 
X-axis & Y-axis & $T_{\mathrm{dust}}$ prescription & $\alpha$ & $\beta$ & Dispersion  \\
\hline                       
$L_{\star}[L_{\odot}]$ & $F_{890\mu m}[\mathrm{mJy}]$ & - & $1.27\pm0.13$ & $2.02\pm0.13$ & $0.60\pm0.07$  \\
$F_{890\mu m}[\mathrm{mJy}]$ & $L_{\mathrm{acc}}[L_{\odot}]$ & - & $0.81\pm0.15$ & $-3.28\pm0.22$ & $0.68\pm0.08$  \\
$M_{\star}[M_{\odot}]$ & $M_{\mathrm{dust}}[M_{\odot}]$ & Constant ($20$ K) & $1.69\pm0.19$ & $-3.89\pm0.13$ & $0.60\pm0.05$  \\
 &  & \cite{andrews+2013} & $0.95\pm0.18$ & $-4.08\pm0.12$ & $0.57\pm0.05$  \\
 &  & \cite{vanderplas+2016} & $1.25\pm0.18$ & $-3.98\pm0.12$ & $0.58\pm0.05$  \\
$R_{68\%}[\mathrm{au}]$ & $F_{890\mu m}[\mathrm{mJy}]$ & - & $1.31\pm0.17$ & $-0.41\pm0.27$ & $0.41\pm0.05$  \\
$100 \cdot M_{\mathrm{dust}}[M_{\odot}]$ & $\dot{M}_{\mathrm{acc}}[M_{\odot}/\mathrm{yr}]$ & Constant ($20$ K) & $0.69\pm0.14$ & $-7.26\pm0.36$ & $0.56\pm0.08$  \\
 &  & \cite{andrews+2013} & $0.63\pm0.17$ & $-7.49\pm0.43$ & $0.64\pm0.09$  \\
 &  & \cite{vanderplas+2016} & $0.67\pm0.16$ & $-7.38\pm0.42$ & $0.62\pm0.09$  \\
\hline
\end{tabular}
\tablefoot{
The values of $\alpha$ and $\beta$ correspond to the slope and intercept of the linear fit, following the linear relation $\log(Y) = \beta + \alpha \cdot \log(X)$. The dispersion is the standard deviation of the regression in dex. $F_{890\mu m}$ has been scaled to the average distance of the region ($158.5$ pc). 
}
\end{table*}
We have inspected the $M_{\mathrm{disk}}/\dot{M}_{\mathrm{acc}}$ ratio for the Lupus disk population to confirm or deny this trend. This ratio can be understood as the accretion depletion timescale (or disk age, as in \citealt{jones+2012}, see also \citealt{rosotti+2017}), and provides an estimate of the survival time of the disk, assuming that accretion onto the central object remains constant and that accretion is the dominant mechanism for the depletion of the disk. 
In Figure~\ref{fig:mdustmacc}, we plotted different lines indicating accretion depletion timescales of $0.1$, $1$ and $10$ Myr.
\begin{figure}
  \resizebox{\hsize}{!}{\includegraphics{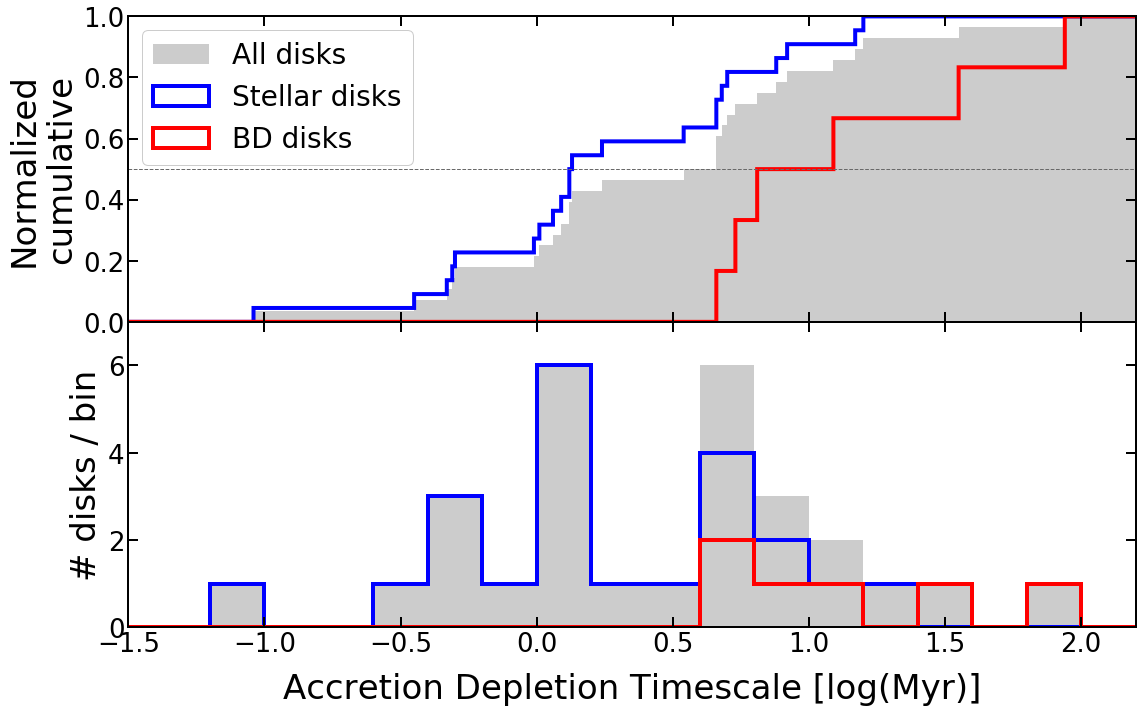}}
  \caption[]{Histograms (bottom) and cumulative distributions (top) of the accretion depletion timescale are shown for both populations (BDs as red, stars as blue). The results summing up both populations are also included (shown as gray).}
  \label{fig:populations_timescale}
\end{figure}

Analogous to the Subsection~\ref{sec:mstarmdust}, we have conducted a statistical analysis of the two populations in order to confirm the behavior of the BD disks. We compare the BD population with the sub-sample of stars with disk masses within the range of the BD disk masses (in other words, all disks with $\log_{10} (M_{\mathrm{disk}})$<$-3$ in Figure~\ref{fig:mdustmacc}). This is done in order to remove the more massive disks of the stellar sample, which may accentuate the difference between populations. Sub-luminous objects and those with upper-limits of the $\dot{M}_{\mathrm{acc}}$ are excluded from the studied samples. The median value of the central object mass is $0.08$ $M_{\star}$ for the BD sample and $0.19$ $M_{\star}$ for the stellar sub-sample. 
The histograms of the accretion depletion timescale and the respective cumulative distributions are shown in Figure~\ref{fig:populations_timescale}. The Anderson-Darling test confirms that the BD disk population has a significantly larger accretion depletion timescale respect to the stellar population, with only a $0.6 \%$ probability that the BD and stellar disks populations are drawn from the same original distribution. The median value of this timescale for the BD sample is $9.5$ Myr, while for the stellar sub-sample considered is $1.4$ Myr. The results hold when using other prescriptions of the disk dust mass for this test, with even lower probabilities ($\sim 0.05\%$).

The result of the accretion depletion timescale is obtained using the total disk mass, which is estimated assuming that the emission of the dust is optically thin, and that dust mass traces the total disk mass. If (sub-)mm emission of BD disks is optically thick, the disk masses are underestimated, and consequently the accretion depletion timescale would be larger than the estimated values. Thus, the difference on the accretion depletion timescale would be even more pronounced if BD disks are optically thick at these wavelengths.

A larger accretion depletion timescale may reflect a difference in the accretion process of BDs with respect to stars. 
If viscous evolution models are invoked to explain accretion onto the central star, a weaker accretion would imply a lower $\alpha$ parameter \citep{ShakuraSunyaev1973} in BD disks. Since the turbulence in viscous disks depends on $\alpha$, a lower $\dot{M}_{\mathrm{acc}}$ implies a less turbulent disk. 
Thus, in viscous disks, our result suggests that the $\alpha$ parameter of disks around BDs is lower than around stars \citep[in contrast with the results of][]{MuldersDominik2012}, and consequently BD disks being less turbulent. A bi-modal behavior of accretion has been suggested observationally in \cite{alcala2017,manara2017a}, and predicted from theoretical modeling in \cite{VorobyovBasu2009}; in those studies the two suggested modes were between VLMs ($M_{\star}<0.2$ $M_{\odot}$) and more massive stars. When performing a statistical comparison in our sample between the VLM population ($0.1$ $M_{\odot}$ < $M_{\star}<0.2$ $M_{\odot}$) and more massive stars, we obtain a likelihood in the Anderson-Darling test of $\sim 9 \%$ (averaged over the three different prescriptions of $M_{\mathrm{dust}}$ used in this work). Thus, VLM stars might show this behavior as well, but less pronounced, and with much lower statistical significance.

A lower viscosity in disks around BDs/VLM stars compared to disks around more massive stars could be explained with a globally lower ionization rate \citep[e.g., see][]{mohanty+2005}. 
As a consequence of the general correlation between X-ray and bolometric luminosities, young BDs/VLM stars have slightly lower X-ray luminosities than more massive young low-mass stars \citep[see, e.g.,][]{gregory+2016}; this might cause slightly lower ionization rates in the disks of the BDs/VLM objects. 
Another tentative explanation of lower ionization rates might be if some BD/VLM disks are flatter than disks around more massive stars \citep[as suggested from SED models and observations by][]{ercolano+2009,pascucci+2003,apai+2004,allers+2006}, which would decrease the irradiation cross-section. However, there is evidence of disks around BDs being flared \citep[e.g.,][]{NattaTesti2001,natta+2002,furlan+2011}, thus this last option would need more detailed and extended investigation.

\subsection{Planet formation around BDs}\label{sec:planetformation}
The exoplanetary systems recently discovered around BDs and VLMs can be used to study the ability of BDs to form planets. Since the planets hosted by Trappist-1 and Proxima Centauri are most likely of rocky composition, the total planetary mass of these systems can be compared to the estimates of the disk dust mass of the BD disk population in Lupus. The summed up mass of the seven known planets \citep{gillon+2017} in Trappist is $4$ $M_{\Earth}$ \citep{wang+2017}. Proxima B, the planet hosted by our closest neighbor \citep{angladaescude+2016}, has a minimum mass of $1.3$ $M_{\Earth}$, and \citep{BixelApai2017} predicted a mass of $1.63^{+1.66}_{-0.72}$ $M_{\Earth}$ with $95\%$ confidence level. The remaining known exoplanets around BDs have estimated masses of at least several times Earth's mass. A considerable fraction of them have been detected via microlensing \citep[e.g.,][]{jung+2018}, which typically much higher masses. Thus the picture for these planets is analogous to the Trappist-1 planets and Proxima B.

From theoretical predictions of planet formation around BDs via core accretion \citep{PayneLodato2007}, disk masses (gas and dust) on the order of a few Jupiter masses are required in order to form Earth-like planets around BDs. 
Not only none of the BD disks in Lupus (this paper) and $\rho$-Oph \citep{testi+2016} have enough mass available at their current stage to 
form a planetary system, but even the available mass in solids is smaller than the total planetary mass in Trappist-1. 
The efficiency to convert the available mass into the final planetary rocky cores might be boosted due to internal recycling of the disk material, but it is unlikely to reach an efficiency close to unity \citep[][and references therein]{manara+2018}. On the other hand, the tentative result from Section~\ref{sec:mdustmacc} of lower viscosity and ionization rates on BD disks might contribute to the presence of an extended dead zone in the disk, which would boost up the planet formation process.

A plausible explanation to alleviate this divergence is that the determination of dust mass from continuum emission flux might be underestimated, as pointed out in \cite{ballering+2019}. This might be the case if the emission at this wavelength is optically thick, then the inferred $M_{\mathrm{dust}}$ provides only a lower limit of the disk dust mass. In Fig.~\ref{fig:r68mdust}, disks laying on the $\tau$ >> $1$ fiducial models are consistent with their emission being fully optically thick. The only Lupus BD disk with well-determined size (J$154518.5$-$342125$) is below these models. This suggests that its emission is optically thin with small regions of the disk being optically thick. The inferred dust mass of this disk is underestimated by an unknown fraction. This can help explaining the mass difference with exoplanetary systems. Nevertheless, it seems unlikely that only this indeterminacy can account for the large solid mass difference in disks with partially optically thin emission.

A likely possibility is that planets might have already formed at this stage of the disk evolution \citep{GreavesRice2010,NajitaKenyon2014,manara+2018,dodds+2015}. If that is indeed the case, the formation of planetary rocky cores would have occurred within the first Myr (considering the estimated ages of Lupus and $\rho$-Ophiuchus). While direct detection of planets embedded in protoplanetary disks is extremely difficult \citep[see][]{sanchis+,johnskrull+2016}, 
the presence of circumplanetary disks sets, if confirmed, a strong indirect evidence of young planets in these disks \citep[][]{keppler+2018,isella+2019,perez+2019}. 
Analysis of the gas kinematics can also be used as an indirect method to study embedded planets \citep{teague+2018,pinte+2018}. 
Other indirect indications, such as the the existence of gaps, spirals, assymmetries, and dust processing are observed frequently, and suggest that planets might already have formed \citep[e.g.,][]{alma2015,zhang+2018,lodato+2019,pinilla+2018b}.

\section{Conclusions}\label{sec:conclusions}
In this work we presented new Band 7 ALMA observations of five protoplanetary disks around BDs in Lupus. Combined with previous observations, we present and analyze the (sub-)mm disk properties of the known population of BDs/VLM objects with infrared excess.

\begin{itemize}
    \item We infer total dust disk masses and characteristic sizes of the disk population, measuring the continuum fluxes and modeling the visibilities.
    \item We update the relations of $M_{\star}$-$M_{\mathrm{dust}}$, size-luminosity and $M_{\mathrm{dust}}$-$\dot{M}_{\mathrm{acc}}$ extending them down to the substellar regime.
    \item BD disks in Lupus follow the relation for stars between stellar mass and dust disk mass. There is no statistical difference with the stellar disk population on the disk mass fraction, however we note the apparent lack of massive BD disks.
    \item BD disks are extremely compact, size determination was only possible on one BD disk, while for the other detected disks we present upper limits on the size.
    \item The accretion depletion timescale (inferred assuming that dust mm continuum emission is a reliable proxy of the total disk mass) of the BD population is significantly longer than for stars ($9.5$ Myr vs. $1.4$ Myr), which in viscously evolving disks may imply a lower $\alpha$ value, possibly linked to a globally lower ionization rate.
    \item Estimated disk dust masses are very low, suggesting that either these systems are unable to form planets, or, more likely, rocky planetary cores are formed within the first Myr. Optically thick emission in BD disks can alleviate this mass discrepancy.
\end{itemize}

%%%%%%%%%%%%%%%%%%%%%%%%%%%%%%%%%%%%%%%%%%%%%%%%%%%%%
%%%%%%%%%%%%%%%%% ACKNOWLEDGEMENTS %%%%%%%%%%%%%%%%%%

\begin{acknowledgements}
This paper makes use of the following ALMA data: ADS/JAO.ALMA$\#2018$.$1$.$00544$.S. ALMA is a partnership of ESO (representing its member states), NSF (USA) and NINS (Japan), together with NRC (Canada) and NSC and ASIAA (Taiwan) and KASI (Republic of  Korea), in cooperation with the Republic of Chile. The Joint ALMA Observatory is operated by ESO, AUI/NRAO and NAOJ. This work was partly supported by the Italian Ministero dell Istruzione, Universit\`a e Ricerca through the grant Progetti Premiali 2012 – iALMA (CUP C$52$I$13000140001$), by the Deutsche Forschungs-gemeinschaft (DFG, German Research Foundation) - Ref no. FOR $2634$/$1$ TE $1024$/$1$-$1$, and by the DFG cluster of excellence Origins (www.origins-cluster.de), and by the European Union's Horizon $2020$ research and innovation program under the Marie Sklodowska-Curie grant agreement No$823823$ (RISE DUSTBUSTERS project). 
T.H. acknowledges support from the European Research Council under the Horizon 2020 Framework Program via the ERC Advanced Grant Origins 83 24 28. 
KM acknowledges funding by the Science and Technology Foundation of Portugal (FCT), grants No. IF/$00194$/$2015$ and PTDC/FISAST/$28731$/$2017$. 
CM, SF, AM acknowledge an ESO Fellowship. 
M.T. has been supported by the UK Science and Technology research Council (STFC).
\end{acknowledgements}

%%%%%%%%%%%%%%%%%%%%%%%%%%%%%%%%%%%%%%%%%%%%%%%%%%%%%
%%%%%%%%%%%%%%%%% BIBLIOGRAPHY %%%%%%%%%%%%%%%%%%%%%%

\bibliographystyle{aa} 
\bibliography{lupus_BDdisks.bib} 

\begin{thebibliography}{115}
\expandafter\ifx\csname natexlab\endcsname\relax\def\natexlab#1{#1}\fi

\bibitem[{{Alcal{\'a}} {et~al.}(2017){Alcal{\'a}}, {Manara}, {Natta}, {Frasca},
  {Testi}, {Nisini}, {Stelzer}, {Williams}, {Antoniucci}, {Biazzo}, {Covino},
  {Esposito}, {Getman}, \& {Rigliaco}}]{alcala2017}
{Alcal{\'a}}, J.~M., {Manara}, C.~F., {Natta}, A., {et~al.} 2017, \aap, 600,
  A20

\bibitem[{{Alcal{\'a}} {et~al.}(2014){Alcal{\'a}}, {Natta}, {Manara}, {Spezzi},
  {Stelzer}, {Frasca}, {Biazzo}, {Covino}, {Randich}, {Rigliaco}, {Testi},
  {Comer{\'o}n}, {Cupani}, \& {D'Elia}}]{alcala2014}
{Alcal{\'a}}, J.~M., {Natta}, A., {Manara}, C.~F., {et~al.} 2014, \aap, 561, A2

\bibitem[{{Allers} {et~al.}(2006){Allers}, {Kessler-Silacci}, {Cieza}, \&
  {Jaffe}}]{allers+2006}
{Allers}, K.~N., {Kessler-Silacci}, J.~E., {Cieza}, L.~A., \& {Jaffe}, D.~T.
  2006, \apj, 644, 364

\bibitem[{{ALMA Partnership} {et~al.}(2015){ALMA Partnership}, {Brogan},
  {P{\'e}rez}, {Hunter}, {Dent}, {Hales}, {Hills}, {Corder}, {Fomalont},
  {Vlahakis}, {Asaki}, {Barkats}, {Hirota}, {Hodge}, {Impellizzeri}, {Kneissl},
  {Liuzzo}, {Lucas}, {Marcelino}, {Matsushita}, {Nakanishi}, {Phillips},
  {Richards}, {Toledo}, {Aladro}, {Broguiere}, {Cortes}, {Cortes}, {Espada},
  {Galarza}, {Garcia-Appadoo}, {Guzman-Ramirez}, {Humphreys}, {Jung}, {Kameno},
  {Laing}, {Leon}, {Marconi}, {Mignano}, {Nikolic}, {Nyman}, {Radiszcz},
  {Remijan}, {Rod{\'o}n}, {Sawada}, {Takahashi}, {Tilanus}, {Vila Vilaro},
  {Watson}, {Wiklind}, {Akiyama}, {Chapillon}, {de Gregorio-Monsalvo}, {Di
  Francesco}, {Gueth}, {Kawamura}, {Lee}, {Nguyen Luong}, {Mangum}, {Pietu},
  {Sanhueza}, {Saigo}, {Takakuwa}, {Ubach}, {van Kempen}, {Wootten},
  {Castro-Carrizo}, {Francke}, {Gallardo}, {Garcia}, {Gonzalez}, {Hill},
  {Kaminski}, {Kurono}, {Liu}, {Lopez}, {Morales}, {Plarre}, {Schieven},
  {Testi}, {Videla}, {Villard}, {Andreani}, {Hibbard}, \&
  {Tatematsu}}]{alma2015}
{ALMA Partnership}, {Brogan}, C.~L., {P{\'e}rez}, L.~M., {et~al.} 2015, \apjl,
  808, L3

\bibitem[{{Andrews}(2015)}]{andrews2015}
{Andrews}, S.~M. 2015, \pasp, 127, 961

\bibitem[{{Andrews} {et~al.}(2013){Andrews}, {Rosenfeld}, {Kraus}, \&
  {Wilner}}]{andrews+2013}
{Andrews}, S.~M., {Rosenfeld}, K.~A., {Kraus}, A.~L., \& {Wilner}, D.~J. 2013,
  \apj, 771, 129

\bibitem[{{Andrews} {et~al.}(2018){Andrews}, {Terrell}, {Tripathi}, {Ansdell},
  {Williams}, \& {Wilner}}]{andrews2018A}
{Andrews}, S.~M., {Terrell}, M., {Tripathi}, A., {et~al.} 2018, \apj, 865, 157

\bibitem[{{Andrews} \& {Williams}(2005)}]{andrewswilliams2005}
{Andrews}, S.~M. \& {Williams}, J.~P. 2005, \apj, 631, 1134

\bibitem[{{Andrews} {et~al.}(2010){Andrews}, {Wilner}, {Hughes}, {Qi}, \&
  {Dullemond}}]{andrews+2010}
{Andrews}, S.~M., {Wilner}, D., {Hughes}, M., {Qi}, C., \& {Dullemond}, C.~P.
  2010, in Bulletin of the American Astronomical Society, Vol.~42, American
  Astronomical Society Meeting Abstracts \#215, 527

\bibitem[{{Anglada-Escud{\'e}} {et~al.}(2016){Anglada-Escud{\'e}}, {Amado},
  {Barnes}, {Berdi{\~n}as}, {Butler}, {Coleman}, {de La Cueva}, {Dreizler},
  {Endl}, {Giesers}, {Jeffers}, {Jenkins}, {Jones}, {Kiraga}, {K{\"u}rster},
  {L{\'o}pez-Gonz{\'a}lez}, {Marvin}, {Morales}, {Morin}, {Nelson}, {Ortiz},
  {Ofir}, {Paardekooper}, {Reiners}, {Rodr{\'{\i}}guez},
  {Rodr{\'{\i}}guez-L{\'o}pez}, {Sarmiento}, {Strachan}, {Tsapras}, {Tuomi}, \&
  {Zechmeister}}]{angladaescude+2016}
{Anglada-Escud{\'e}}, G., {Amado}, P.~J., {Barnes}, J., {et~al.} 2016, \nat,
  536, 437

\bibitem[{{Ansdell} {et~al.}(2018){Ansdell}, {Williams}, {Trapman}, {van
  Terwisga}, {Facchini}, {Manara}, {van der Marel}, {Miotello}, {Tazzari},
  {Hogerheijde}, {Guidi}, {Testi}, \& {van Dishoeck}}]{ansdell+2018}
{Ansdell}, M., {Williams}, J.~P., {Trapman}, L., {et~al.} 2018, \apj, 859, 21

\bibitem[{{Ansdell} {et~al.}(2016){Ansdell}, {Williams}, {van der Marel},
  {Carpenter}, {Guidi}, {Hogerheijde}, {Mathews}, {Manara}, {Miotello},
  {Natta}, {Oliveira}, {Tazzari}, {Testi}, {van Dishoeck}, \& {van
  Terwisga}}]{ansdell+2016}
{Ansdell}, M., {Williams}, J.~P., {van der Marel}, N., {et~al.} 2016, \apj,
  828, 46

\bibitem[{{Apai} {et~al.}(2004){Apai}, {Pascucci}, {Sterzik}, {van der Bliek},
  {Bouwman}, {Dullemond}, \& {Henning}}]{apai+2004}
{Apai}, D., {Pascucci}, I., {Sterzik}, M.~F., {et~al.} 2004, \aap, 426, L53

\bibitem[{{Ballering} \& {Eisner}(2019)}]{ballering+2019}
{Ballering}, N.~P. \& {Eisner}, J.~A. 2019, \aj, 157, 144

\bibitem[{{Baraffe} \& {Chabrier}(2010)}]{BaraffeChabrier2010}
{Baraffe}, I. \& {Chabrier}, G. 2010, \aap, 521, A44

\bibitem[{{Baraffe} {et~al.}(2015){Baraffe}, {Homeier}, {Allard}, \&
  {Chabrier}}]{baraffe+2015}
{Baraffe}, I., {Homeier}, D., {Allard}, F., \& {Chabrier}, G. 2015, \aap, 577,
  A42

\bibitem[{{Barenfeld} {et~al.}(2016){Barenfeld}, {Carpenter}, {Ricci}, \&
  {Isella}}]{barenfeld+2016}
{Barenfeld}, S.~A., {Carpenter}, J.~M., {Ricci}, L., \& {Isella}, A. 2016,
  \apj, 827, 142

\bibitem[{{Bixel} \& {Apai}(2017)}]{BixelApai2017}
{Bixel}, A. \& {Apai}, D. 2017, \apjl, 836, L31

\bibitem[{{Bustamante} {et~al.}(2015){Bustamante}, {Mer{\'{\i}}n}, {Ribas},
  {Bouy}, {Prusti}, {Pilbratt}, \& {Andr{\'e}}}]{bustamante+2015}
{Bustamante}, I., {Mer{\'{\i}}n}, B., {Ribas}, {\'A}., {et~al.} 2015, \aap,
  578, A23

\bibitem[{{Cazzoletti} {et~al.}(2019){Cazzoletti}, {Manara}, {Baobab Liu}, {van
  Dishoeck}, {Facchini}, {Alcal{\`a}}, {Ansdell}, {Testi}, {Williams}, \&
  {Carrasco-Gonz{\'a}lez}}]{cazzoletti+2019}
{Cazzoletti}, P., {Manara}, C.~F., {Baobab Liu}, H., {et~al.} 2019, \aap, 626,
  A11

\bibitem[{{Chauvin} {et~al.}(2004){Chauvin}, {Lagrange}, {Dumas}, {Zuckerman},
  {Mouillet}, {Song}, {Beuzit}, \& {Lowrance}}]{chauvin+2004}
{Chauvin}, G., {Lagrange}, A.~M., {Dumas}, C., {et~al.} 2004, \aap, 425, L29

\bibitem[{{Chiang} \& {Goldreich}(1997)}]{ChiangGoldreich1997}
{Chiang}, E.~I. \& {Goldreich}, P. 1997, \apj, 490, 368

\bibitem[{{Cieza} {et~al.}(2018){Cieza}, {Ru{\'{\i}}z-Rodr{\'{\i}}guez},
  {Perez}, {Casassus}, {Williams}, {Zurlo}, {Principe}, {Hales}, {Prieto},
  {Tobin}, {Zhu}, \& {Marino}}]{cieza+2018}
{Cieza}, L.~A., {Ru{\'{\i}}z-Rodr{\'{\i}}guez}, D., {Perez}, S., {et~al.} 2018,
  \mnras, 474, 4347

\bibitem[{{Comer{\'o}n}(2008)}]{comeron+2008}
{Comer{\'o}n}, F. 2008, {The Lupus Clouds}, ed. B.~{Reipurth}, 295

\bibitem[{{Comeron} {et~al.}(1998){Comeron}, {Rieke}, {Claes}, {Torra}, \&
  {Laureijs}}]{comeron+1998}
{Comeron}, F., {Rieke}, G.~H., {Claes}, P., {Torra}, J., \& {Laureijs}, R.~J.
  1998, \aap, 335, 522

\bibitem[{{Cox} {et~al.}(2017){Cox}, {Harris}, {Looney}, {Chiang}, {Chandler},
  {Kratter}, {Li}, {Perez}, \& {Tobin}}]{cox+2017}
{Cox}, E.~G., {Harris}, R.~J., {Looney}, L.~W., {et~al.} 2017, \apj, 851, 83

\bibitem[{{Daemgen} {et~al.}(2016){Daemgen}, {Natta}, {Scholz}, {Testi},
  {Jayawardhana}, {Greaves}, \& {Eastwood}}]{daemgen+2016}
{Daemgen}, S., {Natta}, A., {Scholz}, A., {et~al.} 2016, \aap, 594, A83

\bibitem[{{Dodds} {et~al.}(2015){Dodds}, {Greaves}, {Scholz}, {Hatchell},
  {Holland}, \& {JCMT Gould Belt Survey Team}}]{dodds+2015}
{Dodds}, P., {Greaves}, J.~S., {Scholz}, A., {et~al.} 2015, \mnras, 447, 722

\bibitem[{{Dullemond} {et~al.}(2001){Dullemond}, {Dominik}, \&
  {Natta}}]{dullemond+2001}
{Dullemond}, C.~P., {Dominik}, C., \& {Natta}, A. 2001, \apj, 560, 957

\bibitem[{{Dullemond} {et~al.}(2006){Dullemond}, {Natta}, \&
  {Testi}}]{dullemond+2006}
{Dullemond}, C.~P., {Natta}, A., \& {Testi}, L. 2006, The Astrophysical
  Journal, 645, L69

\bibitem[{{Dunham} {et~al.}(2015){Dunham}, {Allen}, {Evans},
  {Broekhoven-Fiene}, {Cieza}, {Di Francesco}, {Gutermuth}, {Harvey},
  {Hatchell}, {Heiderman}, {Huard}, {Johnstone}, {Kirk}, {Matthews}, {Miller},
  {Peterson}, \& {Young}}]{dunham+2015}
{Dunham}, M.~M., {Allen}, L.~E., {Evans}, II, N.~J., {et~al.} 2015, \apjs, 220,
  11

\bibitem[{{Ercolano} {et~al.}(2009){Ercolano}, {Clarke}, \&
  {Robitaille}}]{ercolano+2009}
{Ercolano}, B., {Clarke}, C.~J., \& {Robitaille}, T.~P. 2009, \mnras, 394, L141

\bibitem[{{Facchini} {et~al.}(2019){Facchini}, {van Dishoeck}, {Manara},
  {Tazzari}, {Maud}, {Cazzoletti}, {Rosotti}, {van der Marel}, {Pinilla}, \&
  {Clarke}}]{facchini+2019}
{Facchini}, S., {van Dishoeck}, E.~F., {Manara}, C.~F., {et~al.} 2019, \aap,
  626, L2

\bibitem[{{Foreman-Mackey} {et~al.}(2013){Foreman-Mackey}, {Hogg}, {Lang}, \&
  {Goodman}}]{foremanmackey+2013}
{Foreman-Mackey}, D., {Hogg}, D.~W., {Lang}, D., \& {Goodman}, J. 2013, \pasp,
  125, 306

\bibitem[{{Frasca} {et~al.}(2017){Frasca}, {Biazzo}, {Alcal{\'a}}, {Manara},
  {Stelzer}, {Covino}, \& {Antoniucci}}]{frasca+2017}
{Frasca}, A., {Biazzo}, K., {Alcal{\'a}}, J.~M., {et~al.} 2017, \aap, 602, A33

\bibitem[{{Furlan} {et~al.}(2011){Furlan}, {Luhman}, {Espaillat}, {D'Alessio},
  {Adame}, {Manoj}, {Kim}, {Watson}, {Forrest}, {McClure}, {Calvet}, {Sargent},
  {Green}, \& {Fischer}}]{furlan+2011}
{Furlan}, E., {Luhman}, K.~L., {Espaillat}, C., {et~al.} 2011, \apjs, 195, 3

\bibitem[{{Gaia Collaboration} {et~al.}(2018){Gaia Collaboration}, {Brown},
  {Vallenari}, {Prusti}, {de Bruijne}, {Babusiaux}, {Bailer-Jones}, {Biermann},
  {Evans}, {Eyer}, \& et~al.}]{gaiacollaboration2018}
{Gaia Collaboration}, {Brown}, A.~G.~A., {Vallenari}, A., {et~al.} 2018, \aap,
  616, A1

\bibitem[{{Gillon} {et~al.}(2017){Gillon}, {Triaud}, {Demory}, {Jehin}, {Agol},
  {Deck}, {Lederer}, {de Wit}, {Burdanov}, {Ingalls}, {Bolmont}, {Leconte},
  {Raymond}, {Selsis}, {Turbet}, {Barkaoui}, {Burgasser}, {Burleigh}, {Carey},
  {Chaushev}, {Copperwheat}, {Delrez}, {Fernandes}, {Holdsworth}, {Kotze}, {Van
  Grootel}, {Almleaky}, {Benkhaldoun}, {Magain}, \& {Queloz}}]{gillon+2017}
{Gillon}, M., {Triaud}, A.~H.~M.~J., {Demory}, B.-O., {et~al.} 2017, \nat, 542,
  456

\bibitem[{{Goodman} \& {Weare}(2010)}]{GoodmanWeare2010}
{Goodman}, J. \& {Weare}, J. 2010, Communications in Applied Mathematics and
  Computational Science, Vol.~5, No.~1, p.~65-80, 2010, 5, 65

\bibitem[{{Greaves} \& {Rice}(2010)}]{GreavesRice2010}
{Greaves}, J.~S. \& {Rice}, W.~K.~M. 2010, \mnras, 407, 1981

\bibitem[{{Gregory} {et~al.}(2016){Gregory}, {Adams}, \&
  {Davies}}]{gregory+2016}
{Gregory}, S.~G., {Adams}, F.~C., \& {Davies}, C.~L. 2016, \mnras, 457, 3836

\bibitem[{{Guilloteau} {et~al.}(2011){Guilloteau}, {Dutrey}, {Pi{\'e}tu}, \&
  {Boehler}}]{guilloteau+2011}
{Guilloteau}, S., {Dutrey}, A., {Pi{\'e}tu}, V., \& {Boehler}, Y. 2011, \aap,
  529, A105

\bibitem[{{Hendler} {et~al.}(2017){Hendler}, {Mulders}, {Pascucci},
  {Greenwood}, {Kamp}, {Henning}, {M{\'e}nard}, {Dent}, \&
  {Evans}}]{hendler+2017}
{Hendler}, N.~P., {Mulders}, G.~D., {Pascucci}, I., {et~al.} 2017, \apj, 841,
  116

\bibitem[{{Huang} {et~al.}(2018){Huang}, {Andrews}, {Dullemond}, {Isella},
  {P{\'e}rez}, {Guzm{\'a}n}, {{\"O}berg}, {Zhu}, {Zhang}, {Bai}, {Benisty},
  {Birnstiel}, {Carpenter}, {Hughes}, {Ricci}, {Weaver}, \&
  {Wilner}}]{huang+2018}
{Huang}, J., {Andrews}, S.~M., {Dullemond}, C.~P., {et~al.} 2018, \apj, 869,
  L42

\bibitem[{{Hughes} {et~al.}(1994){Hughes}, {Hartigan}, {Krautter}, \&
  {Kelemen}}]{hughes+1994}
{Hughes}, J., {Hartigan}, P., {Krautter}, J., \& {Kelemen}, J. 1994, \aj, 108,
  1071

\bibitem[{{Isella} {et~al.}(2019){Isella}, {Benisty}, {Teague}, {Bae},
  {Keppler}, {Facchini}, \& {P{\'e}rez}}]{isella+2019}
{Isella}, A., {Benisty}, M., {Teague}, R., {et~al.} 2019, arXiv e-prints,
  arXiv:1906.06308

\bibitem[{{Jayawardhana} {et~al.}(2003){Jayawardhana}, {Mohanty}, \&
  {Basri}}]{jayawardhana+2003}
{Jayawardhana}, R., {Mohanty}, S., \& {Basri}, G. 2003, \apj, 592, 282

\bibitem[{{Johns-Krull} {et~al.}(2016){Johns-Krull}, {McLane}, {Prato},
  {Crockett}, {Jaffe}, {Hartigan}, {Beichman}, {Mahmud}, {Chen}, \&
  {Skiff}}]{johnskrull+2016}
{Johns-Krull}, C.~M., {McLane}, J.~N., {Prato}, L., {et~al.} 2016, \apj, 826,
  206

\bibitem[{{Jones} {et~al.}(2012){Jones}, {Pringle}, \&
  {Alexander}}]{jones+2012}
{Jones}, M.~G., {Pringle}, J.~E., \& {Alexander}, R.~D. 2012, \mnras, 419, 925

\bibitem[{{Jung} {et~al.}(2018){Jung}, {Udalski}, {Gould}, {Ryu}, {Yee}, {and},
  {Han}, {Albrow}, {Lee}, {Kim}, {Hwang}, {Chung}, {Shin}, {Zhu}, {Cha}, {Kim},
  {Lee}, {Park}, {Lee}, {Kim}, {Pogge}, {KMTNet Collaboration},
  {Szyma{\'n}ski}, {Mr{\'o}z}, {Poleski}, {Skowron}, {Pietrukowicz},
  {Soszy{\'n}ski}, {Koz{\l}owski}, {Ulaczyk}, {Pawlak}, {Rybicki}, \& {OGLE
  Collaboration}}]{jung+2018}
{Jung}, Y.~K., {Udalski}, A., {Gould}, A., {et~al.} 2018, \aj, 155, 219

\bibitem[{{Kelly}(2007)}]{kelly2007}
{Kelly}, B.~C. 2007, \apj, 665, 1489

\bibitem[{{Keppler} {et~al.}(2018){Keppler}, {Benisty}, {M{\"u}ller},
  {Henning}, {van Boekel}, {Cantalloube}, {Ginski}, {van Holstein}, {Maire},
  {Pohl}, {Samland}, {Avenhaus}, {Baudino}, {Boccaletti}, {de Boer},
  {Bonnefoy}, {Chauvin}, {Desidera}, {Langlois}, {Lazzoni}, {Marleau},
  {Mordasini}, {Pawellek}, {Stolker}, {Vigan}, {Zurlo}, {Birnstiel},
  {Brandner}, {Feldt}, {Flock}, {Girard}, {Gratton}, {Hagelberg}, {Isella},
  {Janson}, {Juhasz}, {Kemmer}, {Kral}, {Lagrange}, {Launhardt}, {Matter},
  {M{\'e}nard}, {Milli}, {Molli{\`e}re}, {Olofsson}, {P{\'e}rez}, {Pinilla},
  {Pinte}, {Quanz}, {Schmidt}, {Udry}, {Wahhaj}, {Williams}, {Buenzli},
  {Cudel}, {Dominik}, {Galicher}, {Kasper}, {Lannier}, {Mesa}, {Mouillet},
  {Peretti}, {Perrot}, {Salter}, {Sissa}, {Wildi}, {Abe}, {Antichi},
  {Augereau}, {Baruffolo}, {Baudoz}, {Bazzon}, {Beuzit}, {Blanchard}, {Brems},
  {Buey}, {De Caprio}, {Carbillet}, {Carle}, {Cascone}, {Cheetham}, {Claudi},
  {Costille}, {Delboulb{\'e}}, {Dohlen}, {Fantinel}, {Feautrier}, {Fusco},
  {Giro}, {Gluck}, {Gry}, {Hubin}, {Hugot}, {Jaquet}, {Le Mignant}, {Llored},
  {Madec}, {Magnard}, {Martinez}, {Maurel}, {Meyer}, {M{\"o}ller-Nilsson},
  {Moulin}, {Mugnier}, {Orign{\'e}}, {Pavlov}, {Perret}, {Petit}, {Pragt},
  {Puget}, {Rabou}, {Ramos}, {Rigal}, {Rochat}, {Roelfsema}, {Rousset}, {Roux},
  {Salasnich}, {Sauvage}, {Sevin}, {Soenke}, {Stadler}, {Suarez}, {Turatto}, \&
  {Weber}}]{keppler+2018}
{Keppler}, M., {Benisty}, M., {M{\"u}ller}, A., {et~al.} 2018, \aap, 617, A44

\bibitem[{{Klein} {et~al.}(2003){Klein}, {Apai}, {Pascucci}, {Henning}, \&
  {Waters}}]{klein+2003}
{Klein}, R., {Apai}, D., {Pascucci}, I., {Henning}, T., \& {Waters},
  L.~B.~F.~M. 2003, \apjl, 593, L57

\bibitem[{{Lauer} {et~al.}(1995){Lauer}, {Ajhar}, {Byun}, {Dressler}, {Faber},
  {Grillmair}, {Kormendy}, {Richstone}, \& {Tremaine}}]{lauer+1995}
{Lauer}, T.~R., {Ajhar}, E.~A., {Byun}, Y.-I., {et~al.} 1995, \aj, 110, 2622

\bibitem[{{Lodato} {et~al.}(2019){Lodato}, {Dipierro}, {Ragusa}, {Long},
  {Herczeg}, {Pascucci}, {Pinilla}, {Manara}, {Tazzari}, {Liu}, {Mulders},
  {Harsono}, {Boehler}, {M{\'e}nard}, {Johnstone}, {Salyk}, {van der Plas},
  {Cabrit}, {Edwards}, {Fischer}, {Hendler}, {Nisini}, {Rigliaco}, {Avenhaus},
  {Banzatti}, \& {Gully-Santiago}}]{lodato+2019}
{Lodato}, G., {Dipierro}, G., {Ragusa}, E., {et~al.} 2019, Monthly Notices of
  the Royal Astronomical Society, 486, 453

\bibitem[{{Lodato} {et~al.}(2017){Lodato}, {Scardoni}, {Manara}, \&
  {Testi}}]{lodato+2017}
{Lodato}, G., {Scardoni}, C.~E., {Manara}, C.~F., \& {Testi}, L. 2017, Monthly
  Notices of the Royal Astronomical Society, 472, 4700

\bibitem[{{Long} {et~al.}(2019){Long}, {Herczeg}, {Harsono}, {Pinilla},
  {Tazzari}, {Manara}, {Pascucci}, {Cabrit}, {Nisini}, {Johnstone}, {Edwards},
  {Salyk}, {Menard}, {Lodato}, {Boehler}, {Mace}, {Liu}, {Mulders}, {Hendler},
  {Ragusa}, {Fischer}, {Banzatti}, {Rigliaco}, {van de Plas}, {Dipierro},
  {Gully-Santiago}, \& {Lopez-Valdivia}}]{long+2019}
{Long}, F., {Herczeg}, G.~J., {Harsono}, D., {et~al.} 2019, \apj, 882, 49

\bibitem[{{Luhman}(2012)}]{luhman2012}
{Luhman}, K.~L. 2012, \araa, 50, 65

\bibitem[{{MacGregor} {et~al.}(2017){MacGregor}, {Wilner}, {Czekala},
  {Andrews}, {Dai}, {Herczeg}, {Kratter}, {Kraus}, {Ricci}, \&
  {Testi}}]{macgregor+2017}
{MacGregor}, M.~A., {Wilner}, D.~J., {Czekala}, I., {et~al.} 2017, \apj, 835,
  17

\bibitem[{{Manara} {et~al.}(2016{\natexlab{a}}){Manara}, {Fedele}, {Herczeg},
  \& {Teixeira}}]{manara+2016a}
{Manara}, C.~F., {Fedele}, D., {Herczeg}, G.~J., \& {Teixeira}, P.~S.
  2016{\natexlab{a}}, \aap, 585, A136

\bibitem[{{Manara} {et~al.}(2018){Manara}, {Morbidelli}, \&
  {Guillot}}]{manara+2018}
{Manara}, C.~F., {Morbidelli}, A., \& {Guillot}, T. 2018, \aap, 618, L3

\bibitem[{{Manara} {et~al.}(2016{\natexlab{b}}){Manara}, {Rosotti}, {Testi},
  {Natta}, {Alcal{\'a}}, {Williams}, {Ansdell}, {Miotello}, {van der Marel},
  {Tazzari}, {Carpenter}, {Guidi}, {Mathews}, {Oliveira}, {Prusti}, \& {van
  Dishoeck}}]{manara+2016}
{Manara}, C.~F., {Rosotti}, G., {Testi}, L., {et~al.} 2016{\natexlab{b}}, \aap,
  591, L3

\bibitem[{{Manara} {et~al.}(2019){Manara}, {Tazzari}, {Long}, {Herczeg},
  {Lodato}, {Rota}, {Cazzoletti}, {van der Plas}, {Pinilla}, {Dipierro},
  {Edwards}, {Harsono}, {Johnstone}, {Liu}, {Menard}, {Nisini}, {Ragusa},
  {Boehler}, \& {Cabrit}}]{manara+2019}
{Manara}, C.~F., {Tazzari}, M., {Long}, F., {et~al.} 2019, \aap, 628, A95

\bibitem[{{Manara} {et~al.}(2017){Manara}, {Testi}, {Herczeg}, {Pascucci},
  {Alcal{\'a}}, {Natta}, {Antoniucci}, {Fedele}, {Mulders}, {Henning},
  {Mohanty}, {Prusti}, \& {Rigliaco}}]{manara2017a}
{Manara}, C.~F., {Testi}, L., {Herczeg}, G.~J., {et~al.} 2017, \aap, 604, A127

\bibitem[{{Mer{\'{\i}}n} {et~al.}(2008){Mer{\'{\i}}n}, {J{\o}rgensen},
  {Spezzi}, {Alcal{\'a}}, {Evans}, {Harvey}, {Prusti}, {Chapman}, {Huard}, {van
  Dishoeck}, \& {Comer{\'o}n}}]{merin+2008}
{Mer{\'{\i}}n}, B., {J{\o}rgensen}, J., {Spezzi}, L., {et~al.} 2008, \apjs,
  177, 551

\bibitem[{{Mohanty} {et~al.}(2005){Mohanty}, {Jayawardhana}, \&
  {Basri}}]{mohanty+2005}
{Mohanty}, S., {Jayawardhana}, R., \& {Basri}, G. 2005, \apj, 626, 498

\bibitem[{{Mortier} {et~al.}(2011){Mortier}, {Oliveira}, \& {van
  Dishoeck}}]{mortier+2011}
{Mortier}, A., {Oliveira}, I., \& {van Dishoeck}, E.~F. 2011, \mnras, 418, 1194

\bibitem[{{Mulders} \& {Dominik}(2012)}]{MuldersDominik2012}
{Mulders}, G.~D. \& {Dominik}, C. 2012, \aap, 539, A9

\bibitem[{{Mulders} {et~al.}(2015){Mulders}, {Pascucci}, \&
  {Apai}}]{mulders+2015}
{Mulders}, G.~D., {Pascucci}, I., \& {Apai}, D. 2015, \apj, 814, 130

\bibitem[{{Mulders} {et~al.}(2017){Mulders}, {Pascucci}, {Manara}, {Testi},
  {Herczeg}, {Henning}, {Mohanty}, \& {Lodato}}]{mulders+2017}
{Mulders}, G.~D., {Pascucci}, I., {Manara}, C.~F., {et~al.} 2017, \apj, 847, 31

\bibitem[{{Mu{\v z}i{\'c}} {et~al.}(2015){Mu{\v z}i{\'c}}, {Scholz}, {Geers},
  \& {Jayawardhana}}]{muzic2015}
{Mu{\v z}i{\'c}}, K., {Scholz}, A., {Geers}, V.~C., \& {Jayawardhana}, R. 2015,
  \apj, 810, 159

\bibitem[{{Mu{\v z}i{\'c}} {et~al.}(2014){Mu{\v z}i{\'c}}, {Scholz}, {Geers},
  {Jayawardhana}, \& {L{\'o}pez Mart{\'{\i}}}}]{muzic+2014}
{Mu{\v z}i{\'c}}, K., {Scholz}, A., {Geers}, V.~C., {Jayawardhana}, R., \&
  {L{\'o}pez Mart{\'{\i}}}, B. 2014, \apj, 785, 159

\bibitem[{{Muzerolle} {et~al.}(2003){Muzerolle}, {Hillenbrand}, {Calvet},
  {Brice{\~n}o}, \& {Hartmann}}]{muzerolle+2003}
{Muzerolle}, J., {Hillenbrand}, L., {Calvet}, N., {Brice{\~n}o}, C., \&
  {Hartmann}, L. 2003, \apj, 592, 266

\bibitem[{{Muzerolle} {et~al.}(2005){Muzerolle}, {Luhman}, {Brice{\~n}o},
  {Hartmann}, \& {Calvet}}]{muzerolle+2005}
{Muzerolle}, J., {Luhman}, K.~L., {Brice{\~n}o}, C., {Hartmann}, L., \&
  {Calvet}, N. 2005, \apj, 625, 906

\bibitem[{{Najita} \& {Kenyon}(2014)}]{NajitaKenyon2014}
{Najita}, J.~R. \& {Kenyon}, S.~J. 2014, \mnras, 445, 3315

\bibitem[{{Natta} \& {Testi}(2001)}]{NattaTesti2001}
{Natta}, A. \& {Testi}, L. 2001, \aap, 376, L22

\bibitem[{{Natta} {et~al.}(2007){Natta}, {Testi}, {Calvet}, {Henning},
  {Waters}, \& {Wilner}}]{NattaTesti2007}
{Natta}, A., {Testi}, L., {Calvet}, N., {et~al.} 2007, in Protostars and
  Planets V, ed. B.~{Reipurth}, D.~{Jewitt}, \& K.~{Keil}, 767

\bibitem[{{Natta} {et~al.}(2002){Natta}, {Testi}, {Comer{\'o}n}, {Oliva},
  {D'Antona}, {Baffa}, {Comoretto}, \& {Gennari}}]{natta+2002}
{Natta}, A., {Testi}, L., {Comer{\'o}n}, F., {et~al.} 2002, \aap, 393, 597

\bibitem[{{Natta} {et~al.}(2004){Natta}, {Testi}, {Muzerolle}, {Randich},
  {Comer{\'o}n}, \& {Persi}}]{natta+2004}
{Natta}, A., {Testi}, L., {Muzerolle}, J., {et~al.} 2004, \aap, 424, 603

\bibitem[{{Pascucci} {et~al.}(2003){Pascucci}, {Apai}, {Henning}, \&
  {Dullemond}}]{pascucci+2003}
{Pascucci}, I., {Apai}, D., {Henning}, T., \& {Dullemond}, C.~P. 2003, \apjl,
  590, L111

\bibitem[{{Pascucci} {et~al.}(2016){Pascucci}, {Testi}, {Herczeg}, {Long},
  {Manara}, {Hendler}, {Mulders}, {Krijt}, {Ciesla}, {Henning}, {Mohanty},
  {Drabek-Maunder}, {Apai}, {Sz{\H u}cs}, {Sacco}, \&
  {Olofsson}}]{pascucci+2016}
{Pascucci}, I., {Testi}, L., {Herczeg}, G.~J., {et~al.} 2016, \apj, 831, 125

\bibitem[{{Payne} \& {Lodato}(2007)}]{PayneLodato2007}
{Payne}, M.~J. \& {Lodato}, G. 2007, \mnras, 381, 1597

\bibitem[{{P{\'e}rez} {et~al.}(2019){P{\'e}rez}, {Casassus}, {Hales}, {Marino},
  {Cheetham}, {Zurlo}, {Cieza}, {Dong}, {Alarc{\'o}n}, \&
  {Ben{\'\i}tez-Llambay}}]{perez+2019}
{P{\'e}rez}, S., {Casassus}, S., {Hales}, A., {et~al.} 2019, arXiv e-prints,
  arXiv:1906.06305

\bibitem[{{Pi{\'e}tu} {et~al.}(2014){Pi{\'e}tu}, {Guilloteau}, {Di Folco},
  {Dutrey}, \& {Boehler}}]{pietu+2014}
{Pi{\'e}tu}, V., {Guilloteau}, S., {Di Folco}, E., {Dutrey}, A., \& {Boehler},
  Y. 2014, \aap, 564, A95

\bibitem[{{Pinilla} {et~al.}(2018){Pinilla}, {Natta}, {Manara}, {Ricci},
  {Scholz}, \& {Testi}}]{pinilla+2018b}
{Pinilla}, P., {Natta}, A., {Manara}, C.~F., {et~al.} 2018, \aap, 615, A95

\bibitem[{{Pinte} {et~al.}(2018){Pinte}, {Price}, {M{\'e}nard}, {Duch{\^e}ne},
  {Dent}, {Hill}, {de Gregorio-Monsalvo}, {Hales}, \& {Mentiplay}}]{pinte+2018}
{Pinte}, C., {Price}, D.~J., {M{\'e}nard}, F., {et~al.} 2018, \apjl, 860, L13

\bibitem[{{Ricci} {et~al.}(2014){Ricci}, {Testi}, {Natta}, {Scholz}, {de
  Gregorio-Monsalvo}, \& {Isella}}]{ricci+2014}
{Ricci}, L., {Testi}, L., {Natta}, A., {et~al.} 2014, \apj, 791, 20

\bibitem[{{Rosotti} {et~al.}(2017){Rosotti}, {Clarke}, {Manara}, \&
  {Facchini}}]{rosotti+2017}
{Rosotti}, G.~P., {Clarke}, C.~J., {Manara}, C.~F., \& {Facchini}, S. 2017,
  \mnras, 468, 1631

\bibitem[{{Rosotti} {et~al.}(2019){Rosotti}, {Tazzari}, {Booth}, {Testi},
  {Lodato}, \& {Clarke}}]{rosotti+2019}
{Rosotti}, G.~P., {Tazzari}, M., {Booth}, R.~A., {et~al.} 2019, \mnras, 486,
  4829

\bibitem[{{Sanchis} {et~al.}(2019){Sanchis}, {Picogna}, {Ercolano}, {Testi}, \&
  {Rosotti}}]{sanchis+}
{Sanchis}, E., {Picogna}, G., {Ercolano}, B., {Testi}, L., \& {Rosotti}, G.
  2019, \mnras, submitted (not yet accepted)

\bibitem[{{Scholz}(2008)}]{scholz2008}
{Scholz}, A. 2008, Reviews in Modern Astronomy, 20, 357

\bibitem[{{Scholz} \& {Eisl{\"o}ffel}(2004)}]{ScholzEisloeffel2004}
{Scholz}, A. \& {Eisl{\"o}ffel}, J. 2004, \aap, 419, 249

\bibitem[{{Shakura} \& {Sunyaev}(1973)}]{ShakuraSunyaev1973}
{Shakura}, N.~I. \& {Sunyaev}, R.~A. 1973, \aap, 500, 33

\bibitem[{{Siess} {et~al.}(2000){Siess}, {Dufour}, \& {Forestini}}]{siess+2000}
{Siess}, L., {Dufour}, E., \& {Forestini}, M. 2000, \aap, 358, 593

\bibitem[{{Sonnhalter} {et~al.}(1995){Sonnhalter}, {Preibisch}, \&
  {Yorke}}]{sonnhalter+1995}
{Sonnhalter}, C., {Preibisch}, T., \& {Yorke}, H.~W. 1995, \aap, 299, 545

\bibitem[{Tazzari(2017)}]{uvplot_mtazzari}
Tazzari, M. 2017, mtazzari/uvplot: v0.1.1

\bibitem[{{Tazzari} {et~al.}(2018){Tazzari}, {Beaujean}, \&
  {Testi}}]{tazzari+2018}
{Tazzari}, M., {Beaujean}, F., \& {Testi}, L. 2018, \mnras, 476, 4527

\bibitem[{{Tazzari} {et~al.}(2016){Tazzari}, {Testi}, {Ercolano}, {Natta},
  {Isella}, {Chandler}, {P{\'e}rez}, {Andrews}, {Wilner}, {Ricci}, {Henning},
  {Linz}, {Kwon}, {Corder}, {Dullemond}, {Carpenter}, {Sargent}, {Mundy},
  {Storm}, {Calvet}, {Greaves}, {Lazio}, \& {Deller}}]{tazzari+2016}
{Tazzari}, M., {Testi}, L., {Ercolano}, B., {et~al.} 2016, \aap, 588, A53

\bibitem[{{Tazzari} {et~al.}(2017){Tazzari}, {Testi}, {Natta}, {Ansdell},
  {Carpenter}, {Guidi}, {Hogerheijde}, {Manara}, {Miotello}, {van der Marel},
  {van Dishoeck}, \& {Williams}}]{tazzari+2017A}
{Tazzari}, M., {Testi}, L., {Natta}, A., {et~al.} 2017, \aap, 606, A88

\bibitem[{{Teague} {et~al.}(2018){Teague}, {Bae}, {Bergin}, {Birnstiel}, \&
  {Foreman-Mackey}}]{teague+2018}
{Teague}, R., {Bae}, J., {Bergin}, E.~A., {Birnstiel}, T., \& {Foreman-Mackey},
  D. 2018, \apjl, 860, L12

\bibitem[{{Testi} {et~al.}(2014){Testi}, {Birnstiel}, {Ricci}, {Andrews},
  {Blum}, {Carpenter}, {Dominik}, {Isella}, {Natta}, {Williams}, \&
  {Wilner}}]{testi+2014}
{Testi}, L., {Birnstiel}, T., {Ricci}, L., {et~al.} 2014, in Protostars and
  Planets VI, ed. H.~{Beuther}, R.~S. {Klessen}, C.~P. {Dullemond}, \&
  T.~{Henning}, 339

\bibitem[{{Testi} {et~al.}(2016){Testi}, {Natta}, {Scholz}, {Tazzari}, {Ricci},
  \& {de Gregorio Monsalvo}}]{testi+2016}
{Testi}, L., {Natta}, A., {Scholz}, A., {et~al.} 2016, \aap, 593, A111

\bibitem[{{Todorov} {et~al.}(2010){Todorov}, {Luhman}, \&
  {McLeod}}]{todorov+2010}
{Todorov}, K., {Luhman}, K.~L., \& {McLeod}, K.~K. 2010, \apjl, 714, L84

\bibitem[{{Tripathi} {et~al.}(2017){Tripathi}, {Andrews}, {Birnstiel}, \&
  {Wilner}}]{tripathi+2017}
{Tripathi}, A., {Andrews}, S.~M., {Birnstiel}, T., \& {Wilner}, D.~J. 2017,
  \apj, 845, 44

\bibitem[{{van der Plas} {et~al.}(2016){van der Plas}, {M{\'e}nard},
  {Ward-Duong}, {Bulger}, {Harvey}, {Pinte}, {Patience}, {Hales}, \&
  {Casassus}}]{vanderplas+2016}
{van der Plas}, G., {M{\'e}nard}, F., {Ward-Duong}, K., {et~al.} 2016, \apj,
  819, 102

\bibitem[{{van Terwisga} {et~al.}(2018){van Terwisga}, {van Dishoeck},
  {Ansdell}, {van der Marel}, {Testi}, {Williams}, {Facchini}, {Tazzari},
  {Hogerheijde}, {Trapman}, {Manara}, {Miotello}, {Maud}, \&
  {Harsono}}]{vanterwisga2018A}
{van Terwisga}, S.~E., {van Dishoeck}, E.~F., {Ansdell}, M., {et~al.} 2018,
  \aap, 616, A88

\bibitem[{{van Terwisga} {et~al.}(2019){van Terwisga}, {van Dishoeck},
  {Cazzoletti}, {Facchini}, {Trapman}, {Williams}, {Manara}, {Miotello}, {van
  der Marel}, {Ansdell}, {Hogerheijde}, {Tazzari}, \&
  {Testi}}]{vanterwisga+2019}
{van Terwisga}, S.~E., {van Dishoeck}, E.~F., {Cazzoletti}, P., {et~al.} 2019,
  \aap, 623, A150

\bibitem[{{Vorobyov} \& {Basu}(2009)}]{VorobyovBasu2009}
{Vorobyov}, E.~I. \& {Basu}, S. 2009, \apj, 703, 922

\bibitem[{{Wang} {et~al.}(2017){Wang}, {Wu}, {Barclay}, \&
  {Laughlin}}]{wang+2017}
{Wang}, S., {Wu}, D.-H., {Barclay}, T., \& {Laughlin}, G.~P. 2017, arXiv
  e-prints, arXiv:1704.04290

\bibitem[{{Ward-Duong} {et~al.}(2018){Ward-Duong}, {Patience}, {Bulger}, {van
  der Plas}, {M{\'e}nard}, {Pinte}, {Jackson}, {Bryden}, {Turner}, {Harvey},
  {Hales}, \& {De Rosa}}]{wardduong+2018}
{Ward-Duong}, K., {Patience}, J., {Bulger}, J., {et~al.} 2018, \aj, 155, 54

\bibitem[{{Whelan} {et~al.}(2005){Whelan}, {Ray}, {Bacciotti}, {Natta},
  {Testi}, \& {Randich}}]{whelan+2005}
{Whelan}, E.~T., {Ray}, T.~P., {Bacciotti}, F., {et~al.} 2005, \nat, 435, 652

\bibitem[{{Williams}(2012)}]{williams2012}
{Williams}, J.~P. 2012, Meteoritics and Planetary Science, 47, 1915

\bibitem[{{Yorke} {et~al.}(1993){Yorke}, {Bodenheimer}, \&
  {Laughlin}}]{yorke+1993}
{Yorke}, H.~W., {Bodenheimer}, P., \& {Laughlin}, G. 1993, \apj, 411, 274

\bibitem[{{Zhang} {et~al.}(2018){Zhang}, {Zhu}, {Huang}, {Guzm{\'a}n},
  {Andrews}, {Birnstiel}, {Dullemond}, {Carpenter}, {Isella}, {P{\'e}rez},
  {Benisty}, {Wilner}, {Baruteau}, {Bai}, \& {Ricci}}]{zhang+2018}
{Zhang}, S., {Zhu}, Z., {Huang}, J., {et~al.} 2018, The Astrophysical Journal,
  869, L47

\bibitem[{{Zhu} {et~al.}(2019){Zhu}, {Zhang}, {Jiang}, {Kataoka}, {Birnstiel},
  {Dullemond}, {Andrews}, {Huang}, {P{\'e}rez}, \& {Carpenter}}]{zhu+2019}
{Zhu}, Z., {Zhang}, S., {Jiang}, Y.-F., {et~al.} 2019, \apj, 877, L18

\end{thebibliography}

%%%%%%%%%%%%%%%%%%%%%%%%%%%%%%%%%%%%%%%%%%%%%%%%%%%%%
%%%%%%%%%%%%%%%%% APPENDIX %%%%%%%%%%%%%%%%%%%%%%%%%%

\begin{appendix}

\section{Tests comparing $R_{68\%}$ and $R_{95\%}$}\label{sec:appendix_radiustest}
We computed the radii enclosing $68\%$ ($R_{68\%}$) and $95\%$ ($R_{95\%}$) for different models fitting a disk with high S/N and well resolved continuum emission, in order to test which radius is better as the characteristic size of the disk. Additionally, we demonstrate that the gaussian function can be used to describe the interferometric data of moderate angular resolution observations. For the calculation of these radii we build the cumulative flux as a function of radius:
\begin{equation}\label{eq:fcumul}
f_{\mathrm{cumul}} (R) = 2 \pi \cdot \int_{0}^{R} I_{\nu}({R}^{\prime}) \cdot {R}^{\prime} \cdot \, \mathrm{d}{R}^{\prime} \mathrm{,}
\end{equation}
which gives us the flux contained within the radius $R$. Therefore, $R_{68\%}$ and $R_{95\%}$ is obtained from $f_{\mathrm{cumul}} (R_{68\%}) = 0.68 \cdot F_{\mathrm{tot}}$ and $f_{\mathrm{cumul}} (R_{95\%}) = 0.95 \cdot F_{\mathrm{tot}}$. The total disk flux density $F_{\mathrm{tot}}$ is one of the free parameters of our fit and is obtained from its PDF. We build the PDFs of the derived radii: the median value of each distribution is used as the inferred value for each radius.

\begin{figure}
  \resizebox{\hsize}{!}{\includegraphics{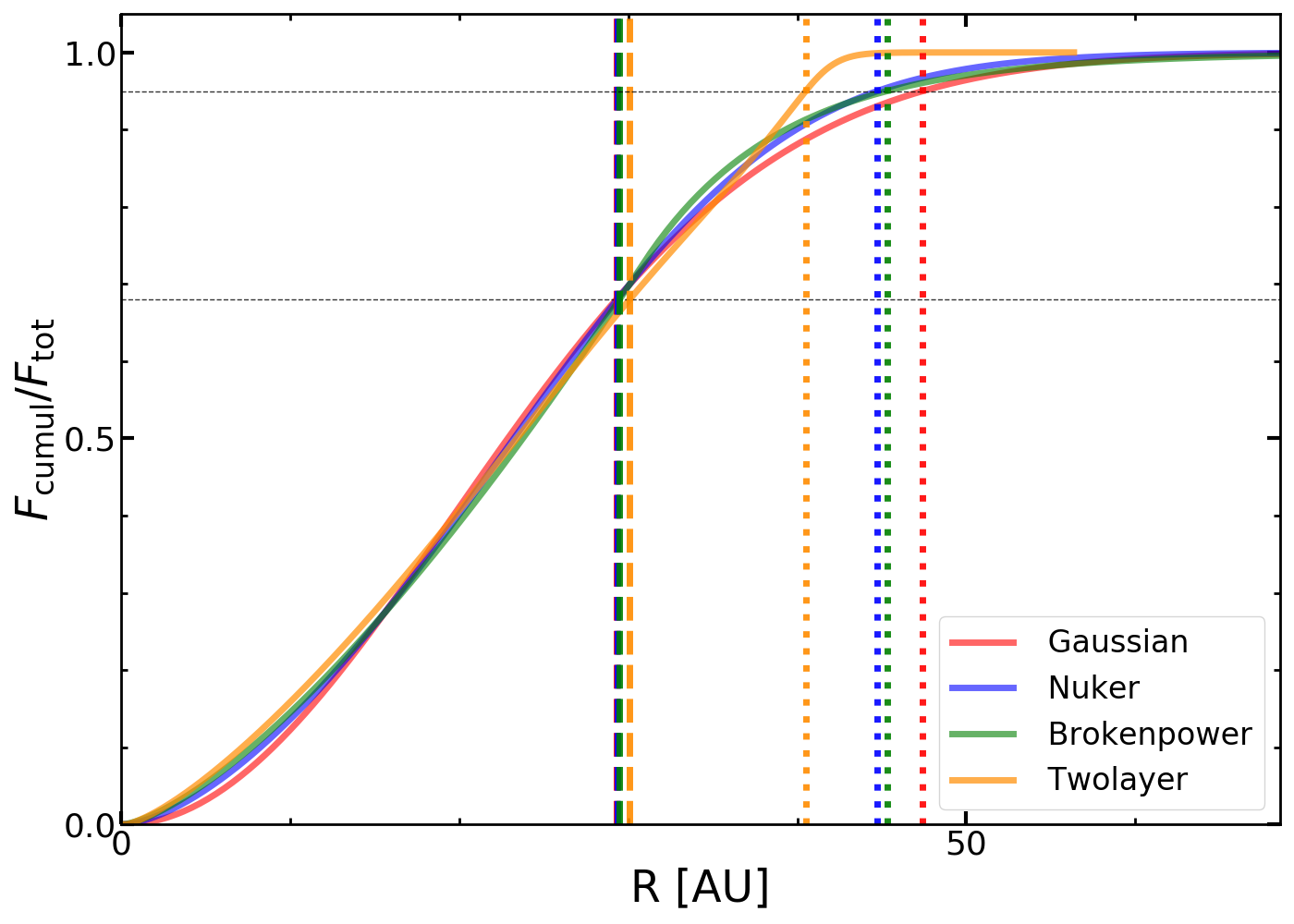}}
  \caption{Normalized cumulative fluxes of Various fits used to model the interferometric visibilities of RXJ$1556.1$-$3655$ disk, used for the size determination. The results are shown for the following models: a gaussian distribution as in Equation~\ref{eq:gaussian} (illustrated in red), the Nuker profile from Equation~\ref{eq:nuker} (blue), a broken power-law (green), and for the "Two-layer" approximation physical model to describe the disk (orange).  The radii enclosing $68 \%$ and $95 \%$ of the total flux for each model are shown as dashed and dotted lines respectively.}
  \label{fig:severalfits}
\end{figure}

The resulting radii using three empirical models (gaussian, function, Nuker function, and broken power-law) and one physical model (two-layer approximation) were used to fit RXJ$1556.1$-$3655$ disk, the results are shown in Figure~\ref{fig:severalfits}. The values of $R_{68\%}$ for the various models range between [$29.3$, $30.1$] $\mathrm{au}$, while the values of $R_{95\%}$ range between [$40.5$, $47.4$] $\mathrm{au}$.

These results show that the difference of $R_{68\%}$ between different models is negligible (thus the value of $R_{68\%}$ is independent of the model used to fit our data) and for the $R_{95\%}$ the values may differ and indeed be considerable. This is in accordance with \cite{tripathi+2017}, and it is a consequence of the low sensitivity of larger scales, thus the noise makes the fitting of the visibilities uncertain. Therefore, the $R_{68\%}$ is favored over the $R_{95\%}$ as the most reliable size definition, since the differences between models are negligible.

On the other hand, this test also shows that the gaussian function is a reliable model to describe emission of moderate angular resolution observations, since the $R_{68\%}$ obtained from the gaussian model is in perfect agreement with the value inferred using other empirical models.

\begin{figure}
  \resizebox{\hsize}{!}{\includegraphics{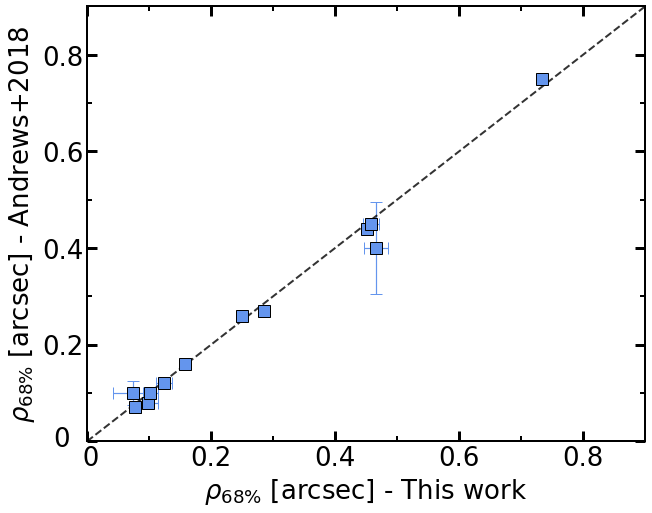}}
  \caption{Comparison of the characteristic size between this work and \cite{andrews2018A}. For this comparison we followed the modeling described in Section~\ref{sec:modeling} and fitted the interferometric data of a random sub-set of Lupus disks that were previously analyzed in \cite{andrews2018A}. The dashed line represents the 1:1 ratio of the radii values.}
  \label{fig:rnukercomparison}
\end{figure}

We also tested the quality of our disk size results, by performing additional fits of disks around stars in Lupus that were already characterized in \cite{andrews2018A}. The comparison between the inferred radii with the radii presented in \cite{andrews2018A} is shown in Figure~\ref{fig:rnukercomparison}; the disk size values are indeed in very good agreement. Therefore the combination of the disk size results of the new disks modeled in this work and the disks modeled in \cite{andrews2018A} can be done adequately for the demographic analysis in Section~\ref{sec:mdustr68}.

\section{Disk properties relations using other dust temperature prescriptions}\label{sec:appendix_otherrelations}
The demographic analysis comparing BD and stellar disks for the relations between disk properties was investigated for various prescriptions of the dust temperature of the disk. The results shown along Section~\ref{sec:discussion} in the text are obtained assuming a constant $T_{\mathrm{dust}}$ of $20$ K. Here we show the disk properties relations for other dust temperature dependence with stellar luminosity. The prescription from \cite{andrews+2013} was designed for disks around central objects of $L_{\star} \in [0.1, 100]$ $L_{\odot}$; the one from \cite{vanderplas+2016} is more suitable for very low-mass objects (VLMs and BDs).

\subsection{Dust temperature from \cite{andrews+2013}}
\begin{figure}
  \resizebox{\hsize}{!}{\includegraphics{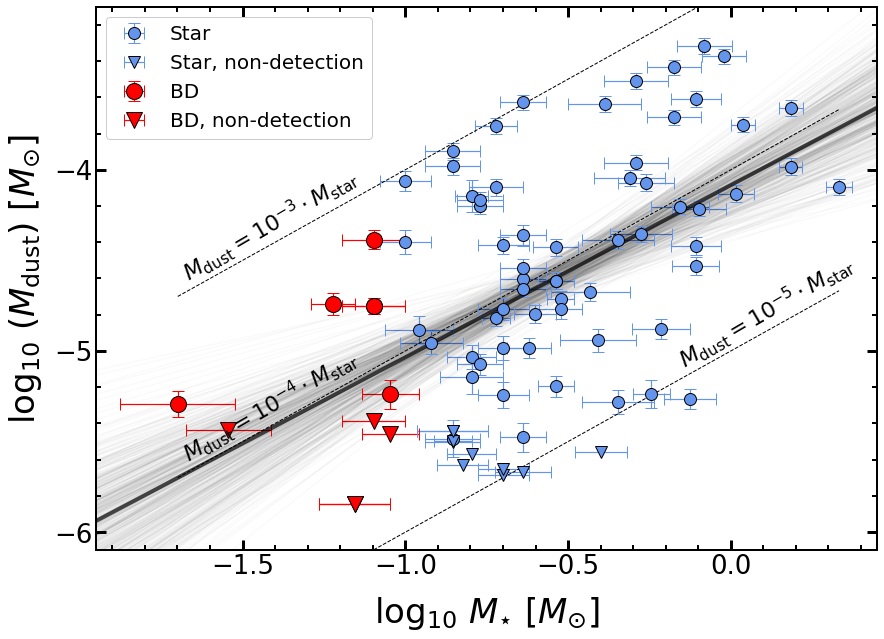}}
  \caption[Lupus disk demographics: stellar mass vs. total disk mass.]{Relation between the stellar mass and the dust disk mass for the BD and stellar populations in Lupus, using the dust temperature dependence with stellar luminosity from \cite{andrews+2013}.}
  \label{fig:mstarmdust4}
\end{figure}
\begin{figure}
  \resizebox{\hsize}{!}{\includegraphics{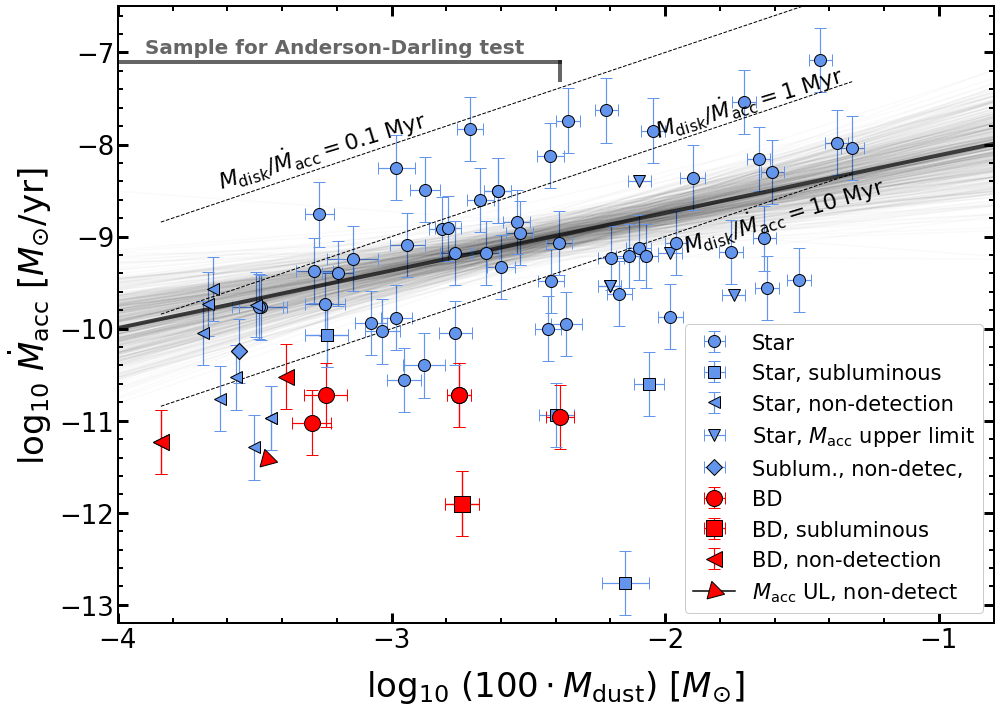}}
  \caption[Lupus disk demographics: mass accretion rate vs. disk mass.]{Relation between the inferred disk mass of the source and the mass accretion rate onto the central object, using the dust temperature dependence with stellar luminosity from \cite{andrews+2013}, and assuming a gas-to-dust ratio of 100.}
  \label{fig:mdustmacc4}
\end{figure}

\subsection{Dust temperature from \cite{vanderplas+2016}}
\begin{figure}
  \resizebox{\hsize}{!}{\includegraphics{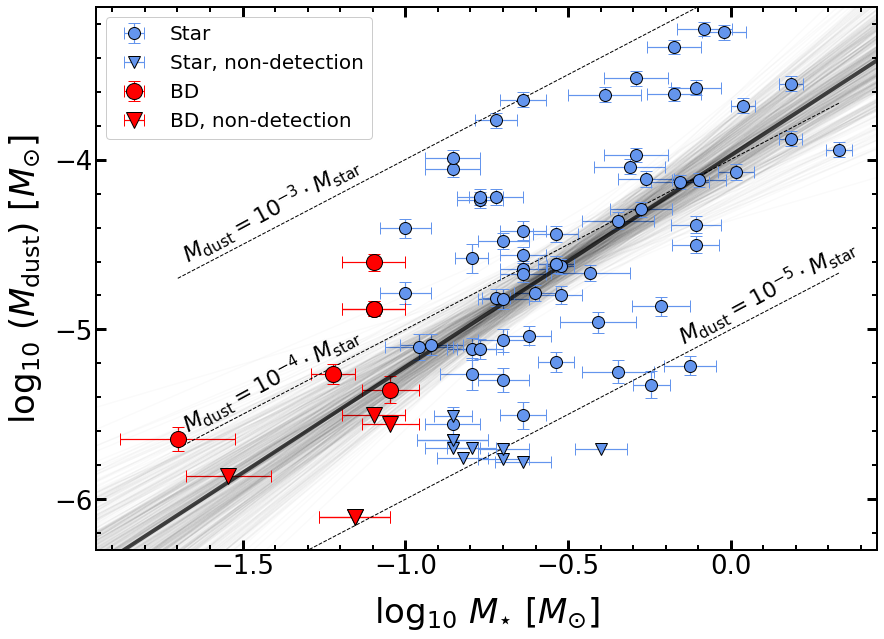}}
  \caption[Lupus disk demographics: stellar mass vs. total disk mass.]{Relation between the stellar mass and the dust disk masses for the BD and stellar populations in Lupus, using the dust temperature dependence with stellar luminosity from \cite{vanderplas+2016}.}
  \label{fig:mstarmdust5}
\end{figure}
\begin{figure}
  \resizebox{\hsize}{!}{\includegraphics{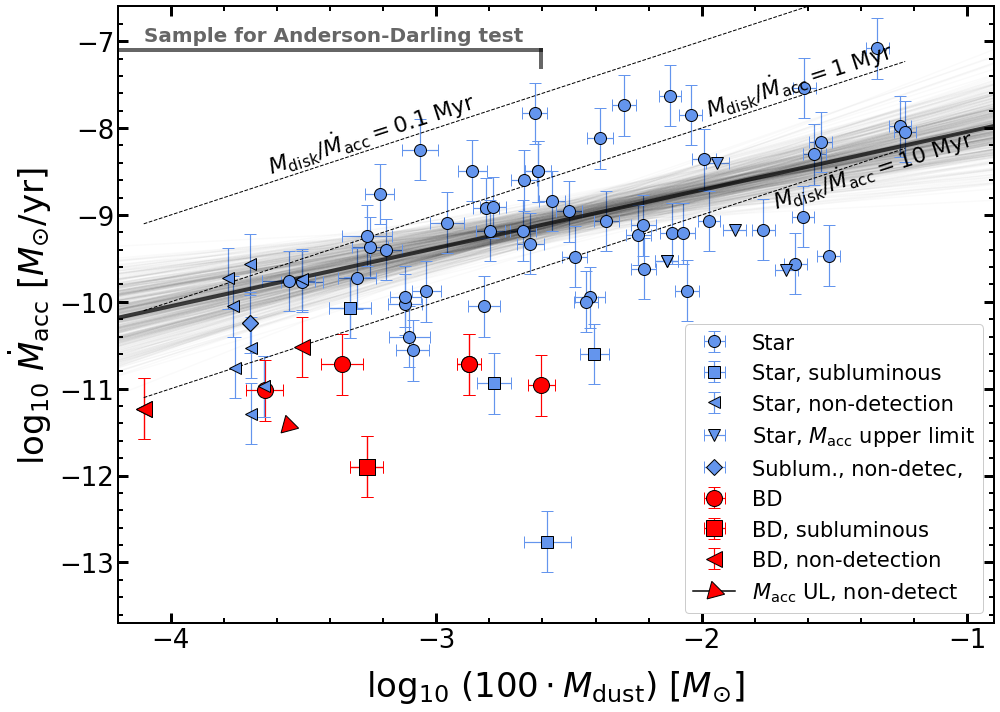}}
  \caption[Lupus disk demographics: mass accretion rate vs. disk mass.]{Relation between the inferred disk mass of the source and the mass accretion rate onto the central object, using the dust temperature dependence with stellar luminosity from \cite{vanderplas+2016}, and assuming a gas-to-dust ratio of 100.}
  \label{fig:mdustmacc5}
\end{figure}

\section{Results from fits}\label{sec:appendix_fits}
The results of the interferometric modeling for the disks that could be characterized in radius are included in this section. The plots shown for each fitted disk are: (top left) the corner figure composed of the 1D and 2D histograms of the parameter investigation, (top right) the model and observed visibilities (real and imaginary part as a function of K$\lambda$), (center right) the modeled brightness profile with its respective cumulative distribution, (bottom) and the observed, modeled and residuals reconstruction in the imaginary plane from the interferometric analysis. Detailed description of the panels can be found in the captions of Figures ~\ref{fig:j1545uvfit}, ~\ref{fig:j1545corner}, ~\ref{fig:j1545fluxes} and ~\ref{fig:j1545obsmodres}.

\subsection{Results for Sz~$102$}
\begin{figure*}
  \resizebox{\hsize}{!}{\includegraphics{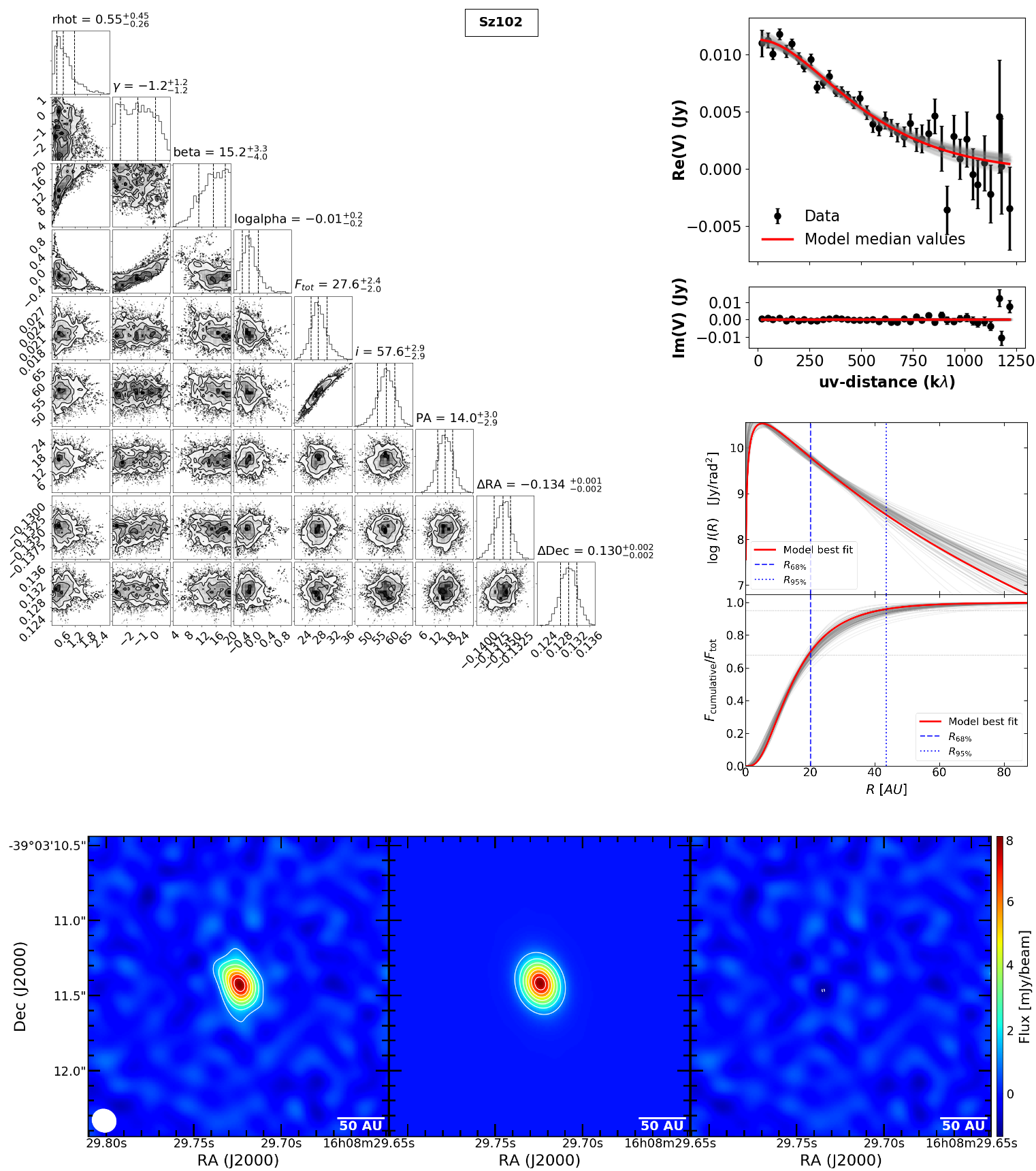}}
  \caption[]{
  Modeling results of the Sz~$102$ disk from ALMA observations. 
  A Nuker profile is fitted to the continuum emission distribution of the disk in the \textit{uv}-plane. 
  The panels show: the real and imaginary part of the observed and modeled visibilities, the 1D and 2D histograms of the Nuker model free parameters, the brightness emission and cumulative distributions, the $R_{68\%}$ and $R_{95\%}$ radii, and the fit results in the image plane.
  }
  \label{fig:fit_sz102}
\end{figure*}

\subsection{Results for V~$1094$ Sco}
\begin{figure*}
  \resizebox{\hsize}{!}{\includegraphics{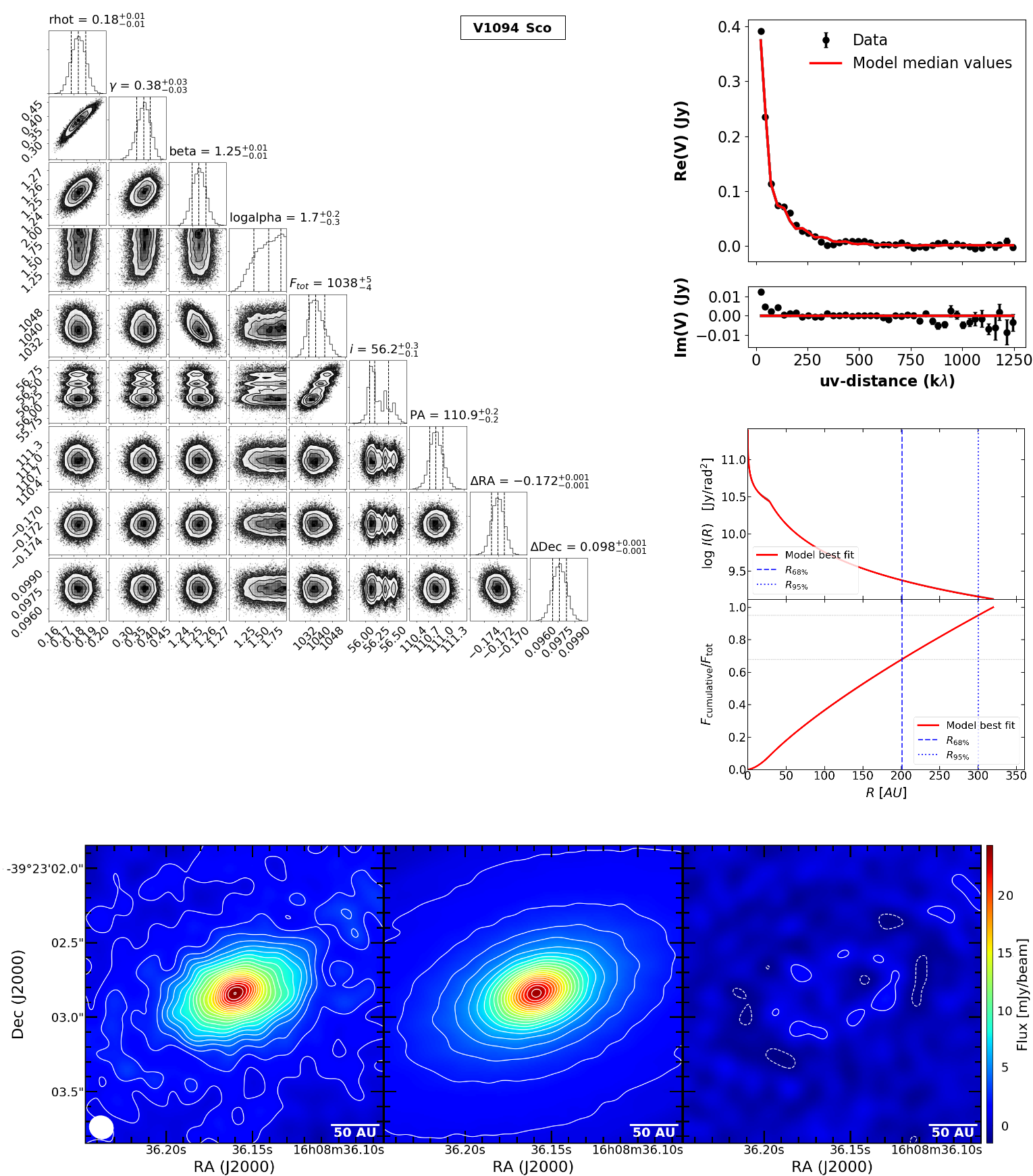}}
  \caption[]{
  Modeling results of the V~$1094$ Sco disk from ALMA observations. 
  A Nuker profile is fitted to the continuum emission distribution of the disk in the \textit{uv}-plane. 
  The panels show: the real and imaginary part of the observed and modeled visibilities, the 1D and 2D histograms of the Nuker model free parameters, the brightness emission and cumulative distributions, the $R_{68\%}$ and $R_{95\%}$ radii, and the fit results in the image plane. 
  The very extended emission of V$1094$ Sco was studied in detail in \cite{vanterwisga2018A}, and fitted with a more detailed function. For a comprehensive characterization of this disk we refer to that work. Nevertheless, our fit using the Nuker profile allows us to infer a characteristic radius consistent with the rest of the Lupus disk population.
  }
  \label{fig:fit_v1094}
\end{figure*}

\subsection{Results for GQ~Lup}
\begin{figure*}
  \resizebox{\hsize}{!}{\includegraphics{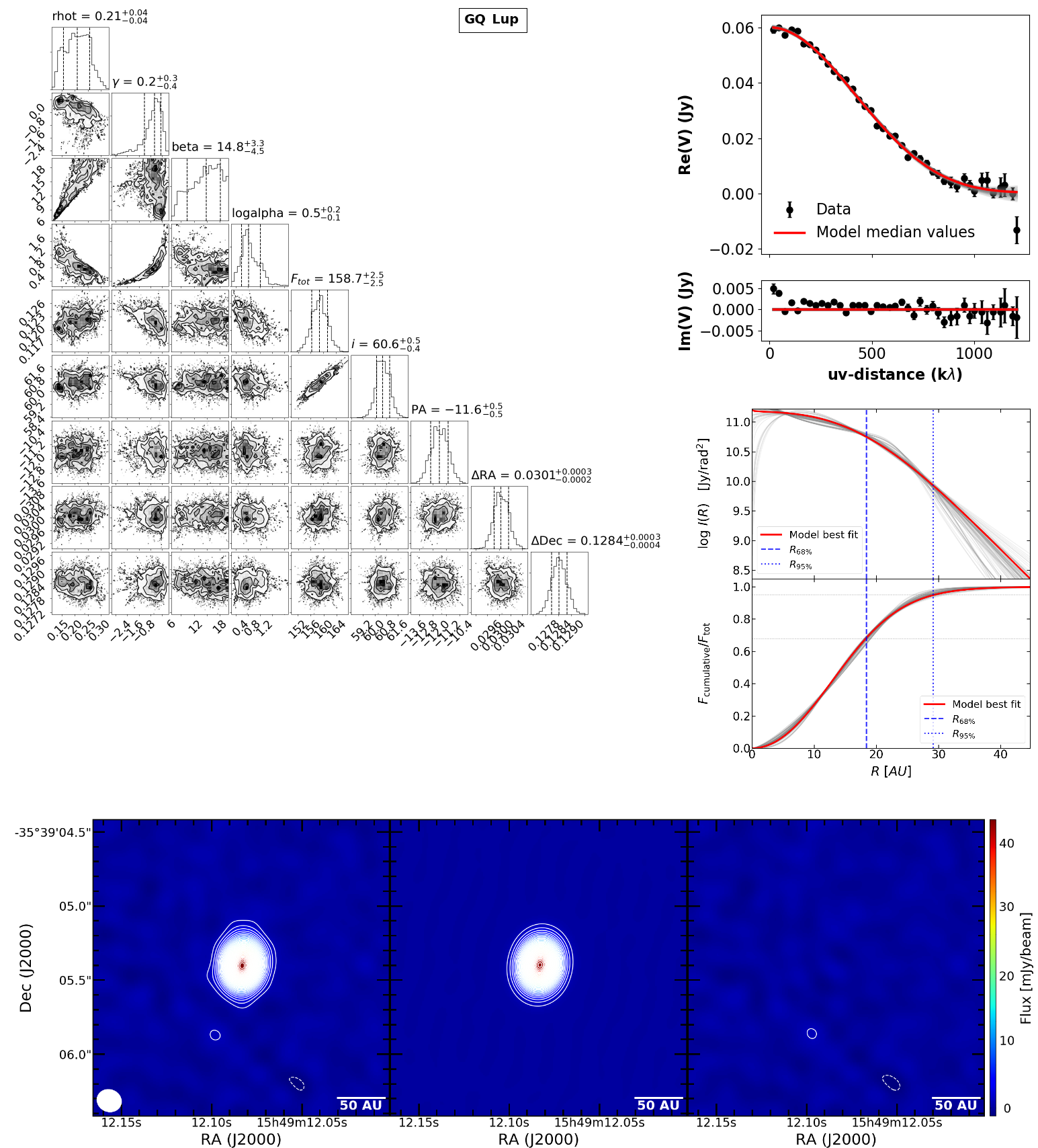}}
  \caption[]{
  Modeling results of the GQ~Lup disk from ALMA observations. 
  A Nuker profile is fitted to the continuum emission distribution of the disk in the \textit{uv}-plane. 
  The panels show: the real and imaginary part of the observed and modeled visibilities, the 1D and 2D histograms of the Nuker model free parameters, the brightness emission and cumulative distributions, the $R_{68\%}$ and $R_{95\%}$ radii, and the fit results in the image plane.
  }
  \label{fig:fit_gqlup}
\end{figure*}

\subsection{Results for Sz~$76$}
\begin{figure*}
  \resizebox{\hsize}{!}{\includegraphics{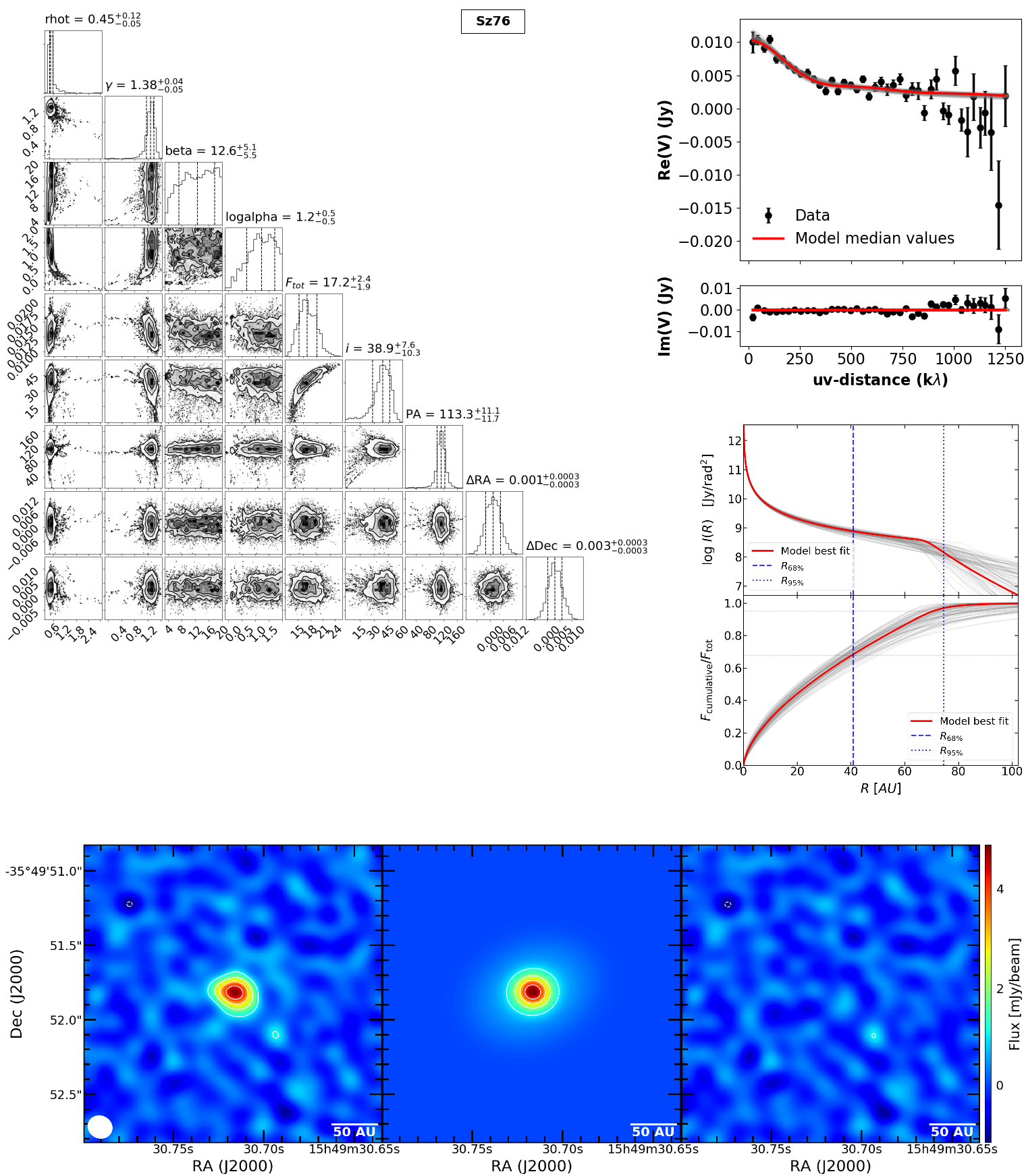}}
  \caption[]{
  Modeling results of the Sz~$76$ disk from ALMA observations. 
  A Nuker profile is fitted to the continuum emission distribution of the disk in the \textit{uv}-plane. 
  The panels show: the real and imaginary part of the observed and modeled visibilities, the 1D and 2D histograms of the Nuker model free parameters, the brightness emission and cumulative distributions, the $R_{68\%}$ and $R_{95\%}$ radii, and the fit results in the image plane.
  }
  \label{fig:fit_sz76}
\end{figure*}

\subsection{Results for RXJ~$1556.1$-$3655$}
\begin{figure*}
  \resizebox{\hsize}{!}{\includegraphics{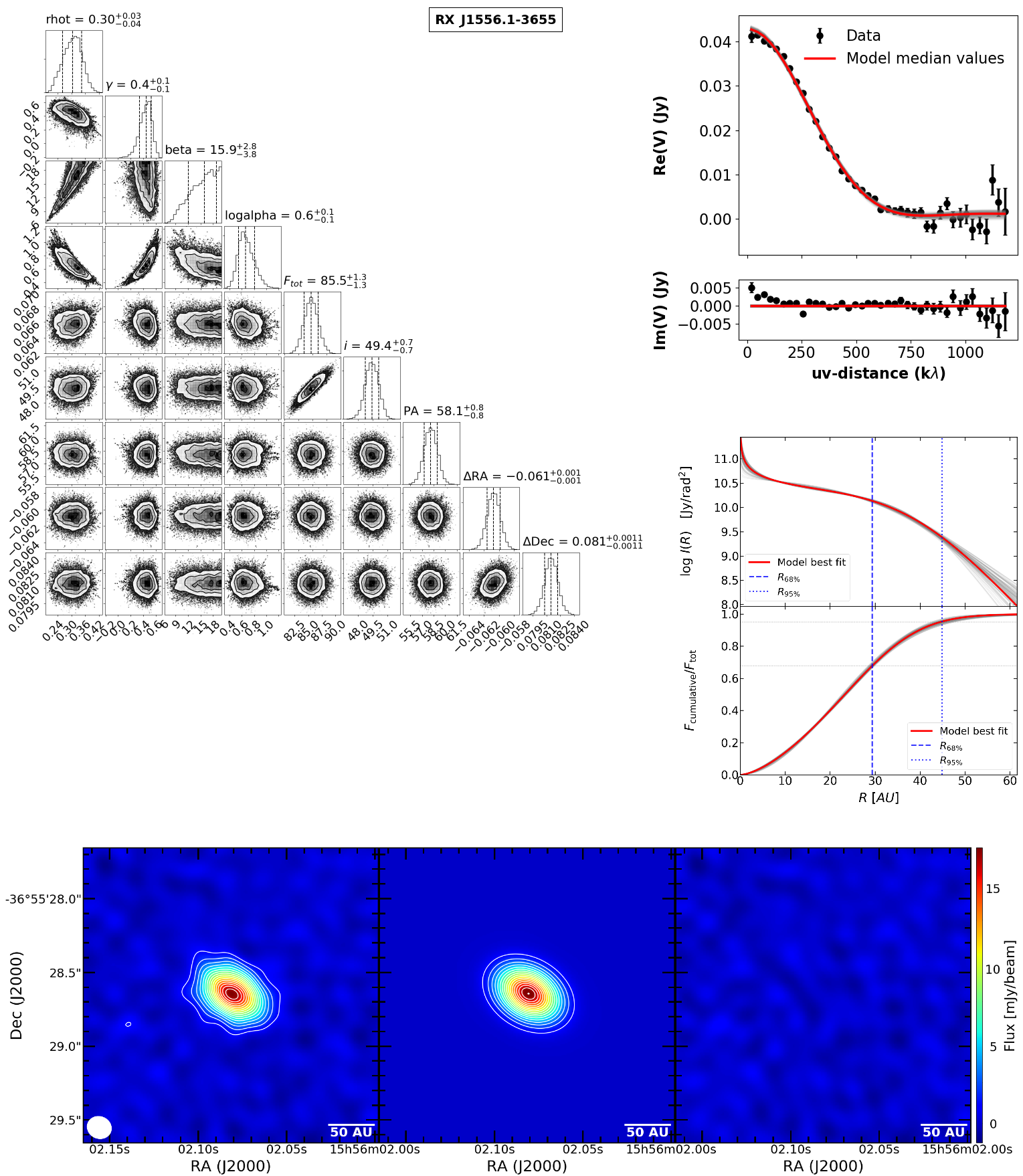}}
  \caption[]{
  Modeling results of the RXJ~$1556.1$-$3655$ disk from ALMA observations. 
  A Nuker profile is fitted to the continuum emission distribution of the disk in the \textit{uv}-plane. 
  The panels show: the real and imaginary part of the observed and modeled visibilities, the 1D and 2D histograms of the Nuker model free parameters, the brightness emission and cumulative distributions, the $R_{68\%}$ and $R_{95\%}$ radii, and the fit results in the image plane.
  }
  \label{fig:fit_rxj}
\end{figure*}

\subsection{Results for EX~Lup}
\begin{figure*}
  \resizebox{\hsize}{!}{\includegraphics{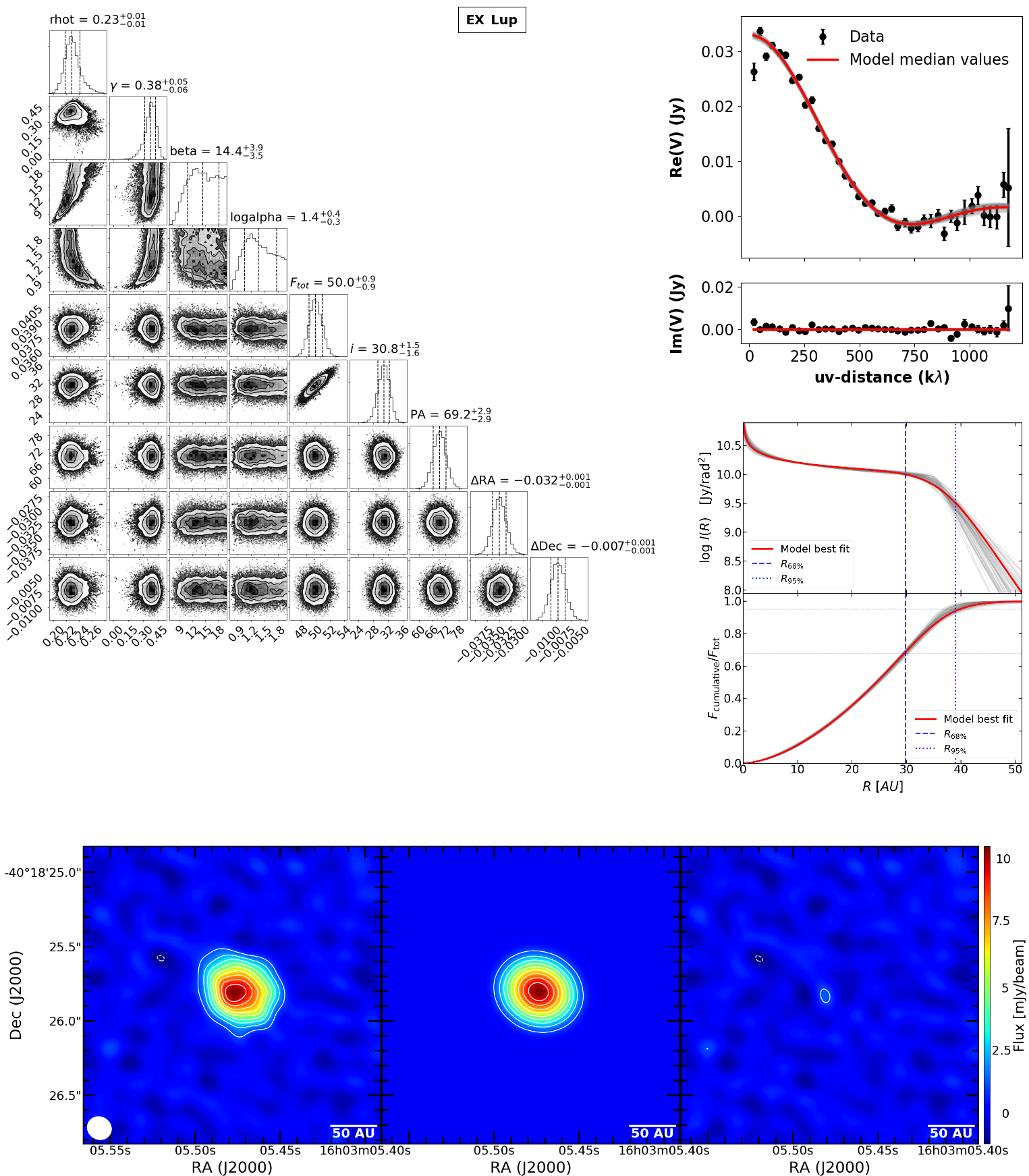}}
  \caption[]{Modeling results of the EX~Lup disk from ALMA observations. 
  A Nuker profile is fitted to the continuum emission distribution of the disk in the \textit{uv}-plane. 
  The panels show: the real and imaginary part of the observed and modeled visibilities, the 1D and 2D histograms of the Nuker model free parameters, the brightness emission and cumulative distributions, the $R_{68\%}$ and $R_{95\%}$ radii, and the fit results in the image plane.}
  \label{fig:fit_exlup}
\end{figure*}

\section{Accretion luminosity vs. scaled continuum flux}\label{sec:appendix_laccfcont}
\begin{figure}
  \resizebox{\hsize}{!}{\includegraphics{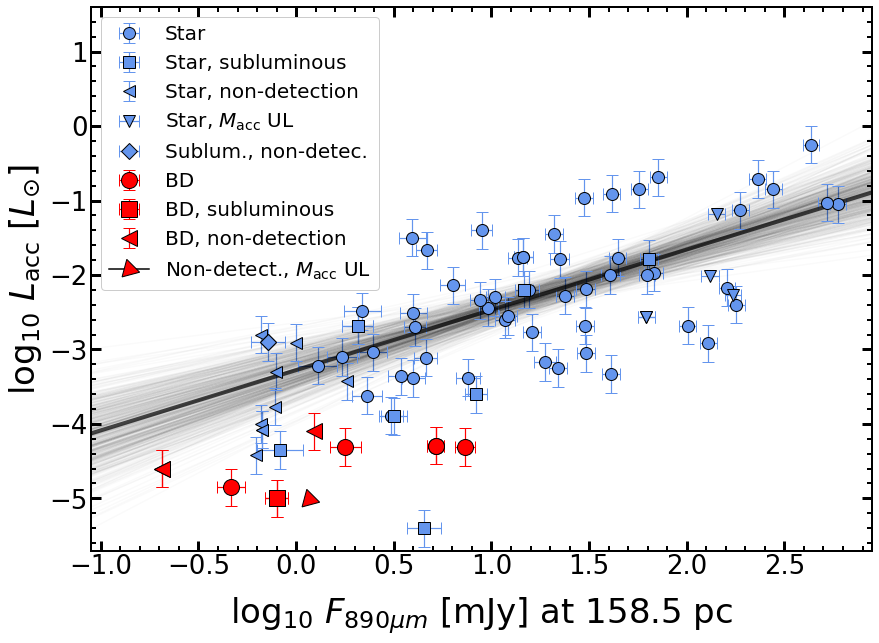}}
  \caption[Lupus disk demographics: mass accretion rate vs. disk mass.]{Relation between the accretion luminosity and the continuum flux of the disk (scaled to the average distance of Lupus). These are the observables used to infer the mass accretion rate and the disk mass as discussed in Section~\ref{sec:discussion}.}
  \label{fig:laccfcont}
\end{figure}

We inspected the relation between accretion luminosity and continuum flux (scaled to $158.5$ pc, the average distance to Lupus region) for the BD and stellar disks population. The demographic analysis is analogous to the different relations studied in Section~\ref{sec:discussion}. The accretion luminosity used is inferred from X-Shooter observations \citep{alcala2014,alcala2017} and corrected with the parallaxes from \cite{gaiacollaboration2018}; continuum flux at $890$ $\mu m$ is obtained from ALMA observations in Band 7 \citep[][, also this work]{ansdell+2016}. The relation is shown in Figure~\ref{fig:laccfcont}. Red datapoints represent the BD population, while blue datapoints indicate the stellar disks. The linear regression of the stellar population is shown in the figure, it has been obtained excluding non-detections and sub-luminous sources.

\begin{figure}
  \resizebox{\hsize}{!}{\includegraphics{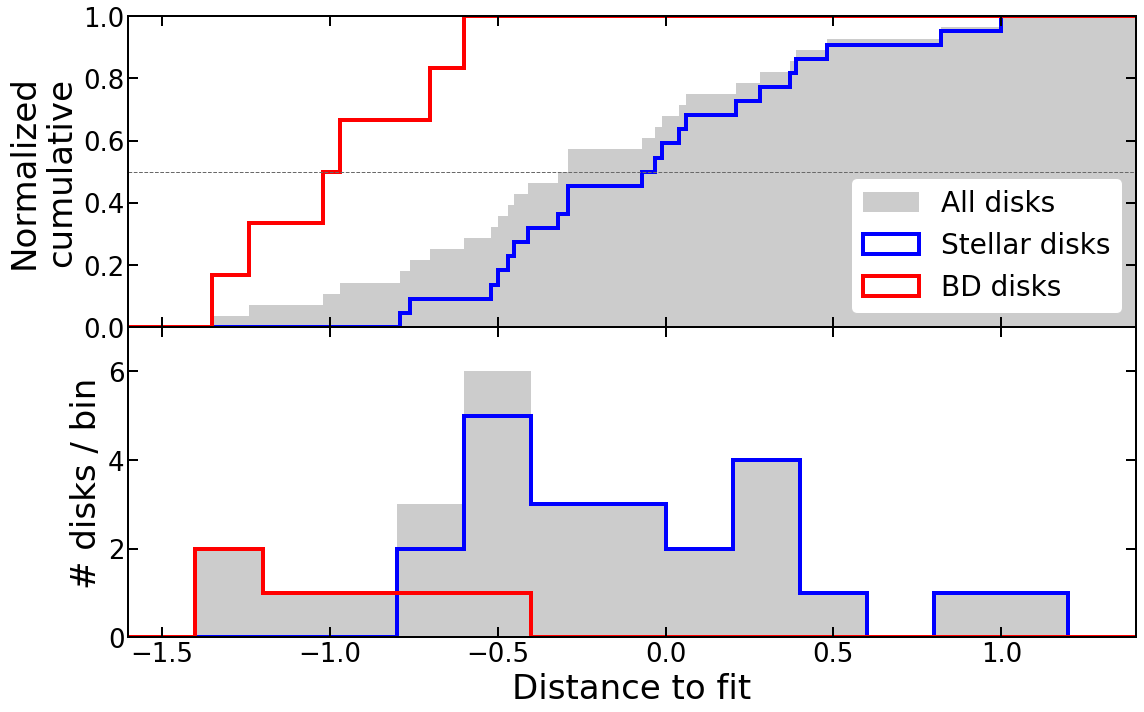}}
  \caption[Lupus disk demographics: mass accretion rate vs. disk mass.]{Histogram of the ratio between accretion luminosity and continuum flux of the disk (scaled to the average distance of Lupus).}
  \label{fig:populations_laccfcont}
\end{figure}

The stellar relation seems to describe poorly the behavior of the BD disk population. In order to verify this result, we build the histograms of the distance to the linear regression of the two populations (Figure~\ref{fig:populations_laccfcont}), and performed the Anderson-Darling test comparing the Bd and stellar populations. This test gives a $0.02\%$ of probability that the BD and the stellar disk populations are drawn from the same distribution. Therefore, the difference of accretion between BD and stars is statistically significant based only on observed quantities.

\end{appendix}

\end{document}